\documentclass[manuscript]{aastex}

\usepackage{natbib}
\usepackage{epsfig}
\usepackage[update,prepend]{epstopdf}
\usepackage{etex}
\reserveinserts{18}
\usepackage{morefloats}

\newcommand{\ik}{{\it Kepler~}}

\newcommand{\pv}{{\it p}-value~}
\newcommand{\pvt}{{\it p}-value}

\shorttitle{TTV catalog}
\shortauthors{Holczer et al.}

\begin{document}

\title{TRANSIT TIMING OBSERVATIONS FROM {\it KEPLER}. IX.
CATALOG OF THE FULL LONG-CADENCE DATA SET}

%% Use \author, \affil, and the \and command to format
%% author and affiliation information.
%% Note that \email has replaced the old \authoremail command
%% from AASTeX v4.0. You can use \email to mark an email address
%% anywhere in the paper, not just in the front matter.
%% As in the title, use \\ to force line breaks.

\author{Tomer Holczer\altaffilmark{1},
Tsevi Mazeh\altaffilmark{1},
Gil Nachmani\altaffilmark{1},
Daniel Jontof-Hutter\altaffilmark{2},
Eric B. Ford\altaffilmark{2,3},
Daniel Fabrycky\altaffilmark{4}
Darin Ragozzine\altaffilmark{5},
Mackenzie Kane\altaffilmark{5}, and
Jason H. Steffen\altaffilmark{6}
%Jason F. Rowe\altaffilmark{2} et al.
}
\altaffiltext{1}{ School of Physics and Astronomy, Raymond and
Beverly Sackler Faculty of Exact Sciences, Tel Aviv University,
Tel Aviv 69978, Israel}
\altaffiltext{2}{Department of Astronomy and Astrophysics, The Pennsylvania
State University, 525 Davey Laboratory, University Park,
PA 16802, USA}
\altaffiltext{3}{Center for Exoplanets and Habitable Worlds, 525 Davey Laboratory, The Pennsylvania State University, University Park, PA, 16802, USA}
\altaffiltext{4}{Department of Astronomy and Astrophysics, University of
Chicago, 5640 South Ellis Avenue, Chicago, IL 60637, USA}
\altaffiltext{5}{Department of Physics and Space Sciences, Florida Institute of Technology, 150 West University Boulevard, Melbourne, FL 32901, USA}
\altaffiltext{6}{CIERA, Northwestern University, 2145 Sheridan Road, Evanston, IL 60208, USA jsteffen@fnal.gov}
%\altaffiltext{2}{NASA Ames Research Center, Moffett Field, CA 94035, USA}

\begin{abstract}
We present a new transit timing catalog of $2599$ {\it Kepler} Objects of Interest (=KOIs), using the PDC-MAP long-cadence light curves that include the full seventeen quarters of the mission (ftp://wise-ftp.tau.ac.il/pub/tauttv/TTV/ver$\_$112).
The goal is to produce an easy-to-use catalog that can stimulate
further analyses of interesting systems.
For $779$ KOIs with high enough SNRs, we derived the timing, duration and depth of 69,914 transits. For 1820 KOIs with lower SNR, we derived only the timing of 225,273 transits. 
After removal of outlier timings,  we derived various statistics for each KOI that were used to
indicate significant variations. 
%+++++++
Including systems found by previous works, we have detected 260 KOIs which showed significant TTVs with long-term variations ($>$100 day), and another fourteen KOIs with periodic modulations shorter than 100 day and small amplitudes.
For five of those, 
the periodicity is probably due to the crossing of rotating stellar spots by the transiting planets.
\end{abstract}

%% Keywords should appear after the \end{abstract} command. The uncommented 
%% example has been keyed in ApJ style. See the instructions to %authors
%% for the journal to which you are submitting your paper to %determine
%% what keyword punctuation is appropriate.

\keywords{planetary systems–--planets and satellites: detection–--techniques: miscellaneous---technique: photometric}

%============
\section{Introduction}
%============
More than four years of almost uninterrupted performance of the \ik mission produced about 190,000 light curves with high precision, leading to the discovery of  more than 8000
planet candidates (=KOIs)\footnote{http://exoplanetarchive.ipac.caltech.edu/} that show periodic shallow transits caused by small eclipsing objects. 
Many of the KOIs display 
shifts of the transit timings (O-Cs) relative to a strict periodicity, which is expected if the transiting planets were to move on a Keplerian orbit. 
These O-Cs, sometimes called transit time variations (=TTVs), can indicate a
dynamical interaction with additional objects in the system, as
was predicted by the seminal works of  \citet{holman05} and
\citet{agol05}. 

Indeed, TTVs turned out to be a crucial tool
in the study of systems with known multiple transiting planets
\citep[e.g.,][]{holman10, cochran11, lissauer11a, fabrycky12, ford12a,
steffen12b, weiss13, xie13,xie14, yang13,hutter14, masuda14, ofir14, agol15}.
Furthermore, observed TTVs may indicate extra {\it non-transiting} planets through their dynamical interaction with the transiting ones
​\citep{ballard11,nesvorny12, nesvorny13, nesvorny14, dawson14}.

Therefore, it can be useful to perform a systematic TTV search
of all KOIs, as was done by \citet{ford11,ford12b} 
and \citet{steffen12a} at the early stages of the mission. 
As \ik released more data, additional systematic analyses were performed, using the longer light curves that became available \citep[see][]{mazeh13, szabo13}. Based on the first 12 \ik\ quarters, a global analysis of all KOIs, including the timing of the transits, was published recently by \citet{rowe15}. In a follow up publication \citet{rowe_thompson15} listed 
258 
KOIs with significant TTVs, based on the first 16 quarters.

The \ik mission in its original mode of operation has been terminated after seventeen quarters, and is now on its K2 mode \citep{howell14}, and we do not expect any additional \ik TTVs for the KOIs identified during the original mission.
Thus, here we analyze the whole data set of the mission and derive a complete catalog of the transit timings. Following the approach of \citet{mazeh13}, we present here an analysis of 2599 KOIs, based on all 17 quarters of the \ik data. 
The goal is to produce an easy-to-use catalog that can stimulate further analysis of interesting systems and a statistical analysis
of the Kepler KOIs with significant long-term TTVs.

After presenting the details of our pipeline and the catalog
itself in Section~\ref{analysis}, we derive in Section 3 a few statistical
characteristics of the timing series of each KOI which
can identify significant variations.
Sections 4 and 5 list and display 274
systems with significant TTVs, and
% and  Section~\ref{comments} comments on some interesting systems, in particular the ones for
%which the derived TTVs could be of non-physical origin.
Section~\ref{discussion} summarizes and discusses briefly the potential of the catalog.

%=========================
\section{Analysis of the transit light curves}
\label{analysis}
%=========================

The analysis presented here is based on the list of $4690$ KOIs in the 
NASA Exoplanet Archives,\footnote[1]{http://exoplanetarchive.ipac.caltech.edu/} as of 2013 November 23rd, ignoring KOIs listed as false positives.
We did not analyze 2091 KOIs, for which at least one of the following is true:

\begin{itemize}
\item The folded light curve did not display a significant transit, either because the folded transit's SNR
(defined as the transit depth of the model, divided by the median
 uncertainty of the individual points of the light curve, and multiplied by the square root of the number of measurements at the folded transit, including its ingress and egress)
was smaller than 7.1, similar to the criterion used by \citet{batalha13}, 
or where  the \pv of the transit model exceeded $ 10^{-4}$, using an  $\mathcal{F}$-test relative to the no-transit assumption.

\item The transit depth was larger than $10\%$; those KOIs were ignored in order to disregard eclipsing binaries in our analysis, with the price of leaving out some ``legitimate" transits such as large planets around M-stars.

\item The orbital period $> 300$ day;
those KOIs were ignored due to too few transits for a significant TTV analysis.

 %because there were too few transits for a significant TTV analysis.
 
\item KOIs identified as EBs, either listed in the Villanova eclipsing binary 
catalog,\footnote{http://keplerebs.villanova.edu/} as of 2014 July, or by  \citet{mcquillan13}.

\item KOIs identified by this study as false alarm, listed in Table~1, with some evidence for stellar binarity or pulsation.
\end{itemize}

Following these cuts we were left with $2599$ KOIs. We started by folding the PDC-MAP {\it Kepler} 
long-cadence\footnote{ftp://archive.stsci.edu/pub/kepler/lightcurves/tarfiles} data, with the BJD$_{TDB}$ timings, using the ephemeris of NASA Exoplanet Archive,
in order to obtain a good template for the transit light curve (see below for
details). We used the best-fit transit model to measure the timing of each individual transit (=TT)  and derived
its O-C---the difference between the TT and the expected time,
based on a linear ephemeris. As in \citet{mazeh13}, for KOIs with high enough SNR (see below), the TT derivation was performed while
allowing the duration and depth of each transit to vary. 

\subsection{Detrending}

The first step of our analysis was finding the continuum around each transit, ignoring the points in or near the transit itself, up to 0.7 transit durations around the expected timing of the transit center. Looking at a more extended region, up to two durations around the expected transit center, we fitted six different polynomials of degrees one to six to that region. The best fit was chosen as the one with the highest degree for which the \pv of all the 
 $\mathcal{F}$-tests with regard to polynomial fits of lower degrees was lower than $10^{-3}$. Finally, we added this polynomial back to the data during transit
and divided the light curve in the entire region by that polynomial.

\subsection{The transit model}
%--------------------------------------------

After detrending, all transits were folded with the transit period and fit with a transit model. As in \citet{mazeh13}, we used three different model templates to fit the folded data: a \citet{mandel02}, a Legendre polynomial and a Fermi function template model. 
We computed these three models for each KOI, and chose the model with the lowest
$\chi^2$ value as the transit template. The Mandel-Agol model was averaged to fit the long exposures of the mission, and used a  quadratic
limb-darkening law of two free parameters. Because we were interested only in getting the best template to fit the light curve and not in analyzing the physical parameters of the system, we accepted limb-darkening parameters even if they were out of the range allowed by the  theory of stellar atmospheres. 

Because of the astrophysical basis of the Mandel-Agol model---in contrast to the
other two which were merely mathematical heuristics, we preferred
the Mandel-Agol model whenever it gave a good enough fit.
Hence, we chose the Mandel-Agol model even in cases where its
rms~exceeded those of the other two models by up to 7\%. More details about the three models can be found in \citet{mazeh13}.

For most KOIs, the pipeline selected the Mandel-Agol
model (2579 KOIs). The pipeline chose the Legendre-based model when there was a
significant asymmetry in the folded light curve of the transit
(17 KOIs). The Fermi-based model was selected only when the SNR of
the folded light curve was too low (3 KOIs).

%----------------------------------------------------------------
\subsection{Deriving the timing, duration and depth of each transit}
%----------------------------------------------------------------

In order to derive the timing of a specific transit, we first searched through a grid of timings around the expected transit time, with a resolution of either one minute, or the estimated transit time uncertainty divided by five---whichever is lower. The transit time uncertainty was approximated to be (100 minute)/(SNR of individual transit) 
\citep[see also Figure~\ref{fig:errors_vsSNR}]{mazeh13}. For each time shift on the grid we fitted the data with the KOI's transit model, while keeping the duration and depth fixed. The grid point with the lowest  $\chi^2$ served as a first guess for the transit time.  

In our next step we divided the transits into two groups.
Group 1 consisted of KOIs whose transits had a duration longer than 1.5 hr and an SNR per transit larger than 10. (The SNR per transit was
 defined as the transit depth of the model divided by the median
 uncertainty of the individual points of the light curve and multiplied by the square root of the typical number of measurements during one transit, including its ingress and egress.)
Group 2 consisted of all other KOIs, which did not follow at least one of these two criteria. For both groups we performed a fine search  for the best estimate of the transit timing, using \begin{footnotesize}MATLAB's FMINSEARCH\end{footnotesize} function, 
which is based on the Nelder-Mead simplex direct search,
 allowing the duration and depth of the transit to vary in group 1, while in group 2 only the timing was varied.

For KOIs of group 1, 
we define the TDV and TPV of each transit as the {\it relative} change of the duration and depth found for that transit, respectively, with respect to the duration and depth of the transit model. 
Deriving the template depends on the folded light curve, which in turn is based on the adopted transit timings. Therefore, fitting the model was performed only after the first  derivation of the timings for each KOI, and then we derived a new set of timings with the new improved template. This process converged after three iterations.

We estimated the uncertainties of the three quantities (when available) from the inverted Hessian matrix, calculated at the identified minimum. The uncertainty of each individual {\it Kepler} measurement of a KOI was derived from the scatter of the light curves around the
polynomial fit before and after all transits of that planet. 
When the
Hessian matrix turned out to be singular, we assigned an uncertainty
that was equal to the median of the other uncertainties derived for
the KOI in question. 
Whenever this was the case, we marked
the error with an asterisk in the table of transit timings (see Table \ref{tab:TTV}).

In order to verify the obtained uncertainties for the transit
timings, we computed for $2339$ KOIs, which had at least
7 transit timings,
the scatter of their O-C values, $s_{\rm O-C}$,
defined as 1.4826 times the median absolute
deviation (MAD) of the O-C series.  
%
%after outlier removal,
%
%(see Figure~\ref{MADvsMed_error}),
We then compared this quantity with the typical error of each KOI, defined as the median of its timing uncertainties---$\overline{\sigma}_{\rm TT}$. 
These two parameters should not be sensitive to timing and uncertainty outliers (see Table \ref{tab:TTV}).

\begin{figure}[t]
\centering
\resizebox{12cm}{10cm}
{\includegraphics{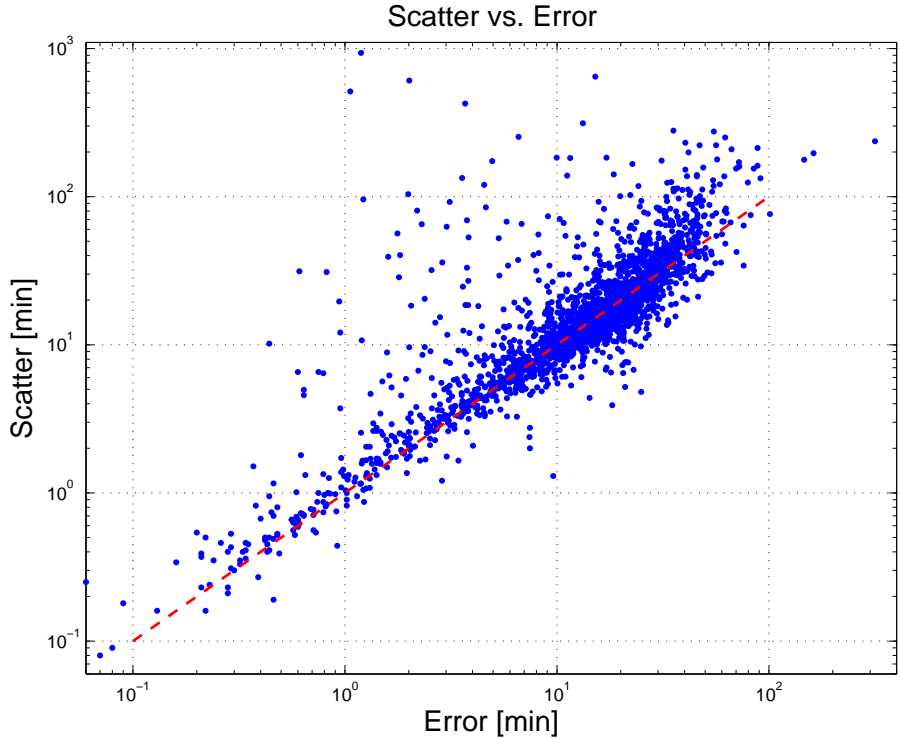}}
%\totalheight=0.5\textheight
\caption{Scatter of the derived O-C timings as a function of their typical uncertainty for $2339$ KOIs with at least 7 transit timings. The dashed line is the locus of points for which the scatter is equal to the typical error.
}
\label{MADvsMed_error}
\end{figure}

\begin{figure}[t]
\centering
\resizebox{12cm}{10cm}
{\includegraphics{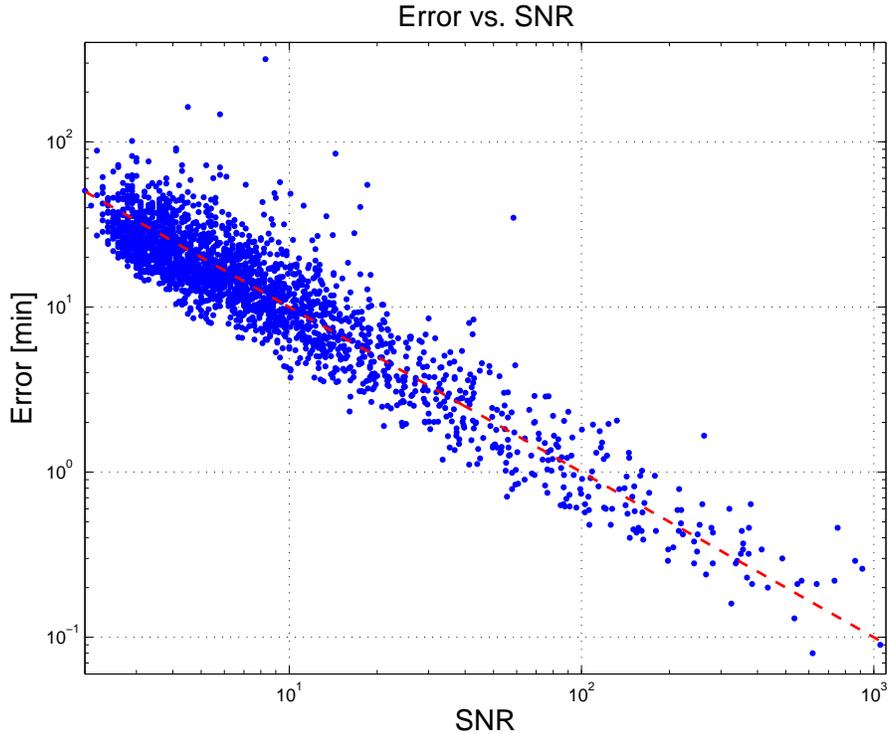}}
%\totalheight=0.5\textheight
\caption{Typical transit timing uncertainty as a function of the 
 typical SNR of a single transit for each KOI.
The dashed red line represents $\overline{\sigma}_{\rm TT}=
(100\, {\rm min})/$SNR.}
\label{fig:errors_vsSNR}
\end{figure}

We expected the estimated scatter and the mean uncertainty values to be
similar for systems with no significant TTV. This was indeed the
case for most KOIs, as seen in Figure~\ref{MADvsMed_error}.
KOIs with O-C scatter substantially larger than the typical timing uncertainty had significant TTVs (see below for a detailed analysis of those systems).

%---------------------------
% Figure 1 %
%---------------------------
%-------------------------------------------------------------

\citet{mazeh13} found that for each KOI $\overline{\sigma}_{\rm TT} \sim$
(100 minute)/SNR. Figure \ref{fig:errors_vsSNR} shows that this relation still holds.

%------------------------
% Figure 2 %
%---------------------------
%-------------------------------------------------------------

%----------------------------------------------------------------
\subsection{Outliers and overlapping transits}
%----------------------------------------------------------------

We labeled a transit as an outlier if it did pass one of the following five tests:

\begin{itemize}

\item  Significance---We performed an $\mathcal{F}$-test to compare the transit model, with the identified timing (and duration and depth when appropriate) found, against
a constant flux assumption (no transit at all). We rejected all transits with an $\mathcal{F}$-test \pv larger than 0.025. We have found 60,856 such transits. 

\item{Moving local outlier}---We considered each transit with respect to its ten neighboring transits, fitting a parabola through five O-Cs before and another five after the transit. We labeled a transit timing as an outlier if its O-C deviated from the parabolic model by more than five times the parabola rms~residuals plus three times the median error of the O-C of that KOI. The scatter was calculated as 1.4826$\times$MAD of the residuals. We have found 3336 such transits. 

\item{Global TTV outlier}---A transit timing was labeled as an outlier in one out of two cases.  The first was when its O-C deviated from the median of the O-Cs of that KOI by at least five times their scatter plus three times their median error. The second case was when its O-C deviated from the median of the O-Cs of that KOI by more that some factor $\eta$ times the scatter of the O-Cs, $\overline{\sigma}$.  The factor $\eta$ we used depended on the total number of derived O-Cs of that KOI, $N$, such that we did not expected even one measurement to fall at random at a distance larger than $\eta\overline{\sigma}$ from the median of the O-Cs of that KOI, if the O-Cs were randomly distributed. We used $N\leq100$, even for systems with larger number of points.
 We have found 23,030 such outlier transits.

\item{TDV global outlier}---A transit was labeled as an outlier when its TDV deviated from zero by at least five times the scatter of all TDVs of that KOI plus three times their median error.  We have found 195 such transits. 

\item{TPV global outlier}---A transit was labeled as an outlier when its TPV deviated from 
zero by at least five times the scatter of all TPVs of that KOI plus three times their median error.   We found 192 such transits. 

\item{TTV, TDV or TPV uncertainty outlier}---A transit was labeled as an outlier when either its TTV, TDV or TPV uncertainty deviated from the corresponding median of the other transit uncertainties of that KOI by at least eight times their scatter. We have found 2851 such transits.   
\end{itemize}
All six types of outliers are noted in the catalog of the transit timings (Table~\ref{tab:TTV}; see below), by numbers that reflect their appearance in the list above. The Global TTV outlier procedure is an efficient approach to remove incorrect identifications, assuming there are no significant TTVs. For KOIs with significant TTVs this assumption does not hold, and therefore we sometimes had to undo this stringent outlier procedure for selected measurements that were obviously correct. We undid a total of 68 TTVs in 36 KOIs, which led to a total of 22,962 measurements that were labeled as global TTV outliers.

We also marked and ignored transits of multiple-planet systems when the transits of two or more planets ``overlapped". 
A transit was considered to be overlapped by another planet whenever the difference between their expected timings was smaller than twice the duration of that transit plus one duration of the interfering transit. We used twice the duration of the analyzed transit since that is the range over which the detrending fit was performed.
However, we did not mark transits that were overlapped by another transit, if the   overlapping transit had a transit duration multiplied by its depth smaller than 0.25 of the corresponding product of the analyzed transit. 
It seemed that in those cases the overlapping transits did not induce large TTV errors.  We have found a total of 11,910 overlapped transits. 

After marking the outliers and overlapped transits, we were left with 150,051 transits with derived timings only,  and 64,121 ones for which we derived the timing, duration and depth. 

%-----------------------------------
\subsection{The catalog}
%-----------------------------------
We present our results in two  tables, available 
at:\\ 
ftp://wise-ftp.tau.ac.il/pub/tauttv/TTV/ver$\_$112.\\
Table~\ref{tab:ephemeris} lists the modified ephemerides of the
KOIs, based on our analysis, together with the durations and depths of their
transits, derived from the folded light curve. The transit duration and depth are quoted in hours and ppm, respectively. 
For each KOI we also list the SNR of the transit (see above), and the 
scatter ratio, defined as the ratio between the data scatter inside the transit, relative to the adopted model, and the scatter of the data outside the transit. The two scatters are derived for the folded light curves. 
Table~\ref{tab:TTV} lists our derived O-Cs,
relative to our modified ephemerides,
 for 295,187 transits of 2599 KOIs. Of those, duration and depth changes, in
units of the transit model duration and depth, are given
for 69,914  transits of 779 KOIs with $ {\rm SNR} > 10$, and a transit duration longer than 1.5 hr. 

%==============================
%Section 3
\section{Identifying KOIs with significant long-term TTVs}
\label{interesting}
%==============================

As the main focus of this study is the TTVs of the KOIs, the
next sections concentrate on the analysis of the derived O-Cs, leaving the analysis of the duration and depth variations for another study. In order to identify KOIs with significant long-term TTVs, we computed a
few statistics, listed in Table~\ref{tab:TTV_master}, of the O-C series  for
2339 KOIs with more than six timing measurements:
\begin{itemize}

\item
The ratio between the median error, $\overline{\sigma}_{\rm TT}$, and the scatter of the O-Cs,  $s_{\rm O-C}$,
(see  discussion above and Figure~\ref{MADvsMed_error}). 
The modified ratio of these two figures,
$s_{\rm O-C}$/$(1.48\overline{\sigma}_{\rm TT}$), 
squared and multiplied by the number of measurements (see Table~3), gave us a modified
 ${\chi}^2$ of the O-C series of each KOI. 
We used this figure to calculate a  ${\chi}^2$ \pv against the no-variation assumption,
listed in the table. Very low \pv due to high
values of  $s_{\rm O-C}$ relative to
$\overline{\sigma}_{\rm TT}$
might indicate a significant TTV, in particular because the MAD
statistic was less sensitive to outliers than the rms~of the O-Cs. 

\item
A modified power spectrum (PS) periodogram of the O-Cs, presenting for each frequency the energy contained in its fundamental and first harmonic. We identified the
highest peak in the periodogram and assigned a \pv to the associated periodicity in the data. This was
done by calculating $10^4$ modified PS periodograms for different
random permutations of the same O-Cs, and counting the number of permutations that had a PS peak higher than the peak of the real data. 
Table~\ref{tab:TTV_master} quotes the estimated period and its \pvt.

 \item
An 'alarm' $\mathcal{A}$ score of the series, following the  statistic of \citet{tamuz06}, which is
sensitive to the correlation between adjacent O-Cs. The value of
$\mathcal{A}$ is sensitive to the number of consecutive O-Cs with the
same sign, without assuming any functional shape of the
modulation \citep[see][for a detailed discussion]{tamuz06}. 
We assigned a false-alarm probability to the occurrence of the obtained score by calculating alarm scores for $10^4$ different random permutations of the same O-C series,
 and counting the number of permutations that had an alarm higher than the peak of the real data.
Table~\ref{tab:TTV_master} quotes the alarm score
and its \pvt.

\item
A long-term polynomial fit to the O-C series. A significant
polynomial fit usually indicates a long-term modulation with a time scale longer
than the data span. We searched for a fit with a polynomial with a degree lower than
four, chose the best polynomial and tested its significance with an $\mathcal{F}$-test.
Table~\ref{tab:TTV_master} quotes the best polynomial fit and its \pvt.
\end{itemize}

%===========================
\section{KOIs with long-term significant TTVs}
\label{long}
%===========================

Each KOI with any of the aforementioned statistics  yielding a \pv lower than $10^{-4}$ 
and a period longer than 100 day was identified as having a significant long-term TTV, provided 
%that it has passed our  "eye test", which confirmed that 
the variation seemed real and not caused by some artifact. Table~\ref{tab:TTV_Interesting} lists 260  KOIs with  long-term  significant  timing variation and summarizes their variability features, while 
Figures~\ref{TTV1}--\ref{TTV26} display their TTVs.
Of those, 73 KOIs did not pass the significance thresholds, but nevertheless have a relatively high significance and seemed to have a real long-term timing variation, and therefore were also included in the table.

Table~\ref{tab:TTV_Interesting} lists the KOI number, the orbital period of the transiting planet and the adopted model, either a Cosine function, ``Cos", or a polynomial ``Pol". For a cosine fit, we list the TTV period and its error, and the amplitude and its error.
For the cases with parabolic fits, we list an estimated figure for the amplitude of the variability, based on half the difference between the maximum and minimum values of the parabola at the times of the TTVs.
For both types of fitting, we list the scatter of the residuals relative
to the adopted fit (which is not plotted).  We also list the
number of TTV measurements, the multiplicity of the system and references to previous studies, when available.

To derive the TTV periods and amplitudes and their errors we fitted the TTVs with a periodic cosine function superimposed on a linear trend. This was performed by Markov chain Monte Carlo runs, each of which used an ensemble of MCMC
samplers \citep{goodman10,forman13}. 
We summarize the marginal posterior distribution
for each period by reporting its median value, along with an uncertainty based on a 68.26\% interval ranging from the 15.87th to the 84.13th percentile. 
Out of $260$ KOIs, $199$ showed
clear significant periodicity, with periods ranging from $100$ to over
$2000$ day, and amplitudes of $1$--$1470$ minutes.
%For each of these $199$ systems, we derived a fit to the TTVs (not
%plotted but given in the table), composed of a straight line,
%which could present a correction to the orbital period of the transiting
%planet, {\it together} with a cosine function with the
%best-found period and phase. 

In six special cases---KOI-142.01, -157.03, -417.01, -474.03, -984.01 and -2283.01, we fitted the TTVs with a straight line and {\it two}
different cosine functions, both identified by the MCMC runs. 
In these cases,  
Table~\ref{tab:TTV_Interesting} lists the two periods, and the scatter of the residuals after the removal of the linear trend and the first periodicity, and also after the removal of the linear trend and the two modulations. The secondary periodicity is very clear only in two cases, KOI-142 and -984, as can be seen by comparing the residuals after removing the first and both periodic modulations. Nevertheless, we suggest that the second modulations in the other four cases are also real, especially because our MCMC samplers clearly showed two periodicities.  

For $61$ KOIs, the cosine function fit was poor, and we therefore fitted instead a  simple parabola to the data. This probably meant that the time scale of the modulation was longer than
the time span of the data. 

The orbital periods listed in Table \ref{tab:ephemeris} for the $2599$ KOIs were corrected by the slope of the linear fit to the TTVs.
This approximation was good when there was no significant long-term TTV modulation. However, in the presence of a strong TTV periodic modulation, the linear fit by itself might not be good enough, and therefore might yield an inaccurate correction to the orbital period. Therefore, for each of the $199$ KOIs that showed a periodic modulation, we have corrected the orbital period using the slope of the linear part of the model in Table~\ref{tab:TTV_Interesting}, which included both a linear slope and a cosine modulation. Using the error of the linear slope from the MCMC model might underestimate the uncertainty of the derived orbital period. Instead, we used the difference between the orbital periods obtained by both methods (linear and linear + cosine models) as an error estimate. The new orbital periods and their estimated uncertainties are given in Table \ref{tab:periods}.

%============================
\section{KOIs with short-period TTVs}
%============================

Following the approach of \citet{mazeh13}, we identified 
10 systems with highly significant short-period TTV modulations, in the
range of 3--80 day. They were found by obtaining a PS peak with \pv lower than 3$\times$10$^{-4}$. 
The modulation amplitudes were relatively small, in the range of $0.07$--$80$ minutes,
and their detection was possible only due to the 
periodic nature of the signal and the long time span of the data relative to the periodicity.

For the sake of comparison, we included in this section plots for KOI-13.01 and -972.01 modulations, which were detected by \citet{mazeh13}, and our analysis showed the same modulation but with a lower significance. 
Another seven KOIs that were listed in  \citet{mazeh13} are not shown here as they now appear to be false positives and not planet candidates.
Two additional systems, KOI-895.01 and -1074.01, are shown because they exhibit TTVs induced by spot crossing events \citep{holczer15}. 

Figures~\ref{TTV_short1}--\ref{TTV_short5} show the PS
periodograms with their prominent peaks, and the phase-folded O-Cs of the 14 systems. Table~\ref{tab:short} lists the periods and
amplitudes found. 

As pointed out by \citet{szabo13} and discussed by \citet{mazeh13},
not all detected short-period modulations are due to physical
TTVs. An {\it apparent} TTV periodicity can be induced either
 by the long-cadence sampling of {\it Kepler}  (the stroboscopic effect), or by an interference with a periodic stellar activity.
To find the frequency of the presumed sampling-induced
periodicity, we used for each KOI its $P_{\rm orb}$ from Table
\ref{tab:ephemeris}, and the pertinent $P_{\rm samp}$. This was
about 29.424
minutes for the long cadence, the exact value taken to be the median of the
differences of the observed timings of that KOI.
We searched for a stellar spot periodicity using the autocorrelation technique
\citep[e.g.,][]{mcquillan13}, and, if present, checked whether its
frequency and/or one of its harmonics or aliases was equal to the TTV frequency. We
mark the pertinent frequencies in
Figures~\ref{TTV_short1}--\ref{TTV_short5} \citep[see][]{holczer15}. 

In each of the figures we marked the frequency of the highest peak, the  stroboscopic frequency (due to sampling) and the rotational frequency,  with its harmonics when relevant. 
For five systems, KOI-203.01, -217.01, -883.01, -895.01, and -1074.01, the rotational frequency and/or one of its harmonics coincided with the highest periodogram peak \citep[see also][]{holczer15}.
The periodicity in those cases could be induced by crossing the rotating stellar spots by the transiting planet.
 Two systems, KOI-883.01 and -13.01, showed a strong stroboscopic effect.

%============================
\section{TTV periodicities induced by an adjacent planet near a first-order resonance}
%============================
\label{multi}

As discussed in the introduction, we expect most TTV periodic modulations to be caused by a dynamical interaction with an adjacent additional planet, near first-order resonance in particular.  For those cases, we have prepared a theoretical infrastructure to calculate the expected TTV period, based on the orbital periods of the two planets
\cite[e.g.,][]{lithwick12,hadden14}. 
In the case that the adjacent planet is also a transiting one, we should be able to know its orbital period, and therefore to compare the predicted TTV periodicity with the observed one. Thus, we searched the known multiple KOIs systems to see if any additional planet candidate can account for the 199 TTV periodicities of Table \ref{tab:TTV_Interesting}.

We found 80 pairs of KOIs for which two KOIs resided in the same planetary system and at least one of them showed a clear periodicity as listed in Table~\ref{tab:TTV_Interesting}. 
Following \citet{xie13,xie14}, we checked for each of these KOI pairs whether it was close to a first-order resonance by calculating for each of them the normalized distance to the closest first-order resonance, $\Delta$:

\begin{equation}
 \Delta \equiv \frac{P_2}{P_1} \frac{j-1}{j} -1, 
\end{equation}
where $P_1$ and $P_2$ are the inner and outer orbital periods of the planets of the pair, respectively, and $j$ is the resonance number. 
 We set a cutoff at $\Delta = 0.1$, ignored the pairs with higher values, and found 75 pairs of KOIs that were close to some first-order resonance.

For those pairs we derived the expected super-period $P^j$, given by
\begin{equation}
  P^j \equiv \frac{1}{\left | j/P_2 - (j-1)/P_1 \right |} \ ,
\end{equation}
and compared it with the observed TTV periodicities (two if available, otherwise only one observed periodicity) of the pair.
We consider this derived super-period as a rough estimate, as the obtained orbital periods of the two transiting planets, $P_1$ and $P_2$, depend on their assumed TTV periodicities. We found 59 pairs, listed  in Table \ref{tab:multi}, for which at least one of the TTV periodicities was consistent with the derived super-period. The table includes the two orbital periods of the pair, 
the resonance found,  the distance to resonance, the calculated super-period, and the observed TTV periodicities with their uncertainties. 

Out of the 59 systems, we have found 26 pairs for which the two observed TTV periodicities agreed with each other and also agreed with the calculated super-period, one pair (KOI-250) for which the two observed TTV periodicities agreed with each other but did not agree with their calculated super-period, and 32 pairs for which only one TTV periodicity agreed with the calculated super-period (the other planet did not show any significant periodicity or its TTV period did not agree with the calculated super-period). 
In one case, KOI-157.03, the derived TTV period was the {\it second} periodicity found.
 
To summarize this section, out of the 199 KOIs with periodic TTVs, 
84 reside in systems where only one transiting planet is evident.
Another 39 KOIs are found in multi-planet systems without an obvious adjacent planet that can explain the detected TTV periodicity. Another 23 KOIs are in multi-planet systems, for which the period of one adjacent planet yields a super-period that is consistent with the observed TTV periodicity. Finally, 53  KOIs show a TTV periodicity which is anti-correlated with TTV modulation of another adjacent planet in the system. 
%++++++++++++++++++++++++++++ 
 %============================
\section{Comparison with the recent catalogs}
%============================
\label{rowe}
Before concluding, we compare our results with the published KOI catalog by \citet[][RoweI]{rowe15}, which also contains the O-C measurements for 5751 KOIs, and with \citet[][RoweII]{rowe_thompson15} catalog,\footnote{http://exoplanetarchive.ipac.caltech.edu/docs/Kepler$\_$KOI$\_$docs.html} which presents timing measurements for 258 KOIs that they identified as having significant TTV signals. (The criteria used were not given.)

A total of 2290 KOIs were analyzed by this paper and by either RoweI or RoweII. The reason is that RoweI and II analyzed an extended set of KOIs, including the new batch of identified planet candidates, while we considered the older list of KOIs released in 2013 November. Further, we excluded from the list a set of KOIs that we found to be EBs. Therefore the intersection of the two studies is only 2290 KOIs. 
 In those KOIs, RoweI obtained 275,551 transit timings, while we had a total of 287,428 measurements, and 208,650 timings that were not rejected as outliers. The TTV error estimates were very similar in both works.  

We now compare the results of RoweII with the KOIs we found to have significant TTVs (Tables \ref{tab:TTV_Interesting} and \ref{tab:short}). Out of the 258 KOIs RoweII found with significant TTVs, we have analyzed only 208, 
out of which 124 KOIs appeared in Table \ref{tab:TTV_Interesting}, 
another 4 KOIs appeared in Table \ref{tab:short}, 
and 80 KOIs that we did not identify as variables. 
On the other hand, we identified 137 KOIs as having significant variability (127/10  in Tables \ref{tab:TTV_Interesting}/\ref{tab:short}), which were analyzed by RoweI and were not marked as variables by RoweII. The three sets of timing series derived by the two analyses were carefully compared.

\begin{itemize}

\item{80 KOIs labeled significant by RoweII only}---we checked again these systems and found most of them not to be significantly variable, both by eye and by  our threshold statistics.
Only two KOIs (KOI-94.02 and -209.02) actually did (barely) pass our statistic threshold, but we had removed them from the list, as the data did not seem significant enough. 
%We found KOI-481.03 to show the strongest TTV signal however we do not think its significaunt enough to enter our table. We believe most of these KOIs were labeled due to power spectrum peak significance at low periodicites so that the TTV signal is not very significant by eye.

\item{127 KOIs labeled significant by this work only}---to check our analysis, we used RoweI timings and still found them to be significantly variable. Just to name a few: KOI-474.03, -841.01, -841.02, -1081.01, -1270.02, -1529.01, -1599.01, and -2061.01 
(see Figures 10, 15, 17, 18, 20, 21 and 25). 

\item
{124 KOIs labeled significant by RoweII and by this work}---in almost all cases the TTVs were comparable. Only KOI-806.03 displayed timing measurements that were significantly different in the two works.  The two sets of derived TTVs are presented in Figure~\ref{806_03}
\citep[see also][]{sanchis12}. RoweI and RoweII probably missed the actual timings of this KOI because of the large TTVs.  

%---------------------------
% Figure 3 %
%---------------------------
\addtocounter{figure}{+31}
\begin{figure}[t]
\centering
\resizebox{12cm}{10cm}
{\includegraphics{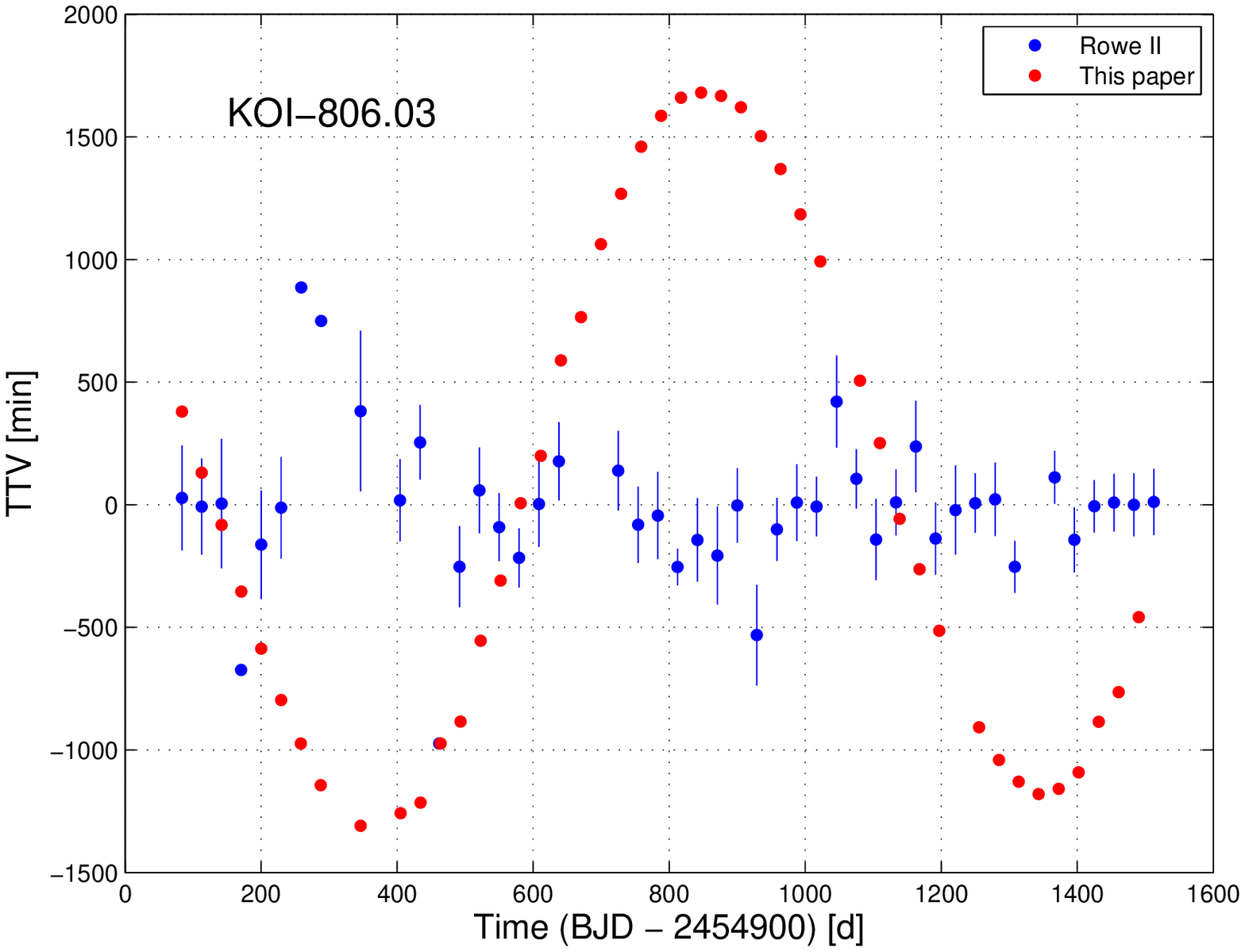}}
%\totalheight=0.5\textheight
\caption{TTV obtained by RoweII (blue) and by this work (red). The error bars of the red points are smaller than the size of their symbols. For each set, the  ordinate presents the derived timings, and the abscissa the deviations from the linear ephemeris. 
}
\label{806_03}
\end{figure}
\addtocounter{figure}{-32}
%-------------------------------------------------------------
\end{itemize}
 
The difference between the works may have resulted from the fact that this work was focused on the TTV variability, while both RoweI and RoweII were studying the general set of KOIs, with the goal of presenting a complete set of planet candidates with analyzed transits.

%============%
\section{Summary}
\label{discussion}
%============%

We presented here a new transit timing catalog of $2599$ {\it Kepler} KOIs, using the PDC-MAP long-cadence light curves that cover the full seventeen quarters of  the mission. The catalog included 69,914 transits of 779 KOIs with high enough SNRs, for which we derived the best fitting timing, duration and depth for each transit, given the KOI's best transit template. For additional 225,273 transits of 1820 KOIs with lower SNRs, we derived only the transit timing, keeping the duration and depth fixed for each KOI. 

The catalog is available at 
ftp://wise-ftp.tau.ac.il/pub/tauttv/TTV/ver$\_$112, 
where separate tables for the duration and depth variations are accessible too. In the future, we plan updating the catalog to include short-cadence data and better analysis of overlapping transits.
% and using Gaussian processes to obtain a better estimated TTVs.

For each KOI we derived various statistics that can be used to
indicate significant variations. Including systems found by previous works, we have found 260 KOIs which showed significant long-term TTVs, with periods longer than 100 day. 
Of those, 199 KOIs displayed well determined TTV periods and amplitudes. 
Another 61 KOIs have periods too long to be established without a doubt, and therefore only a parabola was fitted to the TTV series.

It is interesting to compare the analysis of \citet{mazeh13} with our results,
and to see how the longer time span may change our assessment of
the nature of the modulation. 
For example, after correcting for the TTVs, the deduced orbital periods  of KOI-564.01, -1145.01, -1573.01, and -1884.01 were significantly changed; the TTVs of KOI-984.01 showed sinusoidal behavior in the previous catalog, and now display a non-sinusoidal shape with a sharp linear rise; the TTVs of KOI-1599.01 and -1426.01 showed only parabolic behavior in \citet{mazeh13}, while now allow for a sinusoidal fit.

For most of the KOIs, we need more data before the TTV period can be robustly determined.
However, as the \ik mission in its original mode of operation has been terminated,
we do not expect any additional \ik TTVs. The next space missions that will observe the \ik field in a search for transiting planets  (e.g., TESS, PLATO), or dedicated ground-based follow-up observations \citep[e.g.,][]{raetz14},
may be very useful in determining the true nature and TTV periodicity of these systems.    

Most of the significant TTVs presented here are probably due to dynamical interaction with another planet(s), either transiting or still unknown. The present catalog  includes 121 single KOIs with significant long-term modulations, probably caused by undetected additional planets, and therefore are multi-planet systems. The missing planets can be divided into two classes: 
1.	Small planets that were not detected by the present analysis of  Kepler's data, either because the transits were too shallow and/or the transit timings were shifted by the dynamical interaction with the known planet, so that the folding of the light curve could not reveal the shallow transit.
2.	Planets with high enough inclinations, so they do not pass in front of their parent stars. 

One could hope that the accumulating details of the observed TTVs
could give some hints for the orbital elements of the perturbing
unseen planet. However,
as discussed already by \citet{holman05} and \citet{agol05}, the
amplitude and periodicity of the TTV modulation depends on various parameters,
in particular the mass and the orbital period of the unseen planet, and
how close the orbits of the two planets are to some mean motion resonance
\citep[e.g.,][]{lithwick12}.
Therefore, it is quite difficult to deduce the parameters of the unseen
planet.
In selected cases, some stringent constraints can be derived, as was done
by \citet{ballard11} and \citet{nesvorny12,nesvorny13}. We hope
that our catalog will motivate a similar work on other single-KOI systems, as well as on the multi-planet systems, with significant TTVs.

Ten systems analyzed here showed significant, small amplitude modulations with periods shorter than 100 d. We presented these ten systems together with another two systems, whose modulations were detected by \citet{mazeh13} and still showed periodic, but less significant, modulations, and another two KOIs that displayed photometric slope correlated with the TTVs, found by \citet{holczer15}.
Out of the thirteen systems found by \citet{mazeh13} with short-period modulation, seven now appear to be false positives, and the other six are presented here.  

The short-period modulations might not be due to dynamical interaction with another planet, but could be due to either the long cadence sampling of {\it Kepler}, or
the stellar spots periodic activity. 
For five systems, the stellar rotational frequency and/or one of its harmonics coincide with the highest peak \citep[see also][]{holczer15} of the derived periodogram. Two system, KOI-883.01 and -13.01, show a strong stroboscopic effect. 
The stars with TTVs related to the stellar spot activity are interesting, 
because the transit timings can shed some light on the stellar spot activity, as was shown very recently by \citet[]{mazeh15a,mazeh15b} and \citet{holczer15}.

Finally, one could
expect that the derived TTVs, for the single KOIs in particular,
could help in constructing a {\it statistical} picture of the
frequency and architecture of the population of the planetary multiple systems
of the Kepler KOIs \citep[e.g.,][]{steffen10, ford11, ford12b, lissauer11b, fabrycky14}. 
To perform such a statistical analysis one needs to model the dependence of the
detectability of long-term TTV coherent modulations on the parameters of the
unseen perturbing planet. The present catalog can
be used for such a study.

\acknowledgments 
%We thank the referee for his/her extremely valuable remarks and suggestions.
We are indebted to the referee for careful reading of the paper and for many thoughtful and helpful comments and suggestions.
The research leading to these results has
received funding from the European Research Council under the
EU's Seventh Framework Programme (FP7/(2007-2013)/ERC Grant Agreement
No.~291352), the ISRAEL SCIENCE FOUNDATION (grant
No.~1423/11) and the Israeli Centers of
Research Excellence (I-CORE,
grant No.~1829/12). EBF was supported in part by NASA Kepler Participating Scientist Program award \# NNX14AN76G.
All photometric data presented in this paper were
obtained from the Mikulsky Archive for Space Telescopes (MAST).
STScI is operated by the Association of Universities for Research
in Astronomy, Inc., under NASA contract NAS5-26555. Support for
MAST for non-HST data is provided by the NASA Office of Space
Science via grant NNX09AF08G and by other grants and contracts.

%=========================
%Table 0
\begin{table}
\footnotesize \caption{Additional false positive KOIs}
\begin{tabular}{|r|r|r|r|r|r|r|r|}
\hline \hline
KOI & KOI & KOI & KOI &KOI & KOI & KOI & KOI \\

  \hline
$ 225.01 $ & $ 302.01 $ & $ 823.01 $ & $ 977.01 $ & $ 1351.01 $ & $ 1452.01 $ & $ 1701.01 $ & $ 1771.01 $ \\ 
$ 2014.01 $ & $ 3175.01 $ & $ 3244.01 $ & $ 3272.01 $ & $ 3290.01 $ & $ 3331.01 $ & $ 3467.01 $ & $ 3565.01 $ \\ 
$ 3606.01 $ & $ 3715.01 $ & $ 4135.01 $ & $ 4294.01 $ & $ 4351.01 $ & $ 4925.01 $ & $ 4936.01 $ & $ 4937.01 $ \\ 
$ 4944.01 $ & $ 4947.01 $ & $ 4951.01 $ & $ 4953.01 $ & $ 4968.01 $ & $ 4970.01 $ & $ 5011.01 $ & $ 5015.01 $ \\ 
$ 5025.01 $ & $ 5061.01 $ & $ 5068.01 $ & $ 5076.01 $ & $ 5087.01 $ & $ 5090.01 $ & $ 5111.01 $ & $ 5112.01 $ \\ 
$ 5145.01 $ & $ 5152.01 $ & $ 5171.01 $ & $ 5172.01 $ & $ 5173.01 $ & $ 5218.01 $ & $ 5233.01 $ & $ 5255.01 $ \\ 
$ 5293.01 $ & $ 5295.01 $ & $ 5305.01 $ & $ 5306.01 $ & $ 5353.01 $ & $ 5354.01 $ & $ 5369.01 $ & $ 5392.01 $ \\ 
$ 5448.01 $ & $ 5460.01 $ & $ 5463.01 $ & $ 5542.01 $ & $ 5564.01 $ & $ 5569.01 $ & $ 5587.01 $ & $ 5683.01 $\\
$ 5714.01 $ & $ 5733.01 $ & $ 5774.01 $ & $ 5780.01 $ & $ 5797.01 $ & $ 5828.01 $ & $ 5894.01 $ & $ 5906.01 $ \\
 $ 5976.01$ &                     &                      &                     &                     &                     &                     &                      \\ 

\hline
\end{tabular}
%\tablecomments{1}

\label{tab:EB}
\end{table}

%=====================
% Table 1
%----------------------------------------
\begin{table}
\footnotesize \caption{KOI transits: linear ephemerides, duration and depth}
\begin{tabular}{ccccccc}
\hline \hline
KOI     &  ~~~~~~$T_0$\tablenotemark{a} & ~~~~~Period\tablenotemark{b} &
~~~Duration\tablenotemark{c} &
Depth\tablenotemark{d} & SNR\tablenotemark{e} & Scatter\tablenotemark{f} \\
           &   ~~~~ (day)   & ~~~~~(day) &~~  (hr) & (ppm)& & Ratio \\
\end{tabular}

\begin{tabular}{r l l r r r r}
\hline
$ 1.01 $ & $ 55.763337 $ & $ 2.47061338 $ & $ 1.8638 $ & $ 14210 $ & $ 618.0 $ & $ 1.66 $ \\ 
   & $ \pm 0.000006 $ & $ \pm 0.00000001 $ &    &    &    & \nodata \\ 
$ 2.01 $ & $ 54.358470 $ & $ 2.20473540 $ & $ 4.0398 $ & $ 6694 $ & $ 368.6 $ & $ 1.14 $ \\ 
   & $ \pm 0.000014 $ & $ \pm 0.00000002 $ &    &    &    & \nodata \\ 
$ 3.01 $ & $ 57.813556 $ & $ 4.88780267 $ & $ 2.5819 $ & $ 4361 $ & $ 266.8 $ & $ 2.17 $ \\ 
   & $ \pm 0.000033 $ & $ \pm 0.00000004 $ &    &    &    & \nodata \\ 
$ 5.01 $ & $ 56.414171 $ & $ 4.78032873 $ & $ 2.1266 $ & $ 980 $ & $ 34.9 $ & $ 0.94 $ \\ 
   & $ \pm 0.000158 $ & $ \pm 0.00000018 $ &    &    &    & \nodata \\ 
$ 7.01 $ & $ 56.612028 $ & $ 3.21366739 $ & $ 4.2575 $ & $ 736 $ & $ 24.6 $ & $ 1.00 $ \\ 
   & $ \pm 0.000272 $ & $ \pm 0.00000032 $ &    &    &    & \nodata \\ 
$ 10.01 $ & $ 54.119425 $ & $ 3.52249865 $ & $ 3.3119 $ & $ 9370 $ & $ 127.3 $ & $ 1.17 $ \\ 
   & $ \pm 0.000045 $ & $ \pm 0.00000005 $ &    &    &    & \nodata \\ 
$ 12.01 $ & $ 61.739669 $ & $ 17.85523009 $ & $ 7.3728 $ & $ 9228 $ & $ 357.1 $ & $ 2.26 $ \\ 
   & $ \pm 0.000246 $ & $ \pm 0.00000028 $ &    &    &    & \nodata \\ 
$ 13.01 $ & $ 53.565659 $ & $ 1.76358757 $ & $ 3.3504 $ & $ 4602 $ & $ 535.1 $ & $ 1.30 $ \\ 
   & $ \pm 0.000010 $ & $ \pm 0.00000001 $ &    &    &    & \nodata \\ 
$ 17.01 $ & $ 54.486657 $ & $ 3.23469918 $ & $ 3.7108 $ & $ 10811 $ & $ 248.6 $ & $ 1.19 $ \\ 
   & $ \pm 0.000028 $ & $ \pm 0.00000003 $ &    &    &    & \nodata \\ 
$ 18.01 $ & $ 55.901581 $ & $ 3.54846550 $ & $ 4.6623 $ & $ 7454 $ & $ 161.6 $ & $ 1.16 $ \\ 
   & $ \pm 0.000042 $ & $ \pm 0.00000005 $ &    &    &    & \nodata \\ 

\end{tabular}
\tablecomments{
$^a$$T_0$ (BJD--2454900). 
$^b$Orbital period (day). 
$^c$Transit duration (hr). 
$^d$Transit depth (ppm).
$^e$Median single-transit SNR.
$^f$Scatter inside the transit to scatter outside the transite ratio. 
\\
(This table is available in its entirety in a machine-readable form in\\
 ftp://wise-ftp.tau.ac.il/pub/tauttv/TTV/ver$\_$112. 
 A portion is shown here for guidance regarding its form and content.)}
\label{tab:ephemeris}
\end{table}

%=========================
%Table 3
\begin{table}
\footnotesize \caption{TTV, duration (TDV) and depth (TPV)
changes of the transits}
\begin{tabular}{|r|rr|rr|rr|rr|rr|}
\hline \hline
KOI & $n$\tablenotemark{a}  & $t_n\tablenotemark{b}$~~ &
TTV$_n$\tablenotemark{c} & $\sigma_{n}$\tablenotemark{d}~&
~~TDV$_n$\tablenotemark{e}
&  $\sigma_{n}$\tablenotemark{f} ~& ~~TPV$_n$\tablenotemark{g} & $\sigma_{n}$\tablenotemark{h}~~ & Outlier\tablenotemark{i} &
Overlap\tablenotemark{j}\\
 & &  (day)~~~ & (minute)  & (minute) & & & & &  Flag & Flag \\

  \hline
$ 1.01 $ & $ 0 $ & $ 55.7633 $ & $ -0.050 $ & $ 0.085 $ & $ 0.001 $ & $ 0.003 $ & $ -0.006 $ & $ 0.003 $ & $ 0 $ & $ 0 $ \\ 
$ 1.01 $ & $ 1 $ & $ 58.2340 $ & $ 0.077 $ & $ 0.086 $ & $ -0.002 $ & $ 0.003 $ & $ -0.008 $ & $ 0.003 $ & $ 0 $ & $ 0 $ \\ 
$ 1.01 $ & $ 2 $ & $ 60.7046 $ & $ -0.037 $ & $ 0.086 $ & $ 0.001 $ & $ 0.002 $ & $ -0.011 $ & $ 0.003 $ & $ 0 $ & $ 0 $ \\ 
$ 1.01 $ & $ 4 $ & $ 65.6458 $ & $ -0.271 $ & $ 0.086 $ & $ 0.001 $ & $ 0.002 $ & $ -0.003 $ & $ 0.003 $ & $ 4 $ & $ 0 $ \\ 
$ 1.01 $ & $ 5 $ & $ 68.1164 $ & $ -0.003 $ & $ 0.085 $ & $ -0.001 $ & $ 0.002 $ & $ -0.002 $ & $ 0.003 $ & $ 0 $ & $ 0 $ \\ 
$ 1.01 $ & $ 6 $ & $ 70.5870 $ & $ 0.061 $ & $ 0.085 $ & $ -0.003 $ & $ 0.003 $ & $ -0.002 $ & $ 0.003 $ & $ 0 $ & $ 0 $ \\ 
$ 1.01 $ & $ 7 $ & $ 73.0576 $ & $ 0.175 $ & $ 0.087 $ & $ 0.019 $ & $ 0.003 $ & $ -0.030 $ & $ 0.003 $ & $ 0 $ & $ 0 $ \\ 
$ 1.01 $ & $ 8 $ & $ 75.5282 $ & $ 0.186 $ & $ 0.085 $ & $ 0.004 $ & $ 0.003 $ & $ -0.003 $ & $ 0.003 $ & $ 0 $ & $ 0 $ \\ 
$ 1.01 $ & $ 9 $ & $ 77.9989 $ & $ 0.047 $ & $ 0.082 $ & $ -0.007 $ & $ 0.003 $ & $ 0.005 $ & $ 0.003 $ & $ 0 $ & $ 0 $ \\ 
$ 1.01 $ & $ 10 $ & $ 80.4695 $ & $ -0.082 $ & $ 0.085 $ & $ 0.005 $ & $ 0.003 $ & $ -0.012 $ & $ 0.003 $ & $ 0 $ & $ 0 $ \\

\hline
\end{tabular}
\tablecomments{
$^a$Transit number. 
$^b$Expected transit time of linear ephemeris (BJD-2454900)
$^c$O-C time difference. 
$^d$O-C uncertainty. 
$^e$Fractional duration variation: (duration of transit -- average)/average. 
$^f$TDV uncertainty. 
$^g$Fractional depth variation. 
$^h$TPV uncertainty.  
$^i$Outlier flag. Sum of: 0 = not an outlier, 
1 = no-variation \pv $>$ 0.025, 2= local TTV, 4= global TTV, 8= global TDV, 16 =global TPV, 32 = TTV, TDV or TPV uncertainty outlier (see Section~2.4).
$^j$Overlapping flag:  0= no overlapping, 1 = another relatively large interfering transit ($>$0.25 the ``area'' of the transit) too close (expected time difference $<$ 2 transit durations + 1 interfering transit duration).
\\
(This table is available in its entirety in a machine-readable
form in\\
 ftp://wise-ftp.tau.ac.il/pub/tauttv/TTV/ver$\_$112.\\
 A portion of the table of KOI-$1.01$ is
shown here for guidance regarding its form and content.)}

\label{tab:TTV}
\end{table}

%==============================
%Table 4

\begin{table}[h]
\footnotesize \caption{Statistical parameters of the O-Cs series}
\begin{tabular}{|r|rrr|rcr|rr|rr|}
\hline \hline
KOI & $\overline{\sigma}_{\rm TT}\tablenotemark{a} $ & 
$s_{\rm O-C}\tablenotemark{b} $ & 
{\it p}-s/$\sigma$\tablenotemark{c}  & 
PS~~ & PS& {\it p}-PS\tablenotemark{f}  &
$\mathcal{A}$\tablenotemark{g}~~  &
{\it p}-$\mathcal{A}$\tablenotemark{h}  & 
Pol.  & {\it p}-$\mathcal{F}$\tablenotemark{j} \\
     &  &  &    &    Period\tablenotemark{d}   & Peak\tablenotemark{e}   &   &     &    &  Deg.\tablenotemark{i}  &
\\
     & (minute) & (minute) &  (log) &    (day)  ~~   &  (log)  & (log)        & 
& (log)  &
&  (log)   \\

\hline
$ 1.01 $ & $ 0.08 $ & $ 0.09 $ & $ 0.0 $ & $ 17.40 $ & $ -9.54 $ & $ 0.0 $ & $ 0.188 $ & $ -0.9 $ & $ 1 $ & $ -0.2 $ \\ 
$ 2.01 $ & $ 0.23 $ & $ 0.24 $ & $ 0.0 $ & $ 4.43 $ & $ -8.75 $ & $ -0.6 $ & $ 0.036 $ & $ -0.3 $ & $ 1 $ & $ -1.3 $ \\ 
$ 3.01 $ & $ 0.24 $ & $ 0.35 $ & $ -0.2 $ & $ 13.94 $ & $ -8.11 $ & $ -0.1 $ & $ 0.220 $ & $ -0.6 $ & $ 2 $ & $ -2.7 $ \\ 
$ 5.01 $ & $ 1.65 $ & $ 1.73 $ & $ 0.0 $ & $ 52.05 $ & $ -6.81 $ & $ -0.2 $ & $ -0.064 $ & $ -0.5 $ & $ 1 $ & $ -0.1 $ \\ 
$ 7.01 $ & $ 3.54 $ & $ 3.28 $ & $ 0.0 $ & $ 8.31 $ & $ -6.19 $ & $ -0.2 $ & $ 0.038 $ & $ -1.1 $ & $ 1 $ & $ -0.3 $ \\ 
$ 10.01 $ & $ 0.60 $ & $ 0.59 $ & $ 0.0 $ & $ 26.03 $ & $ -7.90 $ & $ 0.0 $ & $ -0.278 $ & $ -0.4 $ & $ 1 $ & $ -0.2 $ \\ 
$ 12.01 $ & $ 0.37 $ & $ 1.51 $ & $ <-16.0 $ & $ 1176.37 $ & $ -6.24 $ & $ -3.3 $ & $ 3.234 $ & $< -4.0 $ & $ 2 $ & $ -3.9 $ \\ 
$ 13.01 $ & $ 0.13 $ & $ 0.16 $ & $ 0.0 $ & $ 5.72 $ & $ -8.83 $ & $ -1.1 $ & $ -0.104 $ & $ -2.7 $ & $ 1 $ & $ -1.1 $ \\ 
$ 17.01 $ & $ 0.33 $ & $ 0.40 $ & $ 0.0 $ & $ 10.89 $ & $ -8.10 $ & $ -0.6 $ & $ 0.046 $ & $ -0.6 $ & $ 1 $ & $ -0.2 $ \\ 
$ 18.01 $ & $ 0.57 $ & $ 0.56 $ & $ 0.0 $ & $ 23.72 $ & $ -7.83 $ & $ -0.2 $ & $ -0.104 $ & $ -0.5 $ & $ 1 $ & $ -0.1 $ \\ 

\hline
\end{tabular}
\tablecomments{
$^a$O-C uncertainty median. 
$^b$O-C scatter (1.483 times the MAD).
$^c$The {\it p}-value of the scatter divided by the uncertainty (see text).
$^d$The period with the highest peak in the power spectrum. 
$^e$The amplitude of the highest peak in the power spectrum, in days squared. 
$^f$The logarithm of the {\it p}-value of the $\mathcal{F}$-test for the highest PS peak found. 
$^g$Alarm score (see text). 
$^h$The logarithm of the {\it p}-value of the alarm found.  
$^i$The degree of the best fitting polynomial. 
$^j$The logarithm of the {\it p}-value of the $\mathcal{F}$-test for the best polynomial fit. \\
(This table is available in its entirety in a machine-readable form in \\
ftp://wise-ftp.tau.ac.il/pub/tauttv/TTV/ver$\_$112. A portion is shown
here for guidance regarding its form and content.)}

\label{tab:TTV_master}
\end{table}

%==============================
%Table 5

\begin{table}
\footnotesize
%\scriptsize
\caption{KOIs with significant long-term TTV}
\begin{tabular}{|rr|lrrrrr|r|r|l|}
\hline \hline

KOI & Period\tablenotemark{a} & Model\tablenotemark{b}  & Period\tablenotemark{c}   & $\sigma_P\tablenotemark{d} $ & Amp\tablenotemark{e}   &$\sigma_A\tablenotemark{f} $&
Res\tablenotemark{g}  & $N$\tablenotemark{h}  &   Multi-  &  References \\
     &  (day)  &    & (day)      & (day)    &  (minute)    &   (minute)   &
(minute) &  & plicity\tablenotemark{i}  & \\

\hline
$ 12.01 $ & $ 17.86 $ & Cos & $ 1246 $ & $ 60 $ & $ 1.152 $ & $ 0.081 $ & $ 1.2 $ & $ 69 $ & $ 1 $ &  \\ 
$ 42.01 $ & $ 17.83 $ & Cos & $ 960.6 $ & $ 9.2 $ & $ 15.59 $ & $ 0.49 $ & $ 4.1 $ & $ 75 $ & $ 1 $ & \tablenotemark{3}Kepler410Ab \\ 
$ 75.01 $ & $ 105.88 $ & Pol & \nodata & \nodata & $ \sim20 $ & \nodata & $ 4 $ & $ 10 $ & $ 1 $ &  \\ 
$ 84.01 $ & $ 9.29 $ & Cos & $ 313.5 $ & $ 2.1 $ & $ 5.03 $ & $ 0.34 $ & $ 2.9 $ & $ 142 $ & $ 1 $ & \tablenotemark{4}Kepler19b \\ 
$ 89.01 $ & $ 84.69 $ & Cos & $ 550 $ & $ 23 $ & $ 8.9 $ & $ 2.6 $ & $ 5.2 $ & $ 16 $ & $ 2 $ &  \\ 
$ 89.02 $ & $ 207.59 $ & Cos & $ 1440 $ & $ 110 $ & $ 55 $ & $ 13 $ & $ 12 $ & $ 6 $ & $ 2 $ &  \\
\hline

\end{tabular}
\end{table}

\addtocounter{table}{-1}

\begin{table} \footnotesize \caption{ --continued}
\begin{tabular}{|rr|lrrrrr|r|r|l|}
\hline \hline

KOI & Period\tablenotemark{a} & Model\tablenotemark{b}  & Period\tablenotemark{c}   & $\sigma_P\tablenotemark{d} $ & Amp\tablenotemark{e}   &$\sigma_A\tablenotemark{f} $&
Res\tablenotemark{g}  & $N$\tablenotemark{h}  &   Multi-  &  References \\
     &  (day)  &    & (day)      & (day)    &  (minute)    &   (minute)   &
(minute) &  & plicity\tablenotemark{i}  & \\

 \hline 
$ 248.01 $ & $ 7.20 $ & Cos & $ 368 $ & $ 3.2 $ & $ 11.5 $ & $ 0.92 $ & $ 7.6 $ & $ 122 $ & $ 4 $ & \tablenotemark{1,2,15}~~~~~Kepler49b \\ 
$ 248.02 $ & $ 10.91 $ & Cos & $ 369.6 $ & $ 3.5 $ & $ 20.9 $ & $ 1.7 $ & $ 6.9 $ & $ 79 $ & $ 4 $ & \tablenotemark{1,15}~~~Kepler49c \\ 
$ 250.01 $ & $ 12.28 $ & Cos & $ 751 $ & $ 12 $ & $ 9.58 $ & $ 0.65 $ & $ 6.9 $ & $ 98 $ & $ 4 $ & \tablenotemark{14}~Kepler26b \\ 
$ 250.02 $ & $ 17.25 $ & Cos & $ 734 $ & $ 18 $ & $ 9.9 $ & $ 1 $ & $ 7.9 $ & $ 68 $ & $ 4 $ & \tablenotemark{1,14}~~~Kepler26c \\ 
$ 262.01 $ & $ 7.81 $ & Cos & $ 1060 $ & $ 24 $ & $ 34.8 $ & $ 2.1 $ & $ 16 $ & $ 117 $ & $ 2 $ & \tablenotemark{15}~Kepler50b \\ 
$ 262.02 $ & $ 9.38 $ & Cos & $ 664 $ & $ 11 $ & $ 15.4 $ & $ 1.4 $ & $ 12 $ & $ 114 $ & $ 2 $ & \tablenotemark{15}~Kepler50c \\
\hline

\end{tabular}
\end{table}

\addtocounter{table}{-1}

\begin{table}\footnotesize \caption{ --continued}
\begin{tabular}{|rr|lrrrrr|r|r|l|}
\hline \hline

KOI & Period\tablenotemark{a} & Model\tablenotemark{b}  & Period\tablenotemark{c}   & $\sigma_P\tablenotemark{d} $ & Amp\tablenotemark{e}   &$\sigma_A\tablenotemark{f} $&
Res\tablenotemark{g}  & $N$\tablenotemark{h}  &   Multi-  &  References \\
     &  (day)  &    & (day)      & (day)    &  (minute)    &   (minute)   &
(minute) &  & plicity\tablenotemark{i}  & \\

   \hline 
$ 417.01 $ & $ 19.19 $ & Cos & $ 304.3 $ & $ 8.8 $ & $ 1.05 $ & $ 0.8 $ & $ 1.7 $ & $ 64 $ & $ 1 $ &  \\ 
$  $ & $  $ & Cos & $ 202.6 $ & $ 3.2 $ & $ 0.96 $ & $ 0.26 $ & $ 1.7 $ & $  $ & $  $ &  \\ 
$ 448.02 $ & $ 43.58 $ & Pol & \nodata & \nodata & $ \sim200 $ & \nodata & $ 50 $ & $ 26 $ & $ 2 $ & \tablenotemark{2,7}~~Kepler159c \\ 
$ 456.01 $ & $ 13.70 $ & Cos & $ 879 $ & $ 12 $ & $ 20.4 $ & $ 1.3 $ & $ 7.7 $ & $ 94 $ & $ 2 $ & \tablenotemark{2,7}~~Kepler160c \\ 
$ 457.02 $ & $ 7.06 $ & Cos & $ 280 $ & $ 3.4 $ & $ 7.98 $ & $ 0.98 $ & $ 8.1 $ & $ 149 $ & $ 2 $ & \tablenotemark{7}Kepler161c \\ 
$ 464.01 $ & $ 58.36 $ & Cos & $ 482 $ & $ 11 $ & $ 3.93 $ & $ 0.59 $ & $ 0.93 $ & $ 21 $ & $ 2 $ &  \\
\hline 
\end{tabular}
 \end{table}

\addtocounter{table}{-1}
\begin{table} \footnotesize \caption{ --continued}
\begin{tabular}{|rr|lrrrrr|r|r|l|}
\hline \hline

KOI & Period\tablenotemark{a} & Model\tablenotemark{b}  & Period\tablenotemark{c}   & $\sigma_P\tablenotemark{d} $ & Amp\tablenotemark{e}   &$\sigma_A\tablenotemark{f} $&
Res\tablenotemark{g}  & $N$\tablenotemark{h}  &   Multi-  &  References \\
     &  (day)  &    & (day)      & (day)    &  (minute)    &   (minute)   &
(minute) &  & plicity\tablenotemark{i}  & \\

\hline 
$ 709.01 $ & $ 21.39 $ & Cos & $ 1118 $ & $ 57 $ & $ 15.7 $ & $ 3.3 $ & $ 13 $ & $ 59 $ & $ 1 $ &  \\ 
$ 720.03 $ & $ 18.37 $ & Pol & \nodata & \nodata & $ \sim9 $ & \nodata & $ 5 $ & $ 34 $ & $ 4 $ & \tablenotemark{7}Kepler221e \\ 
$ 730.01 $ & $ 14.79 $ & Cos & $ 1292 $ & $ 78 $ & $ 61.6 $ & $ 9.1 $ & $ 20 $ & $ 38 $ & $ 4 $ & \tablenotemark{1,7}~~Kepler223d \\ 
$ 730.03 $ & $ 19.73 $ & Cos & $ 1690 $ & $ 100 $ & $ 108 $ & $ 16 $ & $ 49 $ & $ 40 $ & $ 4 $ & \tablenotemark{1,7}~~Kepler223e \\ 
$ 738.01 $ & $ 10.34 $ & Pol & \nodata & \nodata & $ \sim55 $ & \nodata & $ 20$ & $ 92 $ & $ 2 $ & \tablenotemark{2,21}~~~Kepler29b \\ 
$ 738.02 $ & $ 13.29 $ & Pol & \nodata & \nodata & $ \sim80$ & \nodata & $ 25 $ & $ 67 $ & $ 2 $ & \tablenotemark{21}~Kepler29c \\
\hline 

\end{tabular}
\label{tab:TTV_Interesting}
\end{table}

\addtocounter{table}{-1}
\begin{table} \footnotesize \caption{ --continued}
\begin{tabular}{|rr|lrrrrr|r|r|l|}
\hline \hline

KOI & Period\tablenotemark{a} & Model\tablenotemark{b}  & Period\tablenotemark{c}   & $\sigma_P\tablenotemark{d} $ & Amp\tablenotemark{e}   &$\sigma_A\tablenotemark{f} $&
Res\tablenotemark{g}  & $N$\tablenotemark{h}  &   Multi-  &  References \\
     &  (day)  &    & (day)      & (day)    &  (minute)    &   (minute)   &
(minute) &  & plicity\tablenotemark{i}  & \\

 \hline 
$ 884.02 $ & $ 20.49 $ & Cos & $ 803.6 $ & $ 1 $ & $ 188.25 $ & $ 0.8 $ & $ 19 $ & $ 52 $ & $ 3 $ & \tablenotemark{1,2,7}~~~~Kepler247d \\ 
$ 886.01 $ & $ 8.01 $ & Cos & $ 849.8 $ & $ 4.8 $ & $ 66.4 $ & $ 1.6 $ & $ 11 $ & $ 139 $ & $ 3 $ & \tablenotemark{2,15}~~~Kepler54b \\ 
$ 886.02 $ & $ 12.07 $ & Cos & $ 852.1 $ & $ 5.5 $ & $ 124 $ & $ 3.4 $ & $ 15 $ & $ 78 $ & $ 3 $ & \tablenotemark{15}~Kepler54c \\ 
$ 902.01 $ & $ 83.93 $ & Pol & \nodata & \nodata & $ \sim85 $ & \nodata & $ 5 $ & $ 18 $ & $ 1 $ &  \\ 
$ 904.02 $ & $ 27.97 $ & Pol & \nodata & \nodata & $ \sim255 $ & \nodata & $ 40 $ & $ 48 $ & $ 5 $ & \tablenotemark{15}~Kepler55b \\ 
$ 904.03 $ & $ 42.12 $ & Pol & \nodata & \nodata & $ \sim225 $ & \nodata & $ 60 $ & $ 29 $ & $ 5 $ & \tablenotemark{15}~Kepler55c \\
\hline 

\end{tabular}
\label{tab:TTV_Interesting}
\end{table}
\addtocounter{table}{-1}
\begin{table} \footnotesize \caption{ --continued}
\begin{tabular}{|rr|lrrrrr|r|r|l|}
\hline \hline

KOI & Period\tablenotemark{a} & Model\tablenotemark{b}  & Period\tablenotemark{c}   & $\sigma_P\tablenotemark{d} $ & Amp\tablenotemark{e}   &$\sigma_A\tablenotemark{f} $&
Res\tablenotemark{g}  & $N$\tablenotemark{h}  &   Multi-  &  References \\
     &  (day)  &    & (day)      & (day)    &  (minute)    &   (minute)   &
(minute) &  & plicity\tablenotemark{i}  & \\

 \hline
$ 1335.01 $ & $ 127.83 $ & Cos & $ 1083 $ & $ 91 $ & $ 6 $ & $ 2.1 $ & $ 2.4 $ & $ 10 $ & $ 1 $ &  \\ 
$ 1353.01 $ & $ 125.86 $ & Cos & $ 1277 $ & $ 40 $ & $ 8.86 $ & $ 0.69 $ & $ 0.47 $ & $ 9 $ & $ 2 $ & \tablenotemark{7}Kepler289c \\ 
$ 1355.01 $ & $ 51.93 $ & Cos & $ 124.5 $ & $ 1.3 $ & $ 6 $ & $ 2 $ & $ 6.2 $ & $ 24 $ & $ 1 $ &  \\ 
$ 1426.01 $ & $ 38.87 $ & Cos & $ 1040 $ & $ 20 $ & $ 44 $ & $ 3.5 $ & $ 8.7 $ & $ 30 $ & $ 3 $ & \tablenotemark{7}Kepler297b \\ 
$ 1426.02 $ & $ 74.93 $ & Pol & \nodata & \nodata & $ \sim50 $ & \nodata & $ 10 $ & $ 14 $ & $ 3 $ & \tablenotemark{7}Kepler297c \\ 
$ 1426.03 $ & $ 150.03 $ & Cos & $ 1323 $ & $ 59 $ & $ 34.6 $ & $ 3.5 $ & $ 5.2 $ & $ 10 $ & $ 3 $ &  \\
\hline

\end{tabular}
\label{tab:TTV_Interesting}
\end{table}
\addtocounter{table}{-1}
\begin{table} \footnotesize \caption{ --continued}
\begin{tabular}{|rr|lrrrrr|r|r|l|}
\hline \hline

KOI & Period\tablenotemark{a} & Model\tablenotemark{b}  & Period\tablenotemark{c}   & $\sigma_P\tablenotemark{d} $ & Amp\tablenotemark{e}   &$\sigma_A\tablenotemark{f} $&
Res\tablenotemark{g}  & $N$\tablenotemark{h}  &   Multi-  &  References \\
     &  (day)  &    & (day)      & (day)    &  (minute)    &   (minute)   &
(minute) &  & plicity\tablenotemark{i}  & \\

 \hline 
$ 1707.01 $ & $ 96.11 $ & Pol & \nodata & \nodata & $ \sim60 $ & \nodata & $ 30 $ & $ 12 $ & $ 2 $ & \tablenotemark{7}Kepler315b \\ 
$ 1720.01 $ & $ 59.66 $ & Pol & \nodata & \nodata & $ \sim45 $ & \nodata & $ 20 $ & $ 21 $ & $ 1 $ &  \\ 
$ 1751.02 $ & $ 21.00 $ & Pol & \nodata & \nodata & $ \sim70 $ & \nodata & $ 5 $ & $ 44 $ & $ 2 $ &  \\ 
$ 1781.01 $ & $ 7.83 $ & Pol & \nodata & \nodata & $ \sim3 $ & \nodata & $ 2.1 $ & $ 97 $ & $ 3 $ &  \\ 
$ 1781.03 $ & $ 58.02 $ & Cos & $ 922 $ & $ 17 $ & $ 39.8 $ & $ 2.4 $ & $ 9.8 $ & $ 11 $ & $ 3 $ &  \\ 
$ 1783.01 $ & $ 134.48 $ & Cos & $ 1390 $ & $ 130 $ & $ 6.2 $ & $ 2.2 $ & $ 5.3 $ & $ 9 $ & $ 2 $ &  \\
\hline

\end{tabular}
\label{tab:TTV_Interesting}
\end{table}

\addtocounter{table}{-1}
\begin{table} \footnotesize \caption{ --continued}
\begin{tabular}{|rr|lrrrrr|r|r|l|}
\hline \hline

KOI & Period\tablenotemark{a} & Model\tablenotemark{b}  & Period\tablenotemark{c}   & $\sigma_P\tablenotemark{d} $ & Amp\tablenotemark{e}   &$\sigma_A\tablenotemark{f} $&
Res\tablenotemark{g}  & $N$\tablenotemark{h}  &   Multi-  &  References \\
     &  (day)  &    & (day)      & (day)    &  (minute)    &   (minute)   &
(minute) &  & plicity\tablenotemark{i}  & \\

\hline
$ 2061.01 $ & $ 14.10 $ & Cos & $ 1269 $ & $ 24 $ & $ 151.1 $ & $ 5.6 $ & $ 26 $ & $ 76 $ & $ 2 $ &  \\ 
$ 2086.02 $ & $ 8.92 $ & Cos & $ 715 $ & $ 28 $ & $ 29.4 $ & $ 7.2 $ & $ 35 $ & $ 100 $ & $ 3 $ & \tablenotemark{15}~Kepler60c \\ 
$ 2092.01 $ & $ 57.69 $ & Cos & $ 1270 $ & $ 130 $ & $ 22.6 $ & $ 5.7 $ & $ 10 $ & $ 20 $ & $ 3 $ & \tablenotemark{7}Kepler359c \\ 
$ 2094.01 $ & $ 42.43 $ & Cos & $ 1510 $ & $ 130 $ & $ 43.2 $ & $ 6.8 $ & $ 19 $ & $ 28 $ & $ 2 $ &  \\ 
$ 2199.01 $ & $ 9.03 $ & Pol & \nodata & \nodata & $ \sim50 $ & \nodata & $ 25 $ & $ 69 $ & $ 1 $ &  \\ 
$ 2217.01 $ & $ 37.80 $ & Pol & \nodata & \nodata & $ \sim10 $ & \nodata & $ 30 $ & $ 29 $ & $ 1 $ &  \\
\hline

\end{tabular}
\label{tab:TTV_Interesting}
\end{table}

\addtocounter{table}{-1}
\begin{table} \footnotesize \caption{ --continued}
\begin{tabular}{|rr|lrrrrr|r|r|l|}
\hline \hline

KOI & Period\tablenotemark{a} & Model\tablenotemark{b}  & Period\tablenotemark{c}   & $\sigma_P\tablenotemark{d} $ & Amp\tablenotemark{e}   &$\sigma_A\tablenotemark{f} $&
Res\tablenotemark{g}  & $N$\tablenotemark{h}  &   Multi-  &  References \\
     &  (day)  &    & (day)      & (day)    &  (minute)    &   (minute)   &
(minute) &  & plicity\tablenotemark{i}  & \\

 \hline 
$ 3683.01 $ & $ 214.31 $ & Cos & $ 1219 $ & $ 84 $ & $ 7.83 $ & $ 0.89 $ & $ 3.2 $ & $ 7 $ & $ 1 $ &  \\ 
$ 3692.01 $ & $ 1.16 $ & Pol & \nodata & \nodata & $ \sim5 $ & \nodata & $ 2 $ & $ 1032 $ & $ 1 $ &  \\ 
$ 3720.01 $ & $ 213.40 $ & Pol & \nodata & \nodata & $ \sim1 $ & \nodata & $ 0.4$ & $ 6 $ & $ 1 $ &  \\ 
$ 3781.01 $ & $ 2.88 $ & Cos & $ 836.4 $ & $ 2.1 $ & $ 5.979 $ & $ 0.077 $ & $ 4.9 $ & $ 431 $ & $ 1 $ &  \\ 
$ 4007.01 $ & $ 11.49 $ & Pol & \nodata & \nodata & $ \sim90 $ & \nodata & $ 45 $ & $ 70 $ & $ 1 $ &  \\ 
$ 4014.01 $ & $ 234.24 $ & Pol & \nodata & \nodata & $ \sim35 $ & \nodata & $ 7 $ & $ 6 $ & $ 1 $ &  \\
\hline 
 
\end{tabular}
\label{tab:TTV_Interesting}
\end{table}

\addtocounter{table}{-1}
\begin{table}[t!] 
\footnotesize \caption{ --continued}
\begin{tabular}{rr|lrrrrr|r|r|l}

\tablecomments{
$^a$Orbital Period. 
$^b$Model type: `Cos' represents a cosine superimposed on a linear trend, `Pol' represents a parabolic fit, while `...' means no fit. 
$^c$Best-fit period of the O-C data using model Cos. 
$^d$Period uncertainty. 
$^e$The amplitude of the cosine fit. 
$^f$Amplitude uncertainty. 
$^g$Residual scatter (1.483 times their MAD). 
$^h$Number of TT measurements. 
$^i$Number of detected planets in the system  (see B12). \\
References.
%$^*$Discussed in Section~\ref{comments}.\\
$^1$\citet{ford11}.
$^2$\citet{ford12b}.
$^3$\citet{van-eylen14}. 
$^4$\citet{ballard11}. 
$^5$\citet{weiss13}.
$^6$\citet{marcy14}.
$^7$\citet{rowe14}. 
$^8$\citet{cochran11}. 
$^9$\citet{nesvorny13}. 
$^{10}$\citet{xie14}. 
$^{11}$\citet{lissauer11a}. 
$^{12}$\citet{ford12a}. 
$^{13}$\citet{deleuil14}.
$^{14}$\citet{steffen12a}.
$^{15}$\citet{steffen13}. 
$^{16}$\citet{carter12}.
$^{17}$\citet{wang14}.
$^{18}$\citet{kipping14}.
$^{19}$\citet{holman10}. 
$^{20}$\citet{masuda14}.
$^{21}$\citet{fabrycky12}. 
$^{22}$\citet{nesvorny12}. 
$^{23}$\citet{dawson14}. 
$^{24}$\citet{ofir14}. }
\end{tabular}
\label{tab:TTV_Interesting}
\end{table}
%------------------------------------
\clearpage
\newpage
\begin{table}[]
\scriptsize \caption{Corrected Periods for KOIs with periodic fits}
\begin{tabular}{|rr|rr|rr|rr|rr|}

\hline \hline
KOI & Period & KOI & Period & KOI & Period & KOI & Period  & KOI & Period \\
      & (day)       &       & (day)       &       & (day)       &       & (day)       &       & (day)         \\
\hline
$ 12.01 $ & $ 17.8552221 $ & $ 166.01 $ & $ 12.4934161 $ & $ 303.01 $ & $ 60.9284885 $ & $ 500.02 $ & $ 9.5216221 $ & $ 775.02 $ & $ 7.8774240 $ \\ 
          & $ \pm 0.0000001 $ &           & $ \pm 0.0000009 $ &           & $ \pm 0.0000019 $ &           & $ \pm 0.0000013 $ &           & $ \pm 0.0000014 $ \\ 
$ 42.01 $ & $ 17.8336258 $ & $ 168.01 $ & $ 10.7424696 $ & $ 308.01 $ & $ 35.5970409 $ & $ 520.01 $ & $ 12.7593931 $ & $ 784.01 $ & $ 19.2716570 $ \\ 
          & $ \pm 0.0000005 $ &           & $ \pm 0.0000019 $ &           & $ \pm 0.0000011 $ &           & $ \pm 0.0000017 $ &           & $ \pm 0.0000023 $ \\ 
$ 84.01 $ & $ 9.2869706 $ & $ 168.03 $ & $ 7.1069636 $ & $ 314.01 $ & $ 13.781063 $ & $ 520.03 $ & $ 25.7526459 $ & $ 806.01 $ & $ 143.2110706 $ \\ 
          & $ \pm 0.0000003 $ &           & $ \pm 0.0000095 $ &           & $ \pm 0.000001 $ &           & $ \pm 0.0000029 $ &           & $ \pm 0.0000013 $ \\
\hline 

\end{tabular}
%**********

\label{tab:periods}
\end{table}
\clearpage

\addtocounter{table}{-1}
\begin{table}[]
\scriptsize \caption{--continued}
\begin{tabular}{|rr|rr|rr|rr|rr|}

\hline \hline
KOI & Period & KOI & Period & KOI & Period & KOI & Period  & KOI & Period \\
      & (day)       &       & (day)       &       & (day)       &       & (day)       &       & (day)         \\

\hline
$ 884.02 $ & $ 20.4857821 $ & $ 1236.03 $ & $ 54.411731 $ & $ 1581.01 $ & $ 29.5420643 $ & $ 1934.01 $ & $ 28.7826087 $ & $ 2691.01 $ & $ 97.451375 $ \\ 
          & $ \pm 0.0000009 $ &           & $ \pm 0.000009 $ &           & $ \pm 0.0000057 $ &           & $ \pm 0.0000025 $ &           & $ \pm 0.000011 $ \\ 
$ 886.01 $ & $ 8.0107718 $ & $ 1241.02 $ & $ 10.501201 $ & $ 1581.02 $ & $ 144.5561670 $ & $ 1938.01 $ & $ 96.9153379 $ & $ 2720.01 $ & $ 6.5714474 $ \\ 
          & $ \pm 0.0000016 $ &           & $ \pm 0.000012 $ &           & $ \pm 0.0000094 $ &           & $ \pm 0.0000033 $ &           & $ \pm 0.0000036 $ \\ 
$ 886.02 $ & $ 12.0714725 $ & $ 1270.02 $ & $ 11.6092377 $ & $ 1582.01 $ & $ 186.439063 $ & $ 1955.04 $ & $ 26.2349240 $ & $ 2975.01 $ & $ 175.325064 $ \\ 
          & $ \pm 0.0000036 $ &           & $ \pm 0.0000025 $ &           & $ \pm 0.000013 $ &           & $ \pm 0.0000048 $ &           & $ \pm 0.000012 $ \\
\hline 

\end{tabular}

\label{tab:periods}
\end{table}

\clearpage

%==============================
%Table 6

\begin{table}[h!]
\footnotesize
\caption{KOIs with significant short-period TTVs }
\begin{tabular}{|rr|rrrrr|r|r|l|}

\hline \hline
KOI & Period\tablenotemark{a} & Period\tablenotemark{b}   & $\sigma_P\tablenotemark{c} $ & Amp\tablenotemark{d} ~  &$\sigma_A\tablenotemark{e} $&
Res\tablenotemark{f}  & $N$\tablenotemark{g}  &   Multi-  &  References \\
     &  (day)  & (day)      & (day)    &  (minute)    &   (minute)   &
(minute) &  & plicity\tablenotemark{h}  & \\

\hline
$\tablenotemark{i}13.01 $ & $ 1.76 $ & $ 5.7205 $ & $ 0.0082 $ & $ 0.065 $ & $ 0.0075 $ & $ 0.15 $ & $ 679 $ & $ 1 $ &  \tablenotemark{1}Kepler13b \\ 
$ 203.01 $ & $ 1.49 $ & $ 6.0177 $ & $ 0.009 $ & $ 0.212 $ & $ 0.029 $ & $ 0.51 $ & $ 670 $ & $ 1 $ & \tablenotemark{2,3,4}~~~Kepler17b \\ 
$ 217.01 $ & $ 3.91 $ & $ 9.653 $ & $ 0.033 $ & $ 0.344 $ & $ 0.053 $ & $ 0.67 $ & $ 328 $ & $ 1 $ & \tablenotemark{5}Kepler71b \\ 
$ 312.01 $ & $ 11.58 $ & $ 39.97 $ & $ 0.32 $ & $ 21.7 $ & $ 1.5 $ & $ 9.6 $ & $ 104 $ & $ 2 $ &  \tablenotemark{6}Kepler136b \\ 
$ 341.01 $ & $ 7.17 $ & $ 15.562 $ & $ 0.06 $ & $ 14.5 $ & $ 1.4 $ & $ 8.5 $ & $ 110 $ & $ 2 $ & \tablenotemark{7}Kepler414c \\ 
$ 341.02 $ & $ 4.70 $ & $ 22.66 $ & $ 0.1 $ & $ 51.7 $ & $ 2.2 $ & $ 13 $ & $ 142 $ & $ 2 $ & \tablenotemark{7}Kepler414b \\
\hline 

\end{tabular}
%**********
\tablecomments{ 
$^a$Orbital Period. 
$^b$Best-fit period of the O-C data. 
$^c$Period uncertainty. 
$^d$The amplitude of the cosine model. 
$^e$Amplitude uncertainty. 
$^f$Residual scatter (1.483 times their MAD). 
$^g$Number of TT measurements. 
$^h$Number of planets in the system. 
$^i$Added for comparison with \citet{mazeh13}. 
$^j$Added due to spot-crossing related TTVs. \citep{holczer15}. \\
References.
$^{1}$\citet{shporer11}.
$^{2}$\citet{desert11}.  
$^{3}$\citet{szabo13}.  
$^{4}$\citet{bonomo12}.
$^{5}$\citet{howell10}.
$^{6}$\citet{rowe14}.
$^{7}$\citet{hadden14}.
}

\label{tab:short}
\end{table}

%---------------------------
% Figure 1+1 %
%---------------------------
\begin{figure*}[p]
\centering
\resizebox{16cm}{11cm}
{\includegraphics{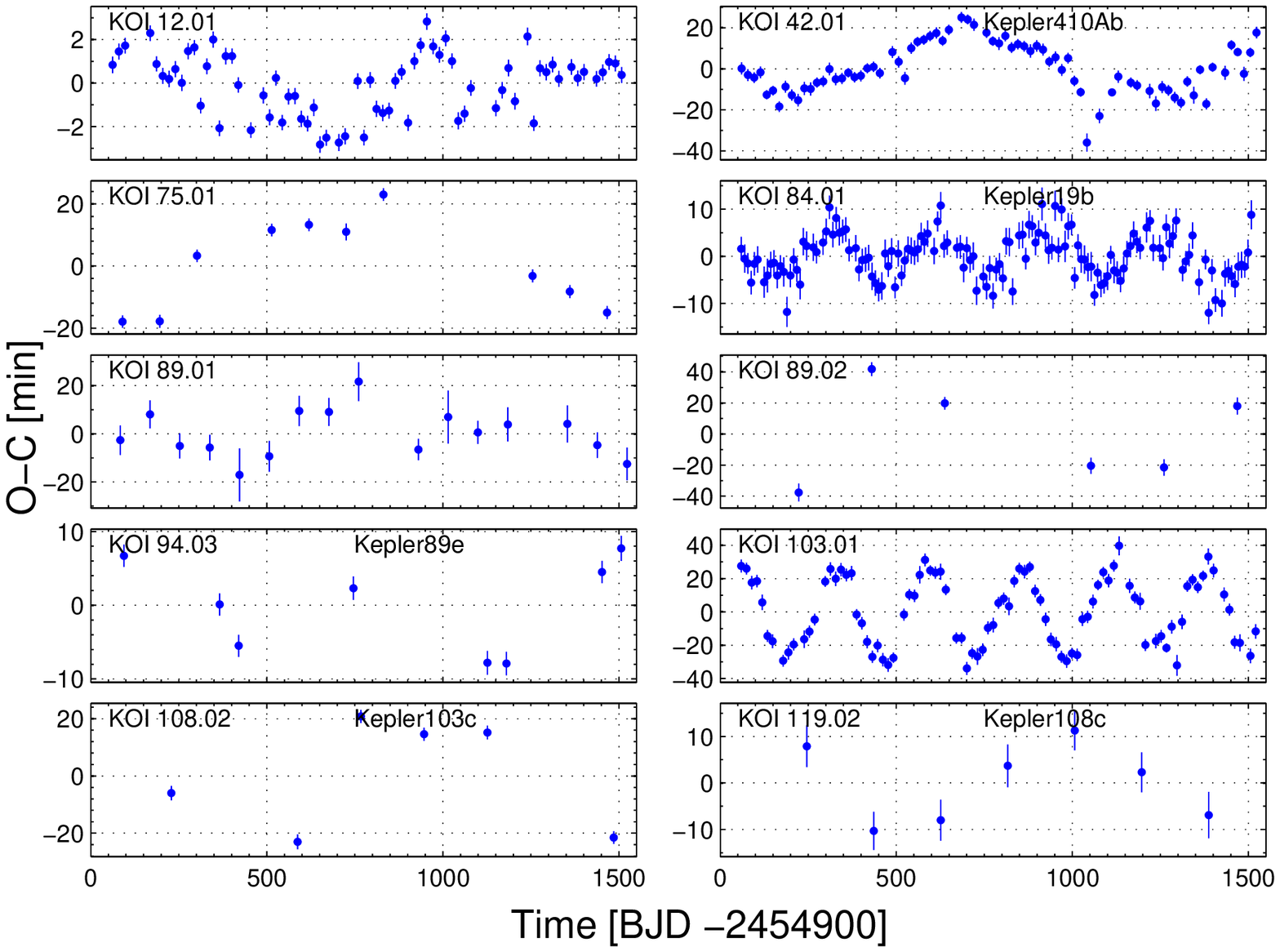}}
\caption{KOIs with significant long-term TTVs.
}
\label{TTV1}
\end{figure*}
%-------------------------------------------------------------
%---------------------------
% Figure  1+2
%---------------------------
\begin{figure*}[p]
\centering
%\resizebox{16cm}{11cm}
\resizebox{16cm}{11cm}
{\includegraphics{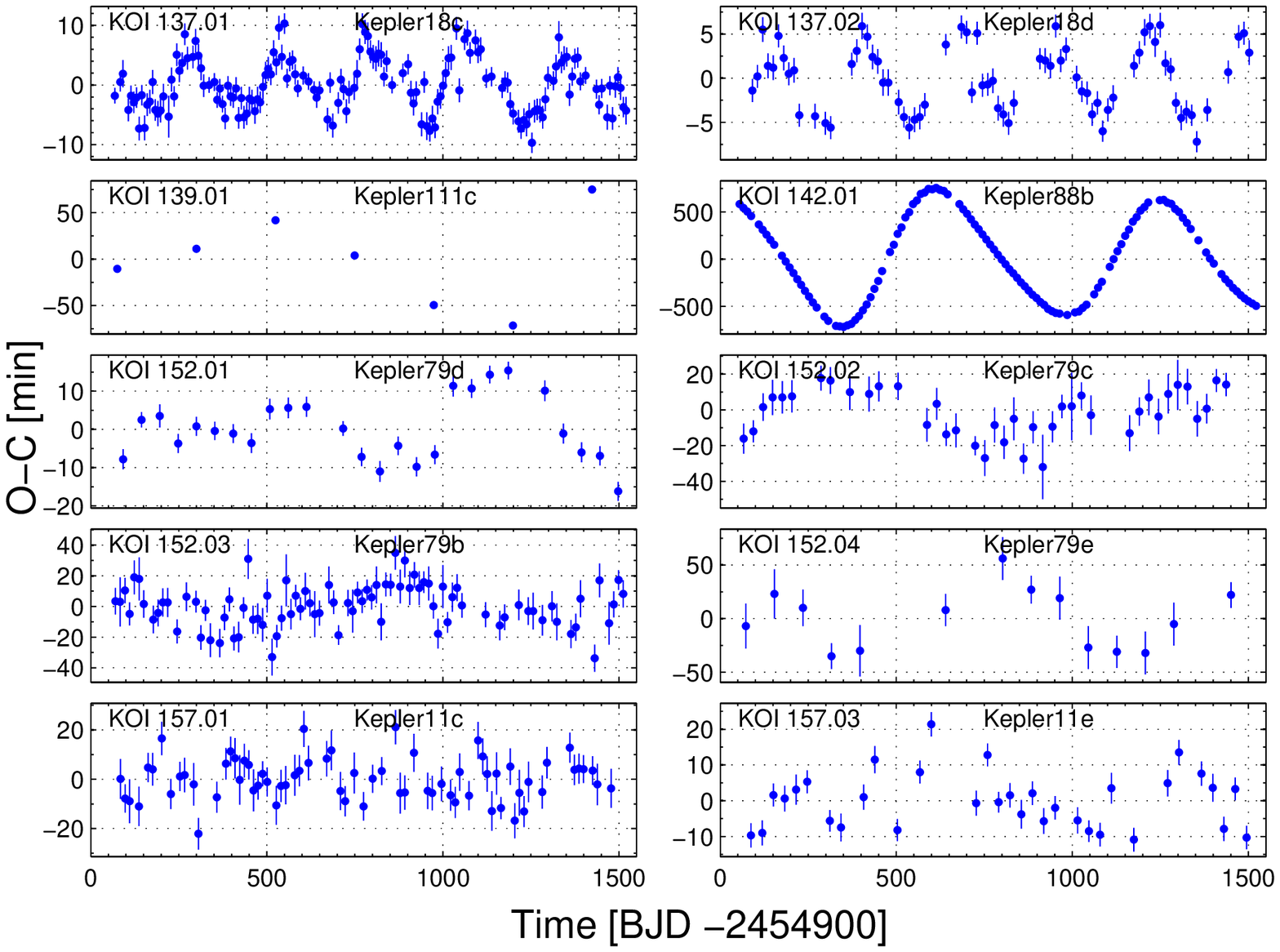}}
\caption{KOIs with significant long-term TTVs.
%See Figure~\ref{TTV1} for details.
}
\label{TTV2}
\end{figure*}
%-------------------------------------------------------------

%---------------------------
% Figure 1+3
%---------------------------
\begin{figure*}[p]
\centering
\resizebox{16cm}{11cm}
{\includegraphics{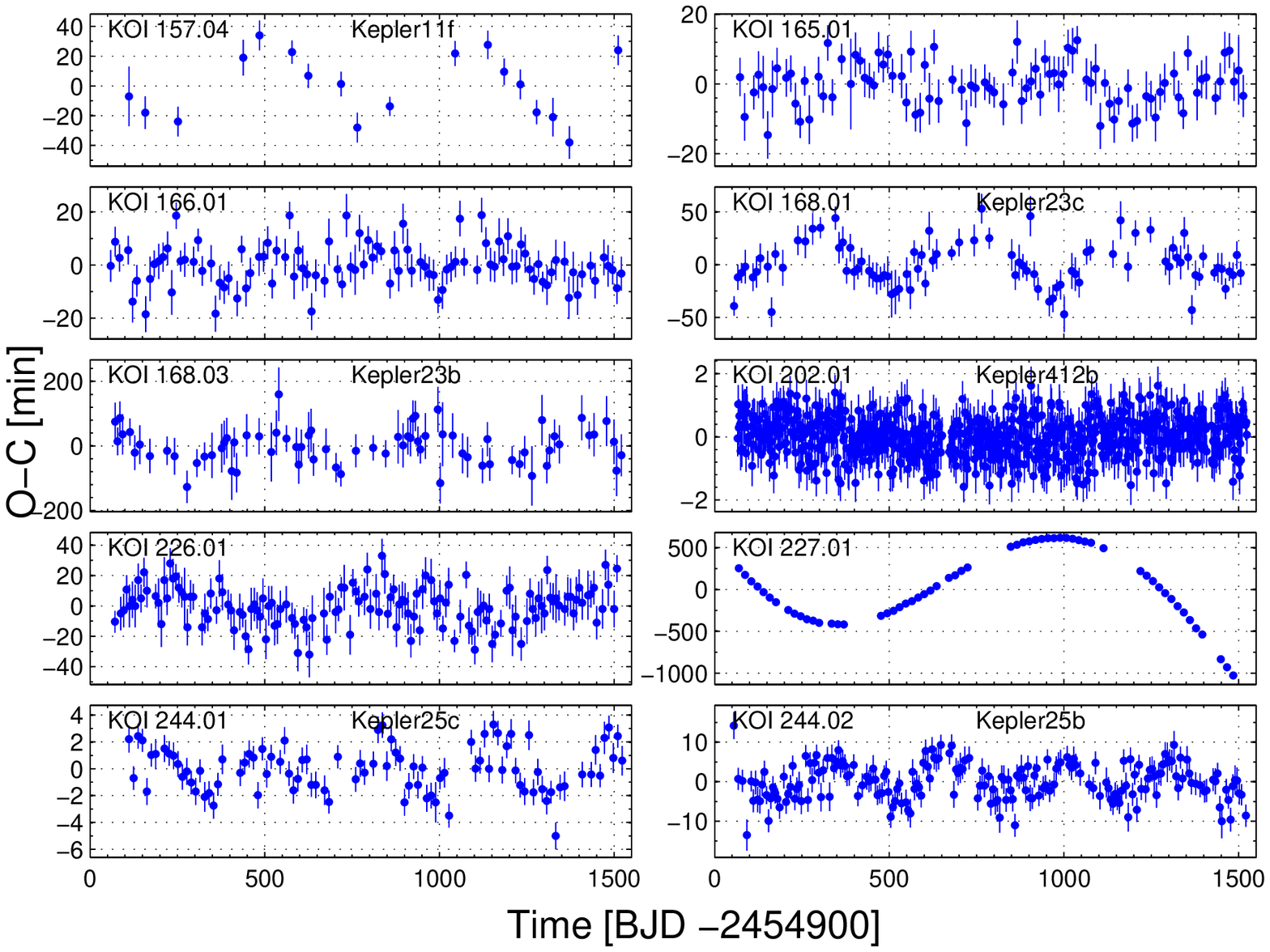}}
\caption{KOIs with significant long-term TTVs.
%See Figure~\ref{TTV1} for details.
}
\label{TTV3}
\end{figure*}
%-------------------------------------------------------------
%---------------------------
% Figure  1+4
%---------------------------
\begin{figure*}[p]
\centering
%\resizebox{16cm}{11cm}
\resizebox{16cm}{11cm}
{\includegraphics{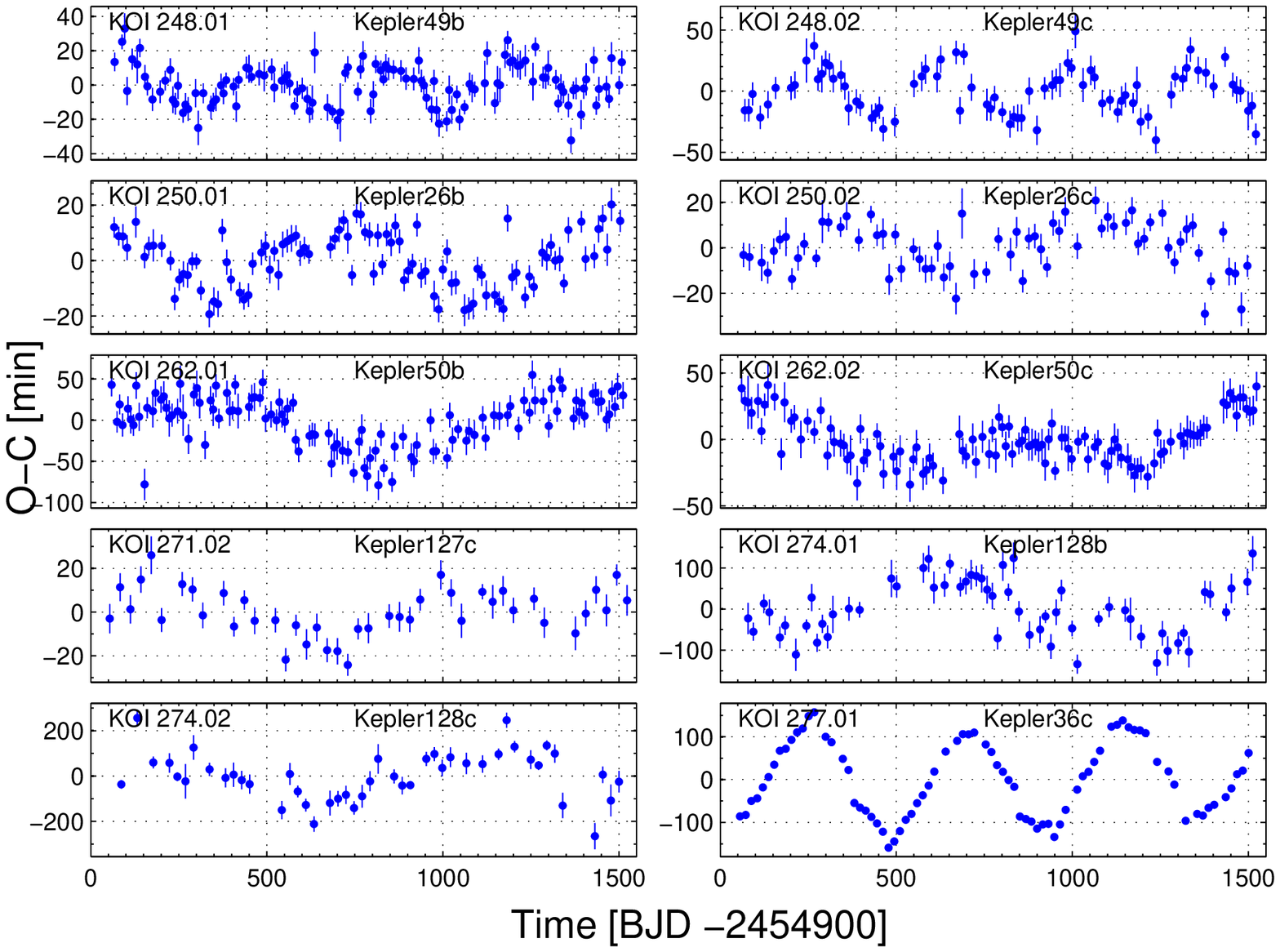}}
\caption{KOIs with significant long-term TTVs.
%See Figure~\ref{TTV1} for details.
}
\label{TTV4}
\end{figure*}
%-------------------------------------------------------------
%---------------------------
% Figure 1+5
%---------------------------
\begin{figure*}[p]
\centering
%\resizebox{16cm}{11cm}
\resizebox{16cm}{11cm}
{\includegraphics{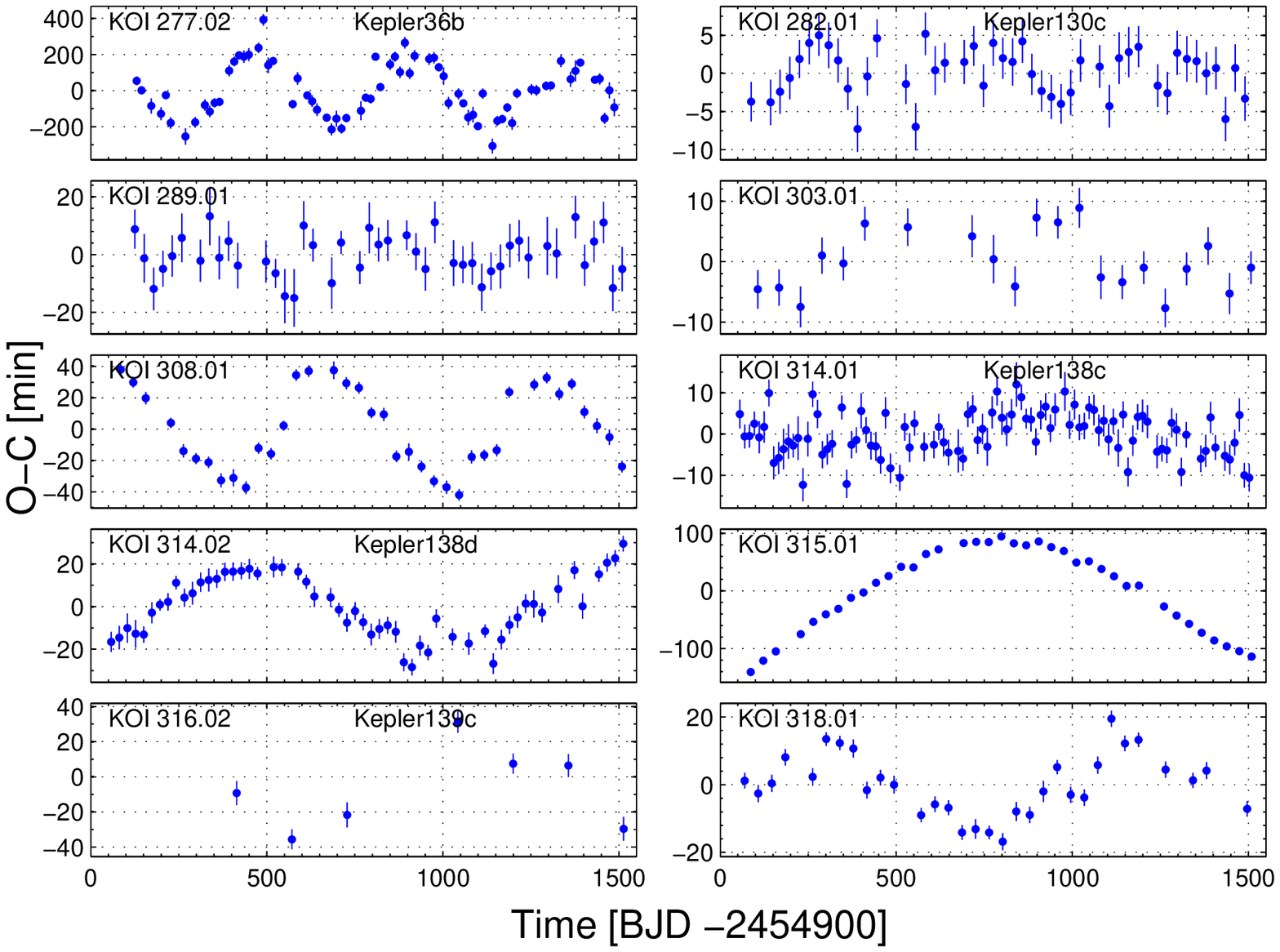}}
\caption{KOIs with significant long-term TTVs.
%See Figure~\ref{TTV1} for details.
}
\label{TTV5}
\end{figure*}
%-------------------------------------------------------------
%---------------------------
% Figure  1+6
%---------------------------
\begin{figure*}[p]
\centering
%\resizebox{16cm}{11cm}
\resizebox{16cm}{11cm}
{\includegraphics{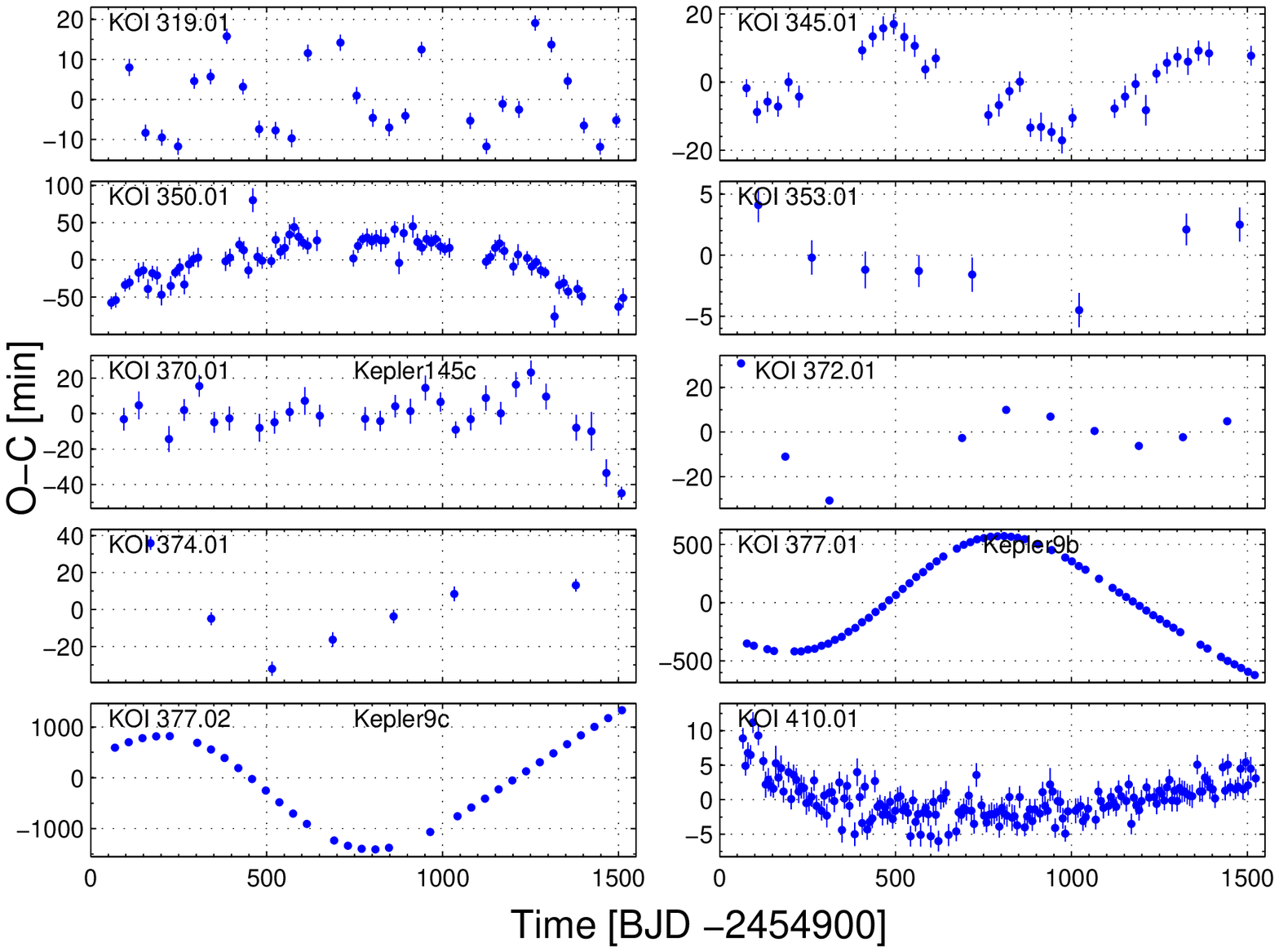}}
\caption{KOIs with significant long-term TTVs.
%See Figure~\ref{TTV1} for details.
}
\label{TTV6}
\end{figure*}
%-------------------------------------------------------------

%---------------------------
% Figure  1+7
%---------------------------
\begin{figure*}[p]
\centering
%\resizebox{16cm}{11cm}
\resizebox{16cm}{11cm}
{\includegraphics{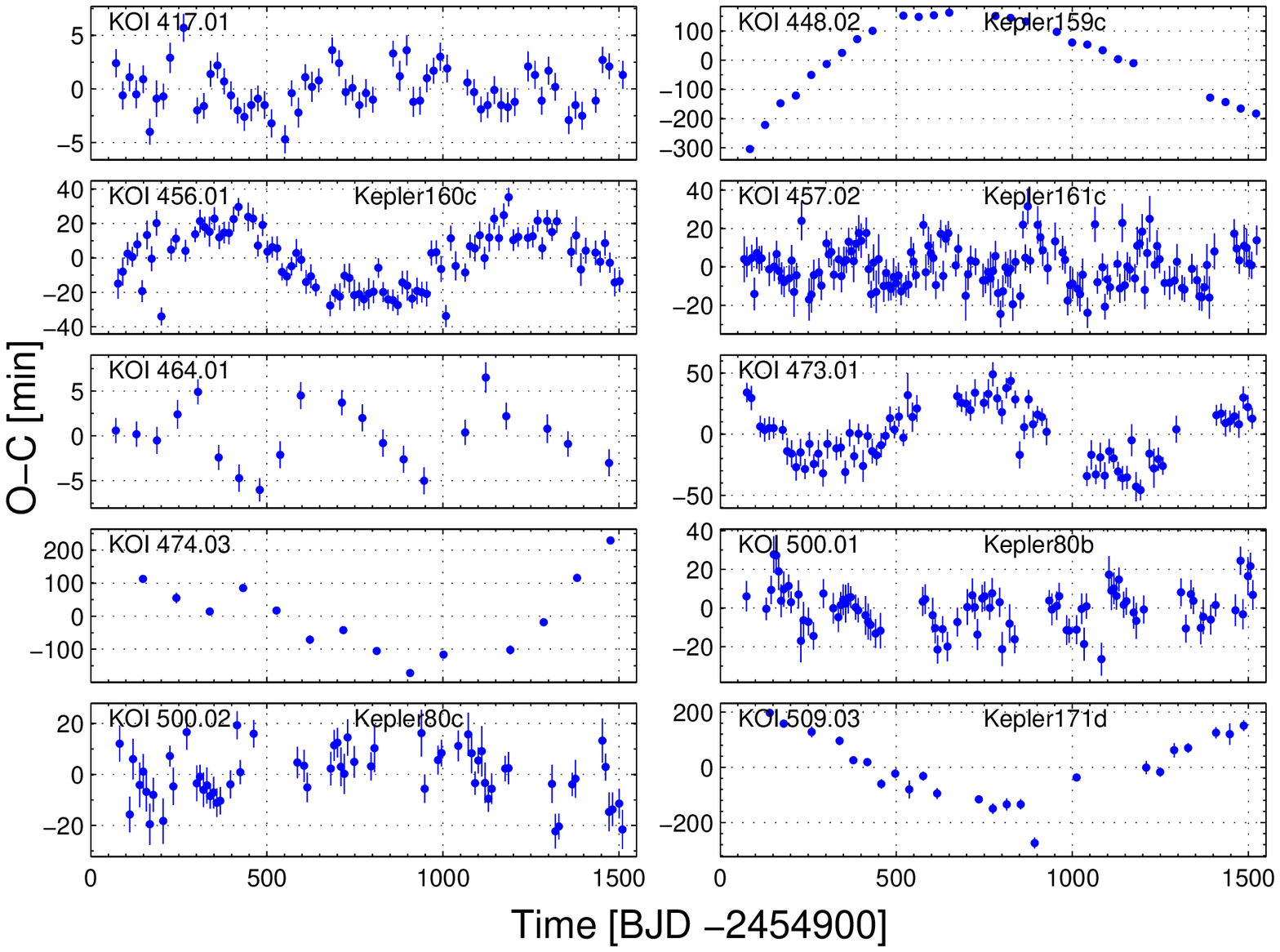}}
\caption{KOIs with significant long-term TTVs.
%See Figure~\ref{TTV1} for details.
}
\label{TTV7}
\end{figure*}
%-------------------------------------------------------------

%---------------------------
% Figure  1+8
%---------------------------
\begin{figure*}[p]
\centering
%\resizebox{16cm}{11cm}
\resizebox{16cm}{11cm}
{\includegraphics{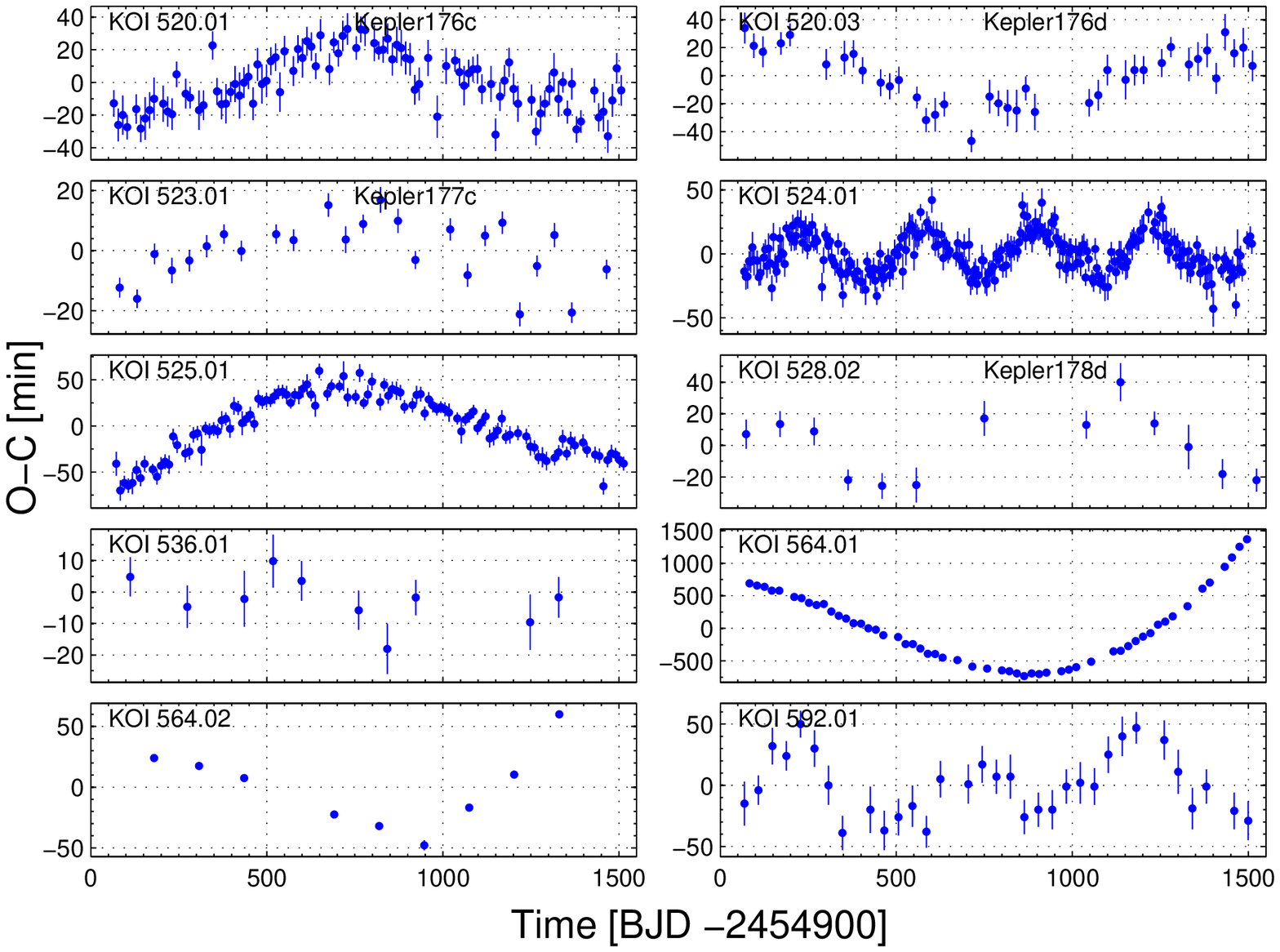}}
\caption{KOIs with significant long-term TTVs.
%See Figure~\ref{TTV1} for details.
}
\label{TTV8}
\end{figure*}
%-------------------------------------------------------------

%---------------------------
% Figure  1+9
%---------------------------
\begin{figure*}[p]
\centering
%\resizebox{16cm}{11cm}
\resizebox{16cm}{11cm}
{\includegraphics{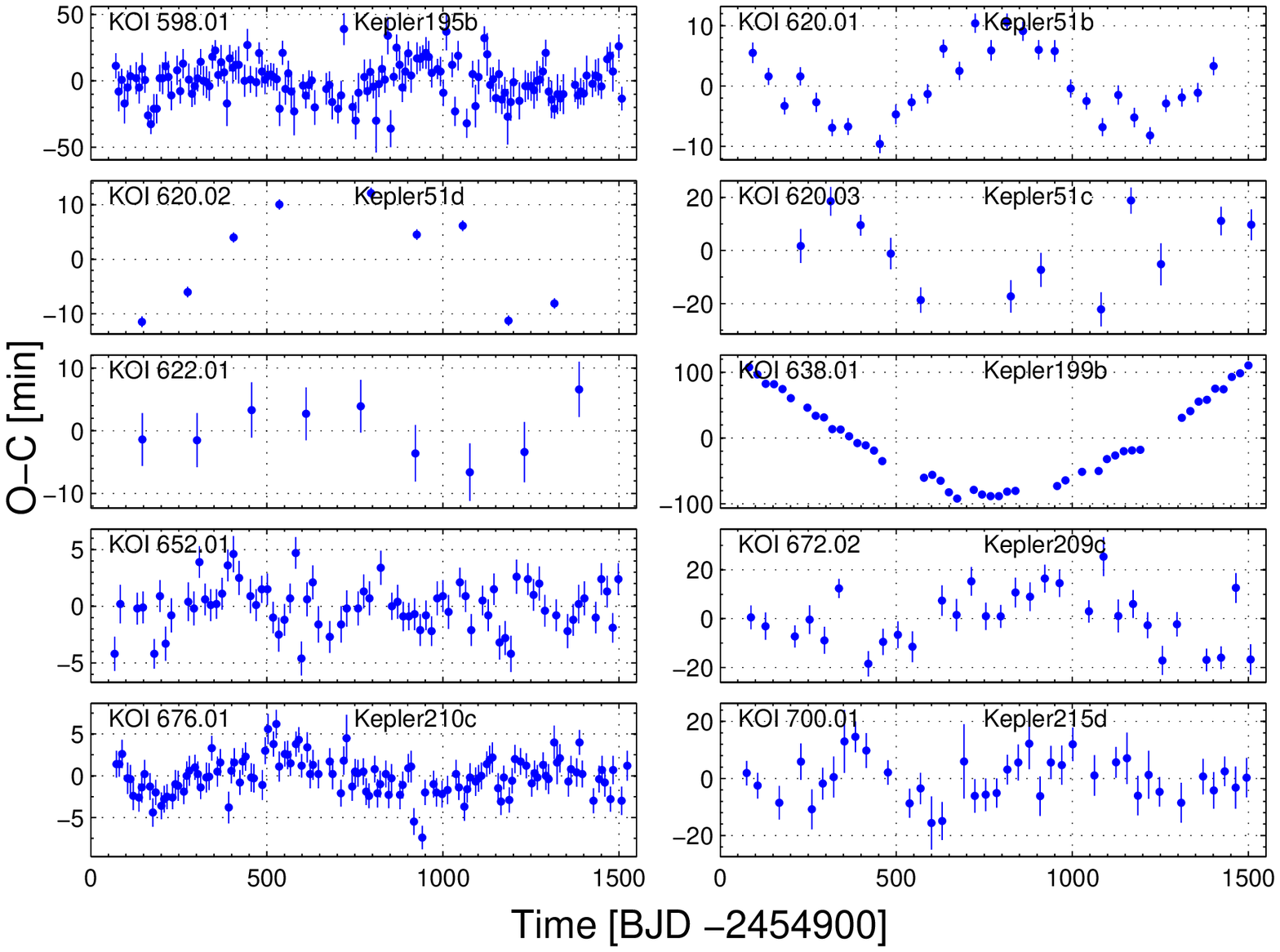}}
\caption{KOIs with significant long-term TTVs.
%See Figure~\ref{TTV1} for details.
}
\label{TTV9}
\end{figure*}
%-------------------------------------------------------------
%---------------------------
% Figure  1+10
%---------------------------
\begin{figure*}[p]
\centering
%\resizebox{16cm}{11cm}
\resizebox{16cm}{11cm}
{\includegraphics{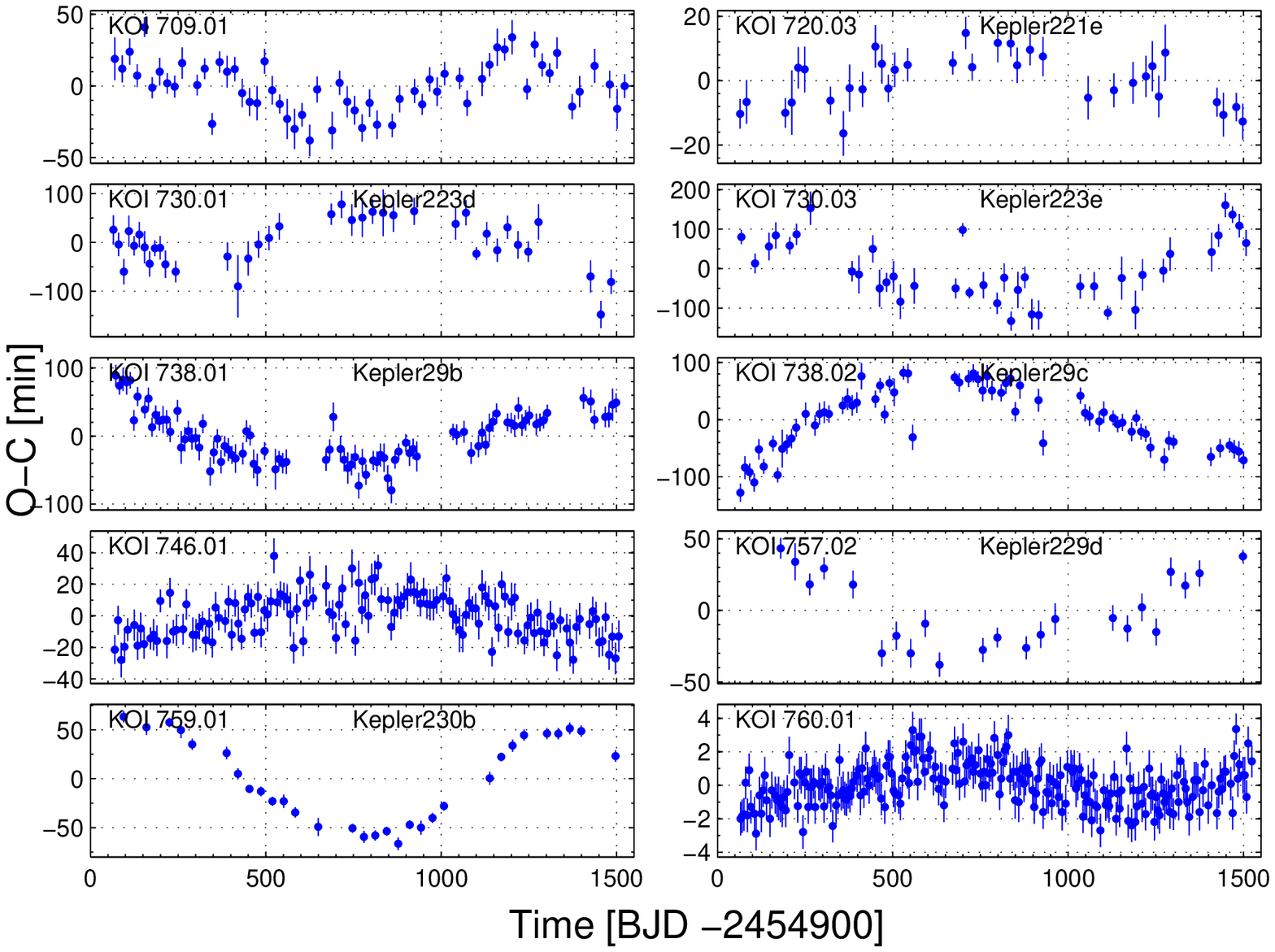}}
\caption{KOIs with significant long-term TTVs.
%See Figure~\ref{TTV1} for details.
}
\label{TTV10}
\end{figure*}
%-------------------------------------------------------------

%---------------------------
% Figure  1+11
%---------------------------
\begin{figure*}[p]
\centering
%\resizebox{16cm}{11cm}
\resizebox{16cm}{11cm}
{\includegraphics{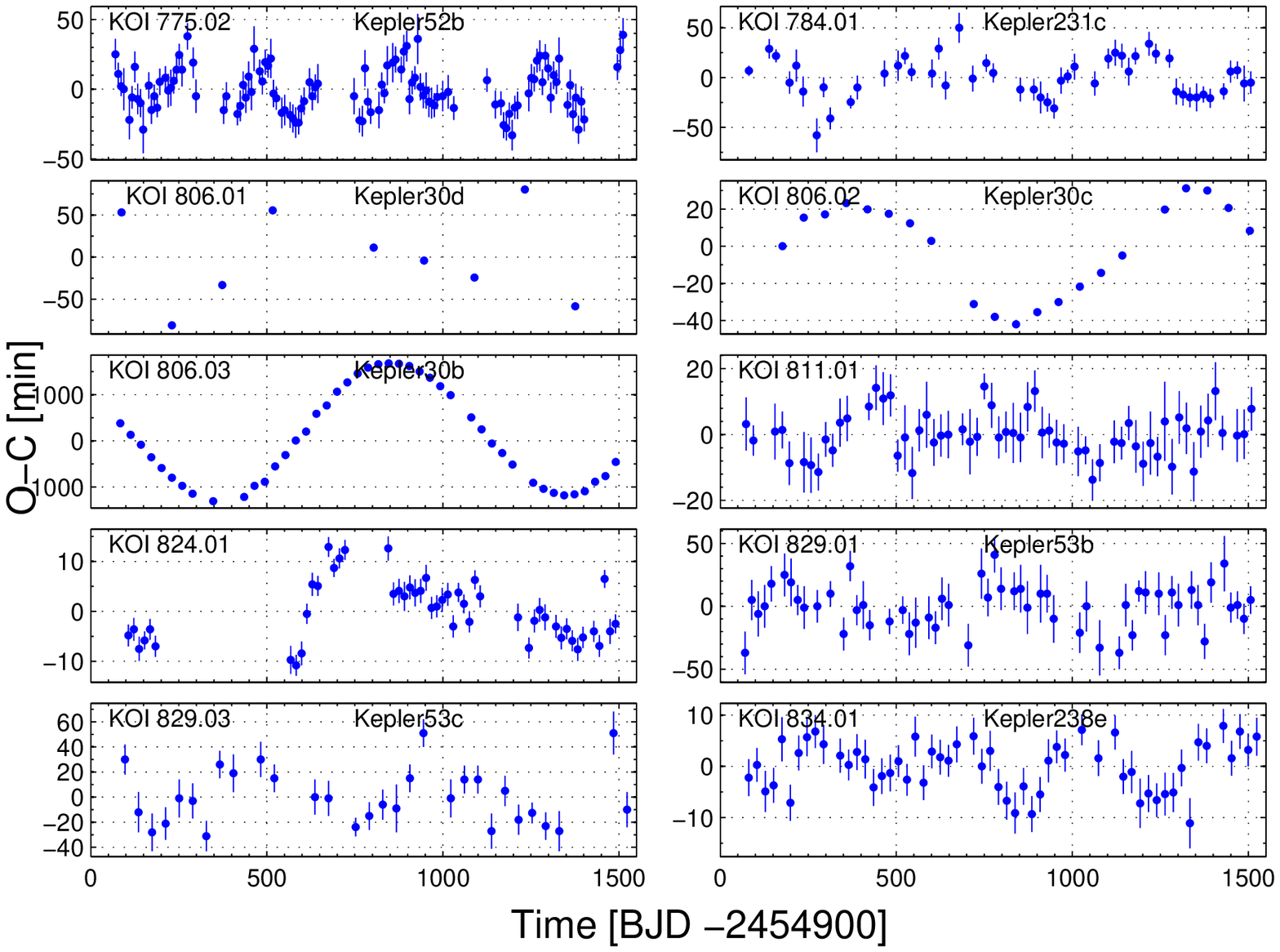}}
\caption{KOIs with significant long-term TTVs.
%See Figure~\ref{TTV1} for details.
}
\label{TTV11}
\end{figure*}
%-------------------------------------------------------------

%---------------------------
% Figure  1+12
%---------------------------
\begin{figure*}[p]
\centering
%\resizebox{16cm}{11cm}
\resizebox{16cm}{11cm} {\includegraphics{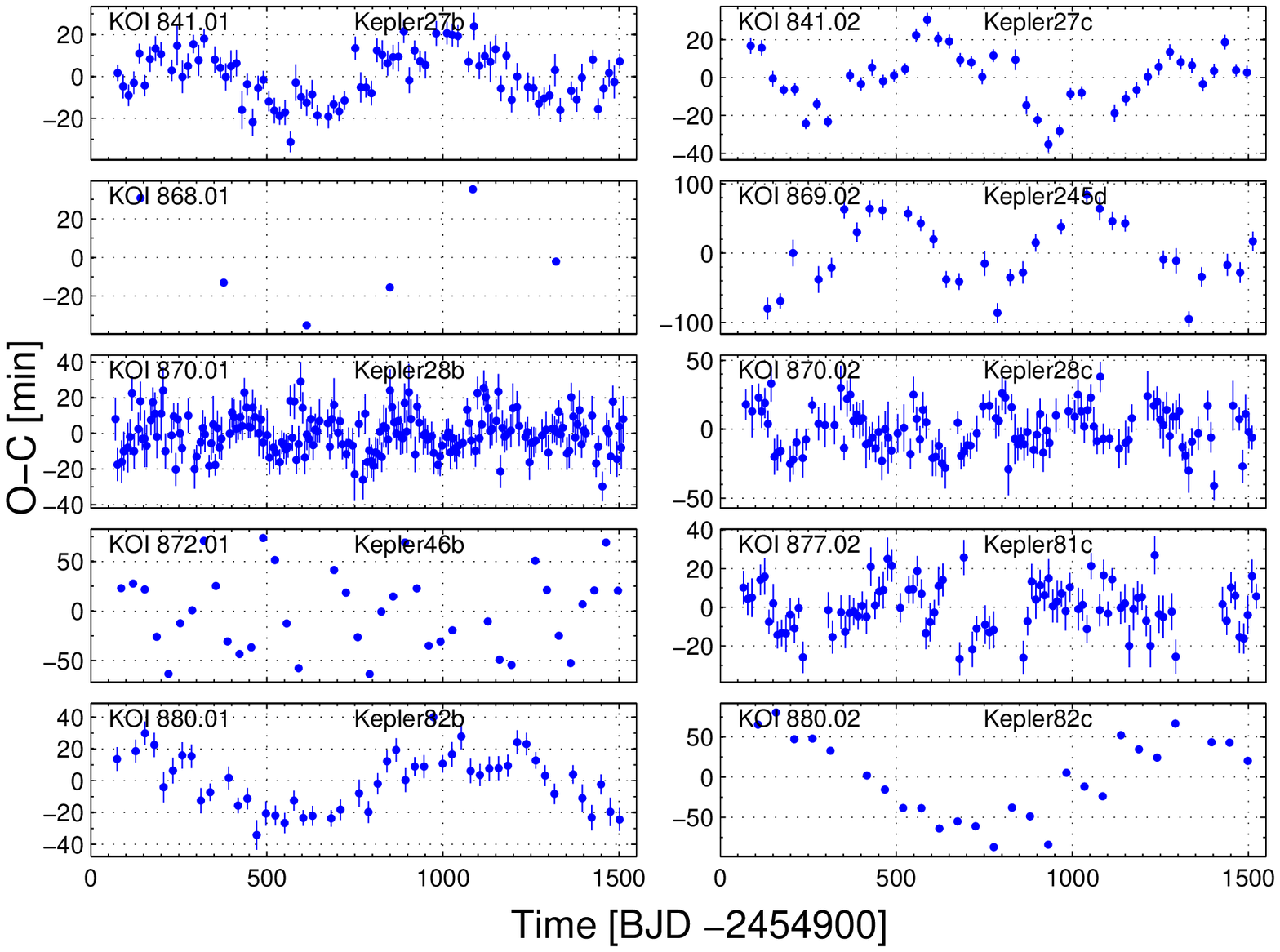}}
\caption{KOIs with significant long-term TTVs.
%See Figure~\ref{TTV1} for details.
}
\label{TTV12}
\end{figure*}
%-------------------------------------------------------------

%---------------------------
% Figure  1+13
%---------------------------
\begin{figure*}[p]
\centering
%\resizebox{16cm}{11cm}
\resizebox{16cm}{11cm} {\includegraphics{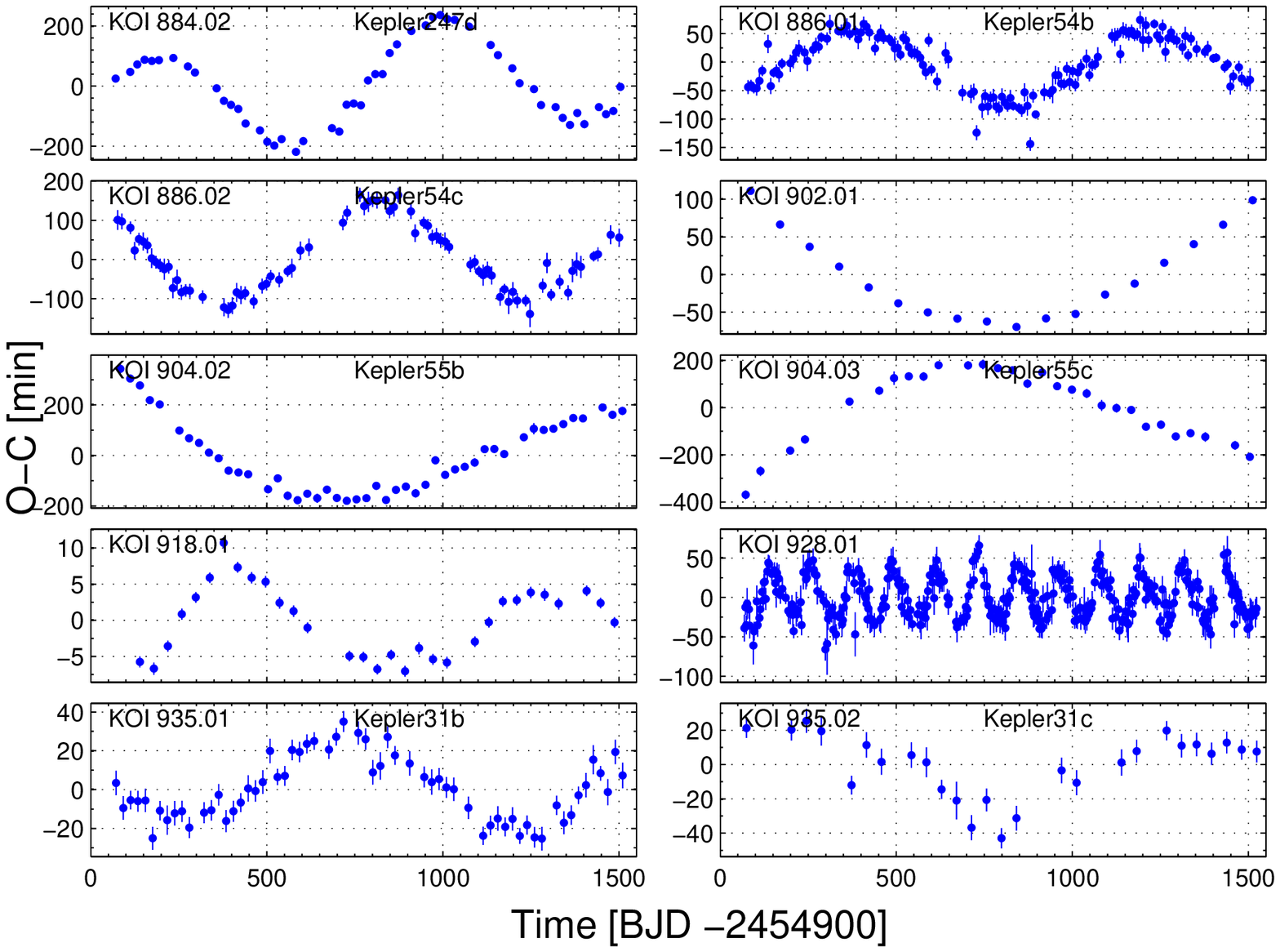}}
\caption{KOIs with significant long-term TTVs.
%See Figure~\ref{TTV1} for details.
}
\label{TTV13}
\end{figure*}
%-------------------------------------------------------------

%---------------------------
% Figure  1+14
%---------------------------
\begin{figure*}[p]
\centering
%\resizebox{16cm}{11cm}
\resizebox{16cm}{11cm} {\includegraphics{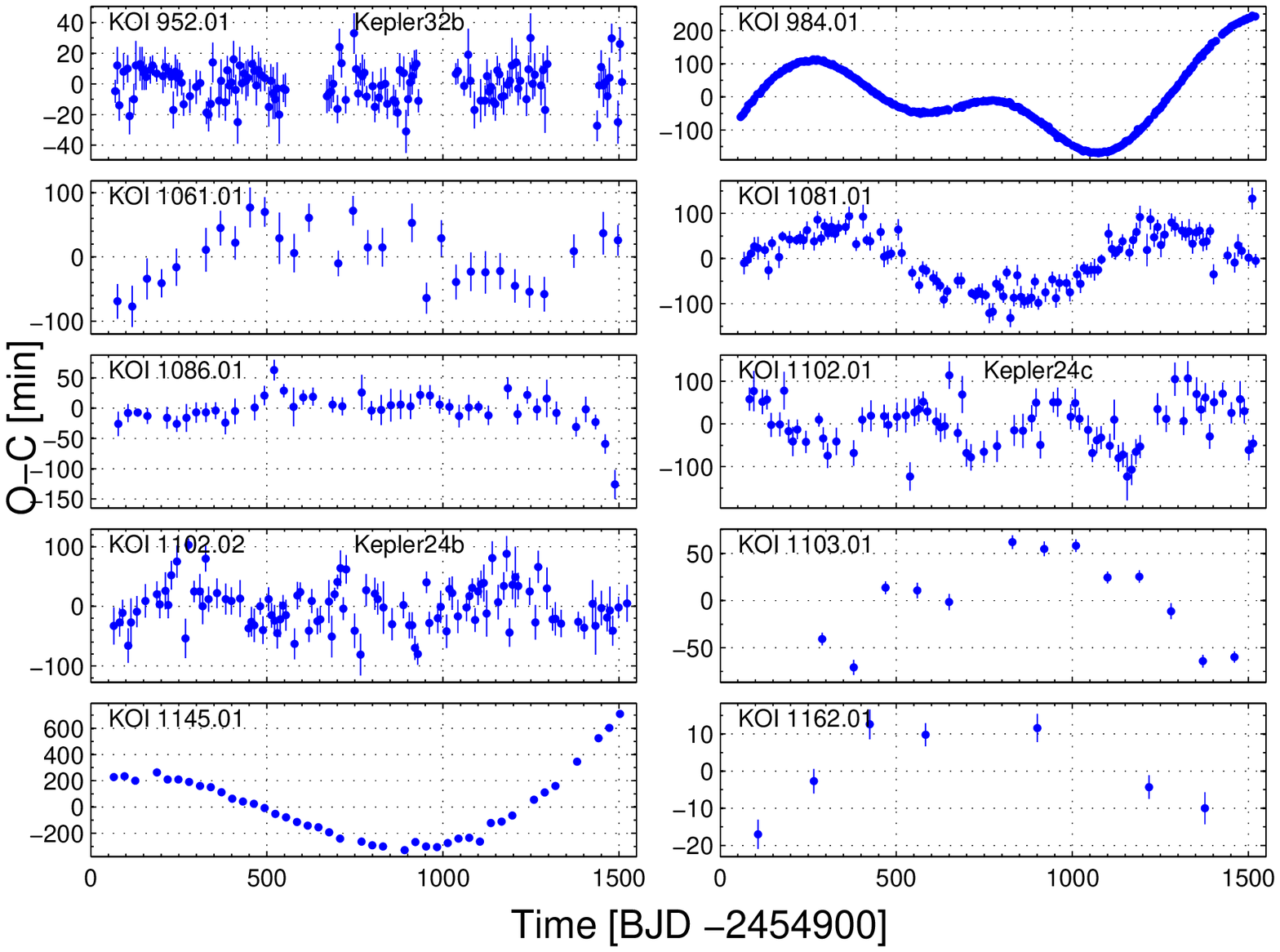}}
\caption{KOIs with significant long-term TTVs.
%See Figure~\ref{TTV1} for details.
}
\label{TTV14}
\end{figure*}
%-------------------------------------------------------------

%---------------------------
% Figure  1+15
%---------------------------
\begin{figure*}[p]
\centering
%\resizebox{16cm}{11cm}
\resizebox{16cm}{11cm} {\includegraphics{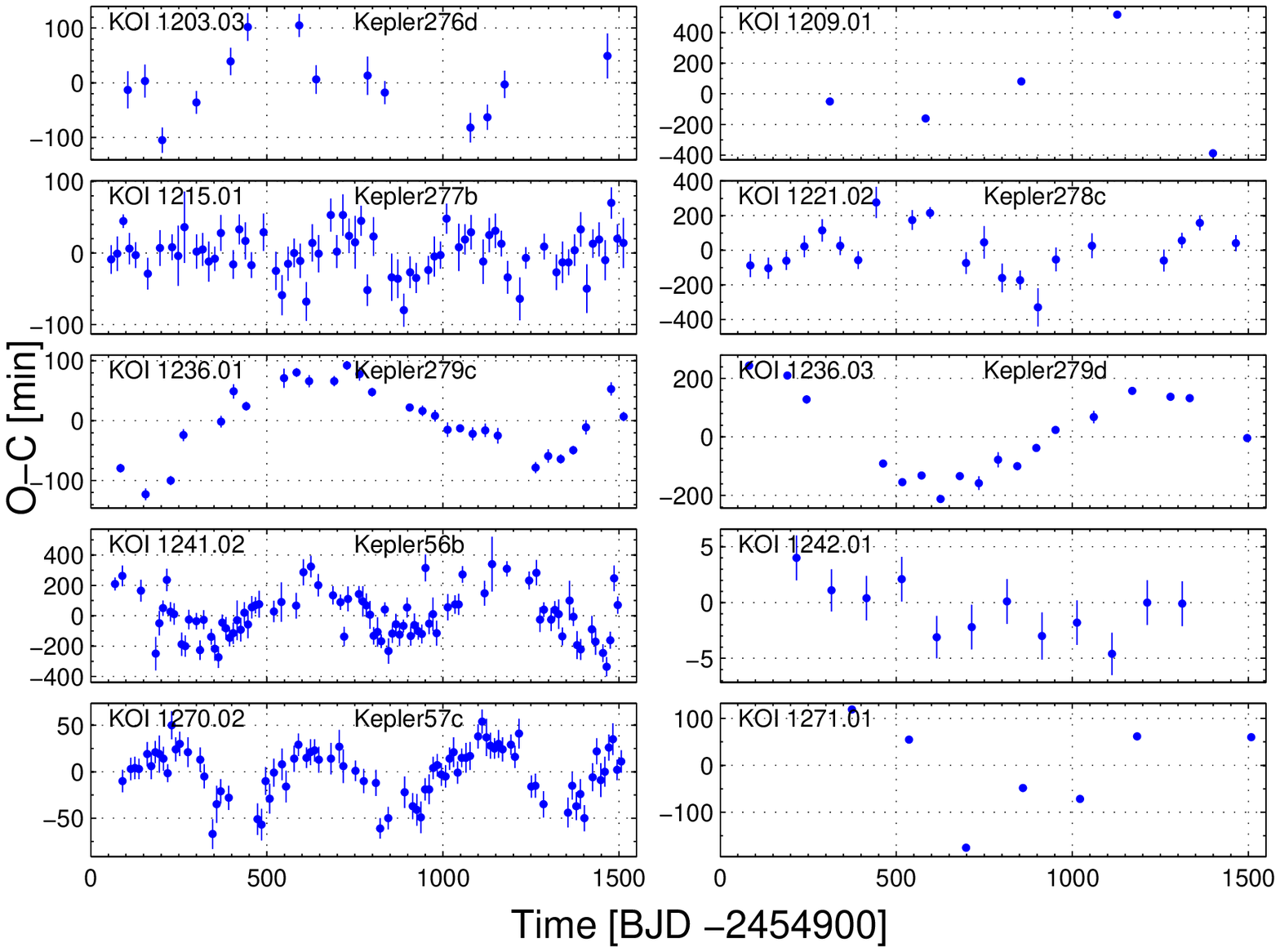}}
\caption{KOIs with significant long-term TTVs.
%See Figure~\ref{TTV1} for details.
}
\label{TTV15}
\end{figure*}
%-------------------------------------------------------------

%---------------------------
% Figure  1+16
%---------------------------
\begin{figure*}[p]
\centering
%\resizebox{16cm}{11cm}
\resizebox{16cm}{11cm} {\includegraphics{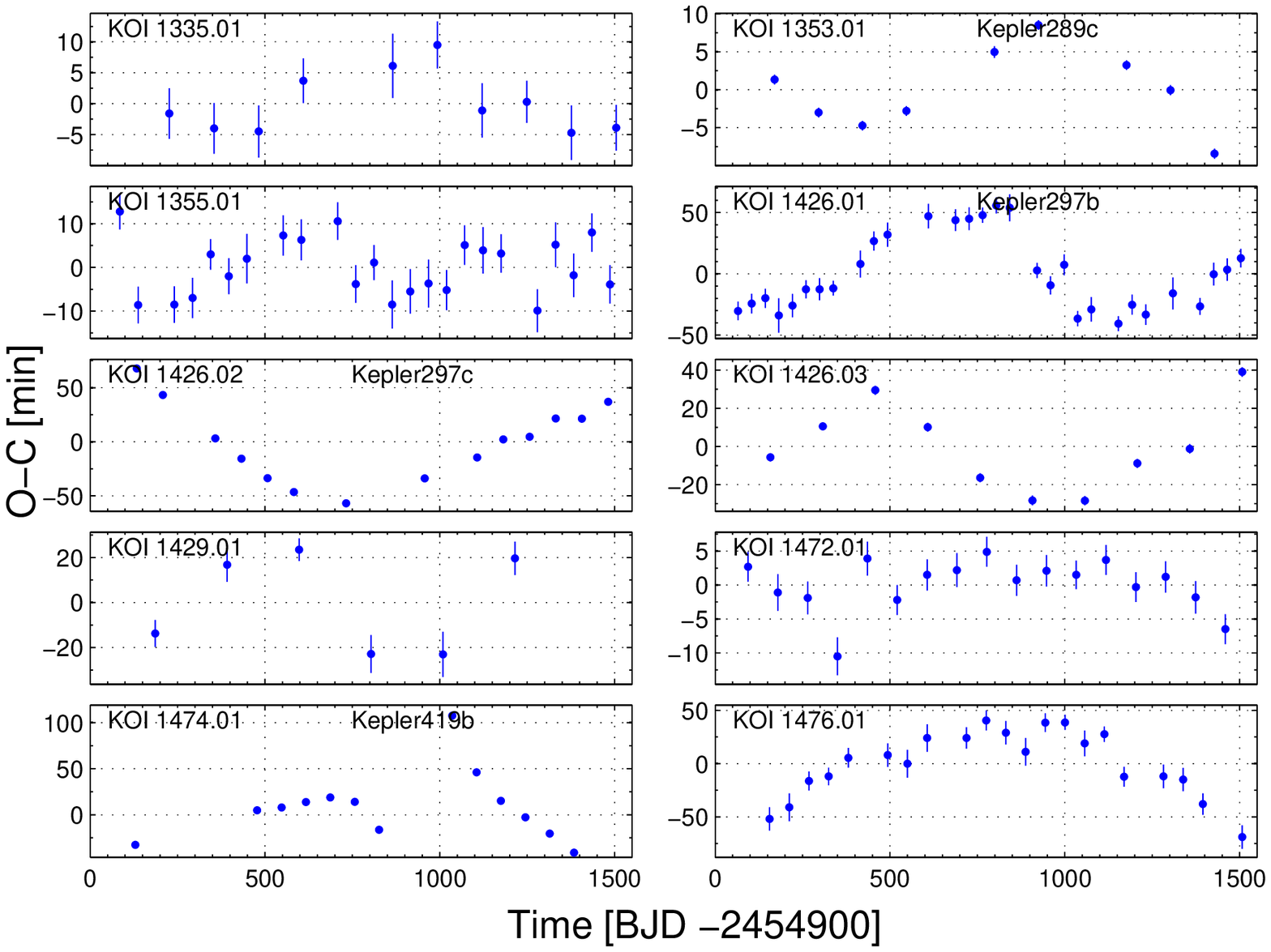}}
\caption{KOIs with significant long-term TTVs.
%See Figure~\ref{TTV1} for details.
}
\label{TTV16}
\end{figure*}
%-------------------------------------------------------------

%---------------------------
% Figure  1+17
%---------------------------
\begin{figure*}[p]
\centering
%\resizebox{16cm}{11cm}
\resizebox{16cm}{11cm} {\includegraphics{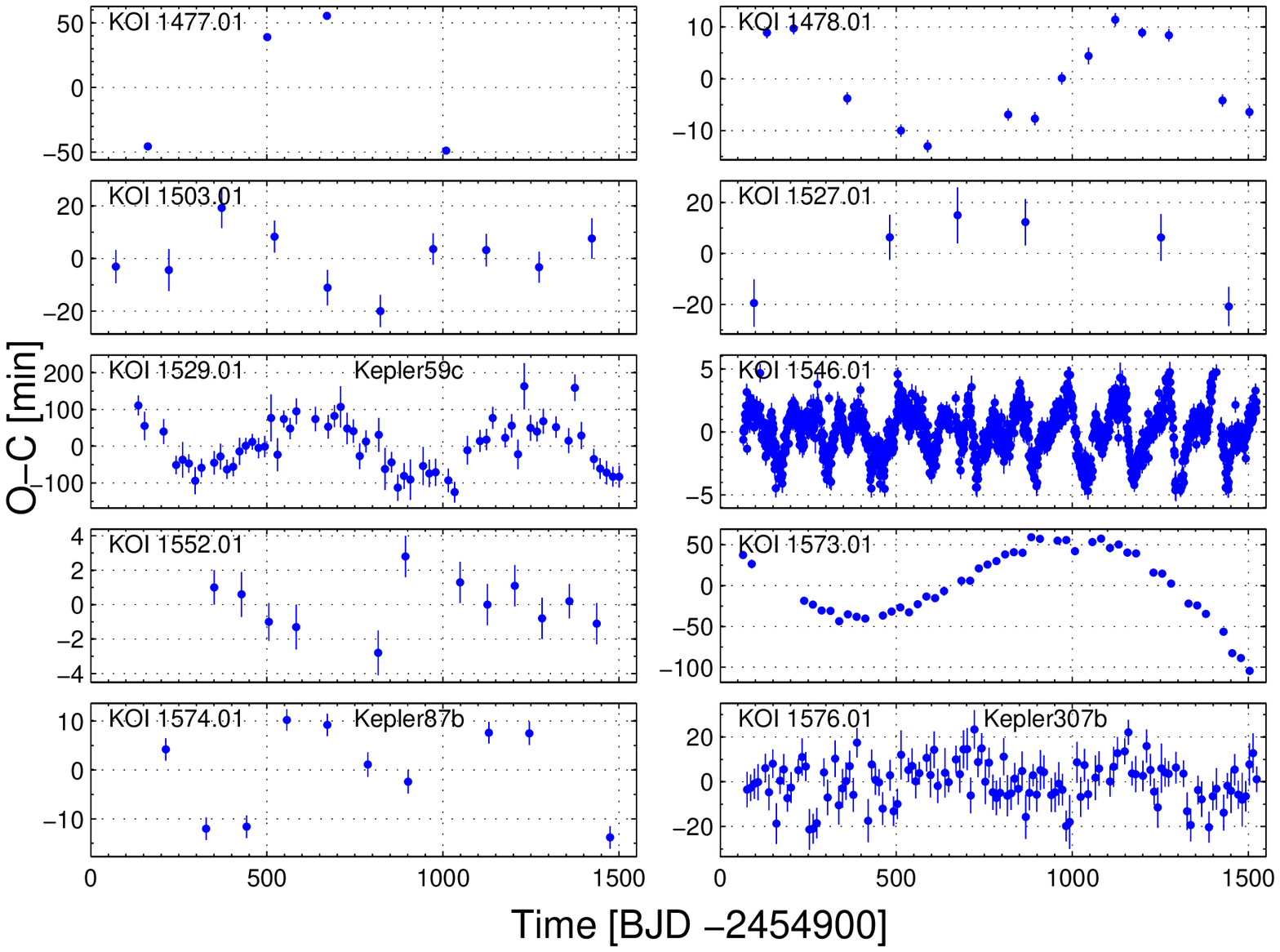}}
\caption{KOIs with significant long-term TTVs.
%See Figure~\ref{TTV1} for details.
}
\label{TTV17}
\end{figure*}
%-------------------------------------------------------------

%---------------------------
% Figure  1+18
%---------------------------
\begin{figure*}[p]
\centering
%\resizebox{16cm}{11cm}
\resizebox{16cm}{11cm} {\includegraphics{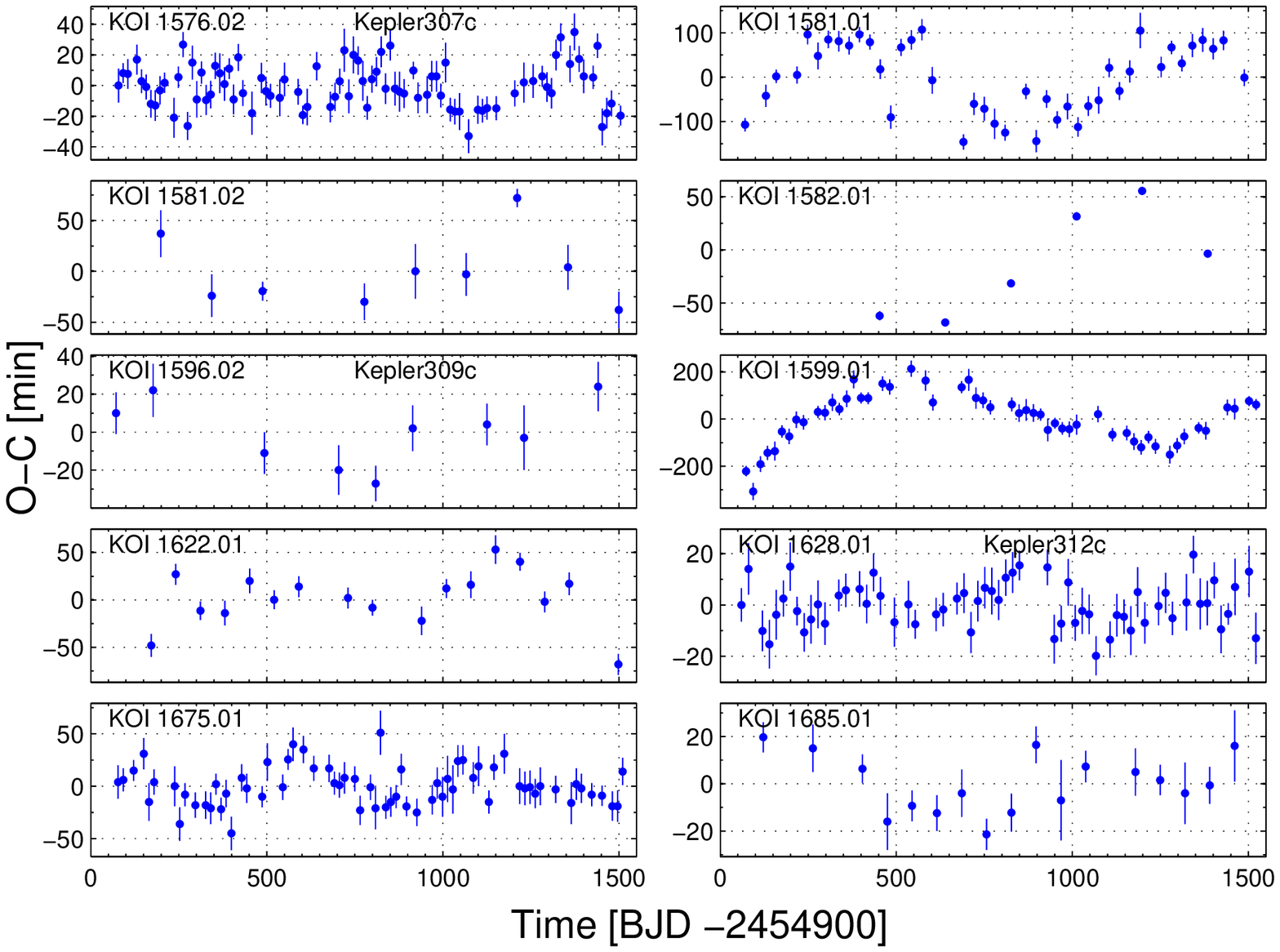}}
\caption{KOIs with significant long-term TTVs.
%See Figure~\ref{TTV1} for details.
}
\label{TTV18}
\end{figure*}
%-------------------------------------------------------------

%---------------------------
% Figure  1+19
%---------------------------
\begin{figure*}[p]
\centering
%\resizebox{16cm}{11cm}
\resizebox{16cm}{11cm} {\includegraphics{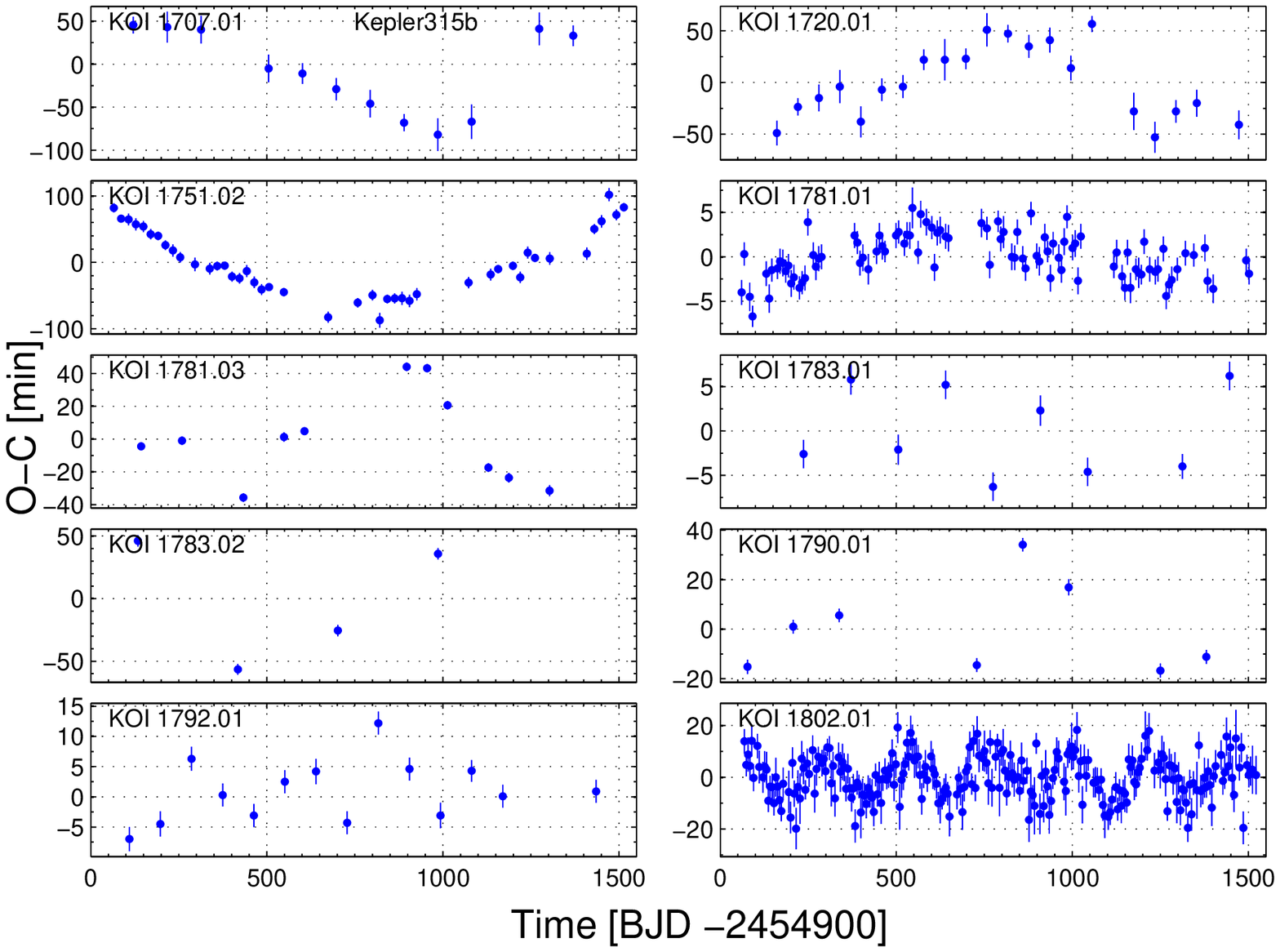}}
\caption{KOIs with significant long-term TTVs.
%See Figure~\ref{TTV1} for details.
}
\label{TTV19}
\end{figure*}
%-------------------------------------------------------------

%---------------------------
% Figure  1+20
%---------------------------
\begin{figure*}[p]
\centering
%\resizebox{16cm}{11cm}
\resizebox{16cm}{11cm} {\includegraphics{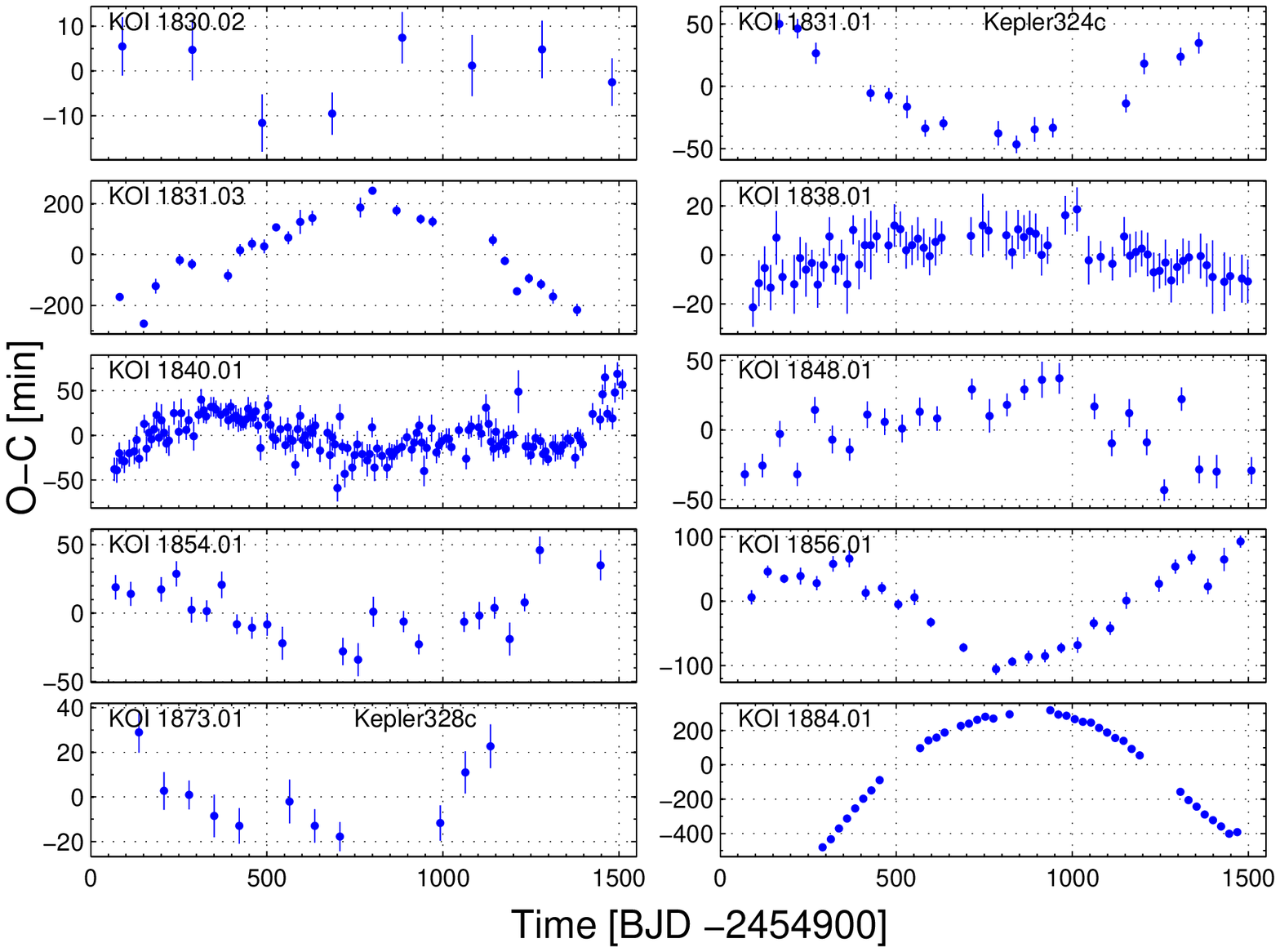}}
\caption{KOIs with significant long-term TTVs.
%See Figure~\ref{TTV1} for details.
}
\label{TTV20}
\end{figure*}
%-------------------------------------------------------------

%---------------------------
% Figure  1+21
%---------------------------
\begin{figure*}[p]
\centering
%\resizebox{16cm}{11cm}
\resizebox{16cm}{11cm} {\includegraphics{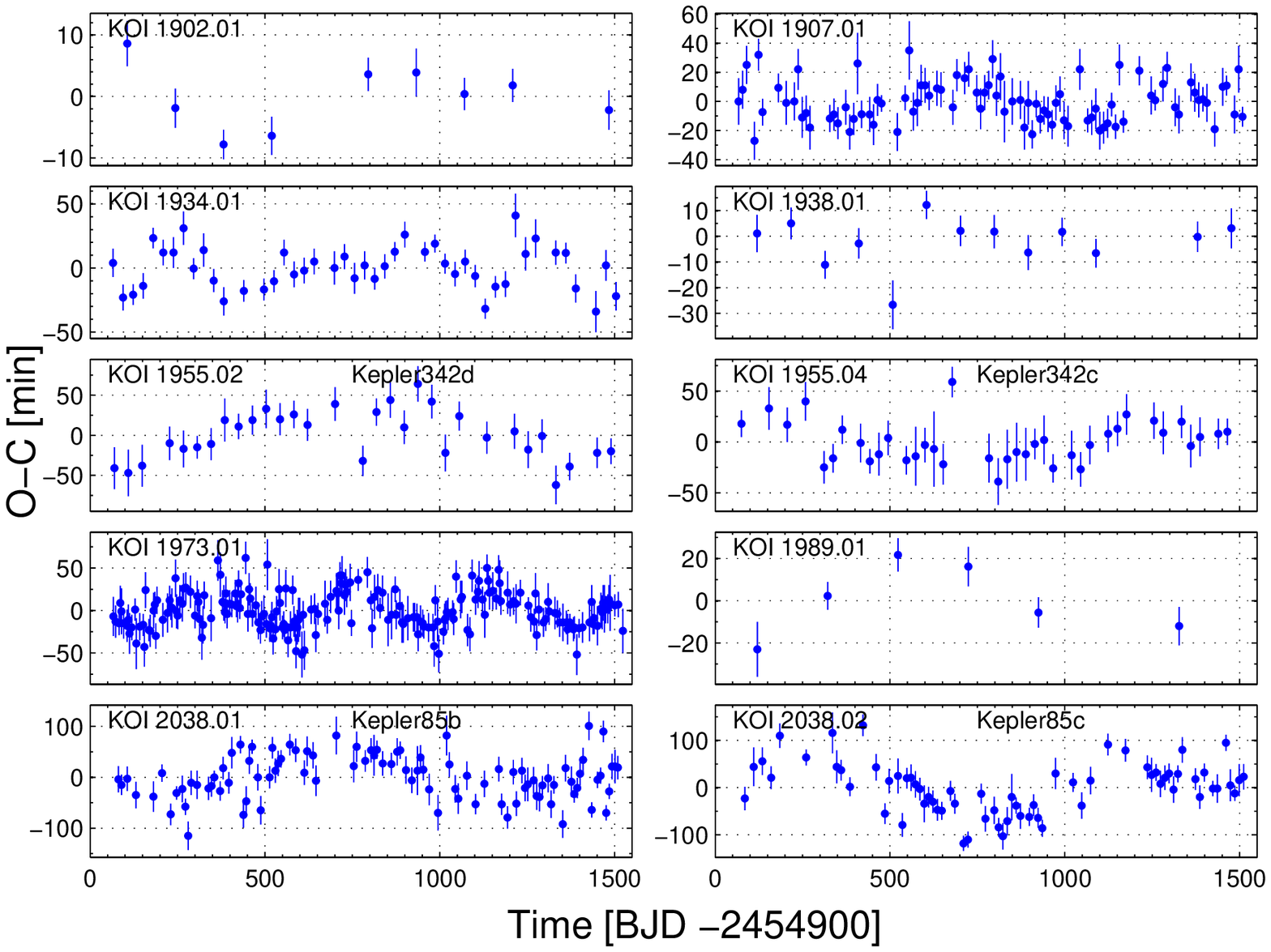}}
\caption{KOIs with significant long-term TTVs.
%See Figure~\ref{TTV1} for details.
}
\label{TTV21}
\end{figure*}
%-------------------------------------------------------------

%---------------------------
% Figure  1+22
%---------------------------
\begin{figure*}[p]
\centering
%\resizebox{16cm}{11cm}
\resizebox{16cm}{11cm} {\includegraphics{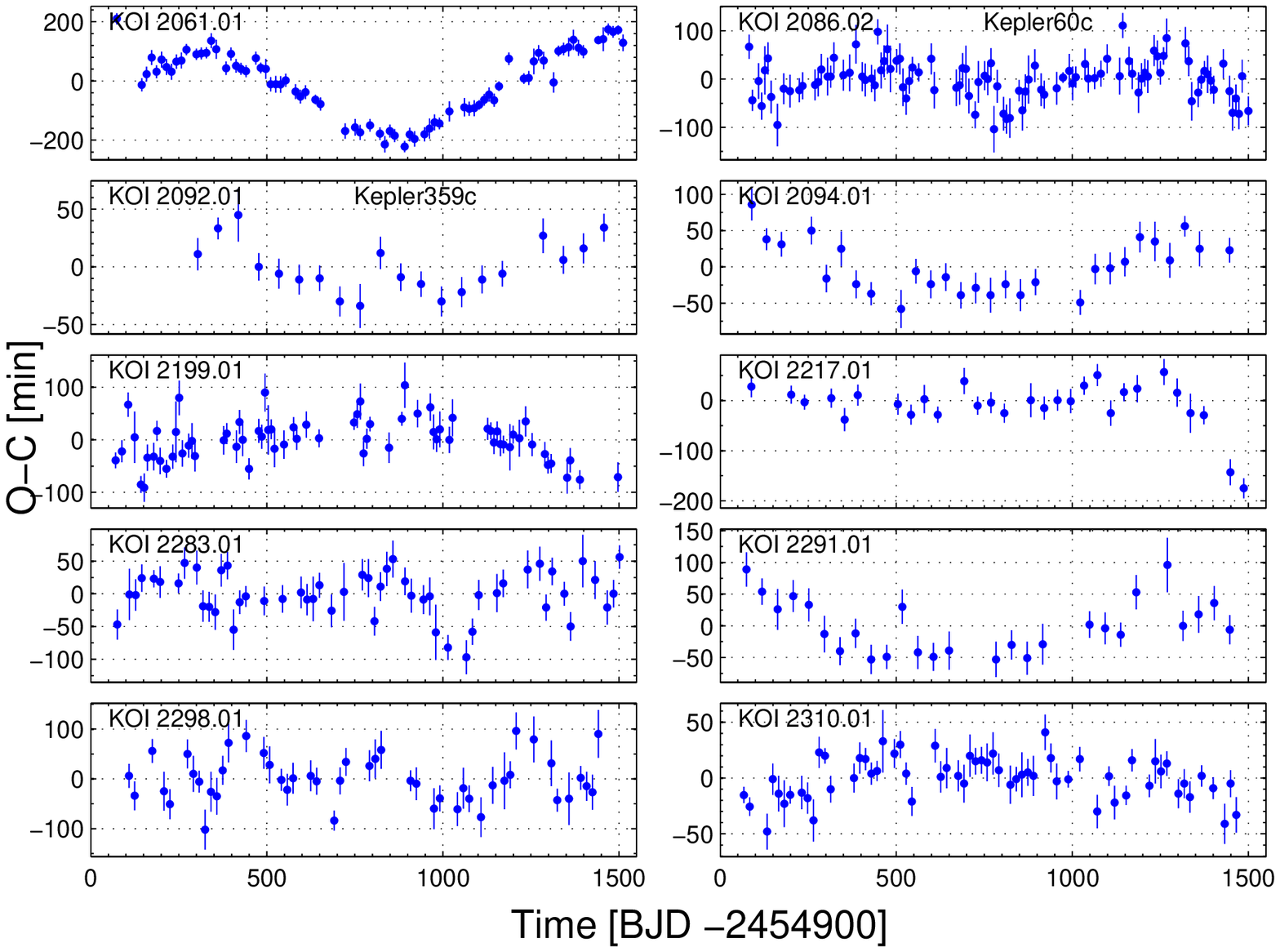}}
\caption{KOIs with significant long-term TTVs.
%See Figure~\ref{TTV1} for details.
}
\label{TTV22}
\end{figure*}
%-------------------------------------------------------------

%---------------------------
% Figure  1+23
%---------------------------
\begin{figure*}[p]
\centering
%\resizebox{16cm}{11cm}
\resizebox{16cm}{11cm} {\includegraphics{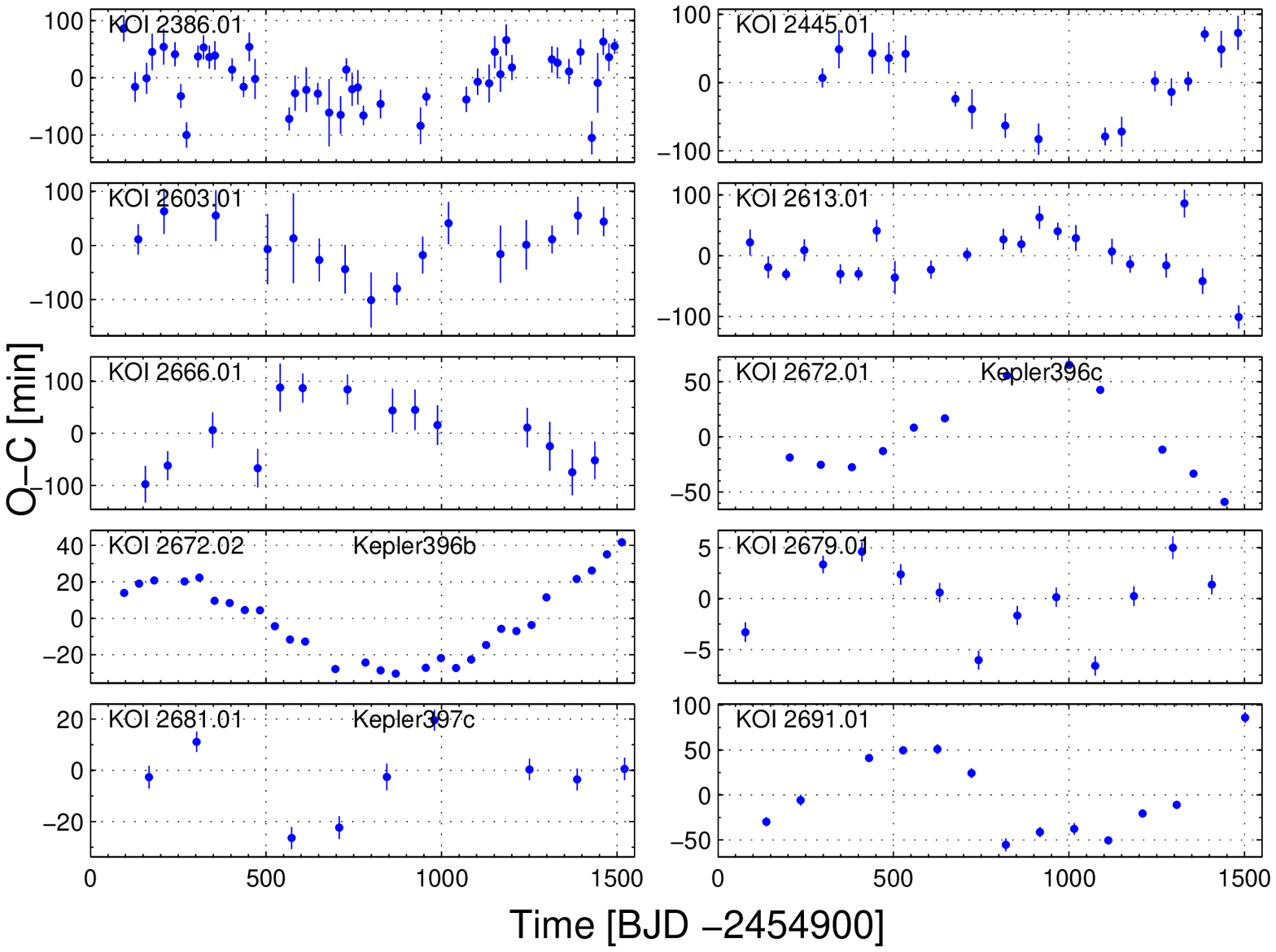}}
\caption{KOIs with significant long-term TTVs.
%See Figure~\ref{TTV1} for details.
}
\label{TTV23}
\end{figure*}
%-------------------------------------------------------------

%---------------------------
% Figure  1+24
%---------------------------
\begin{figure*}[p]
\centering
%\resizebox{16cm}{11cm}
\resizebox{16cm}{11cm} {\includegraphics{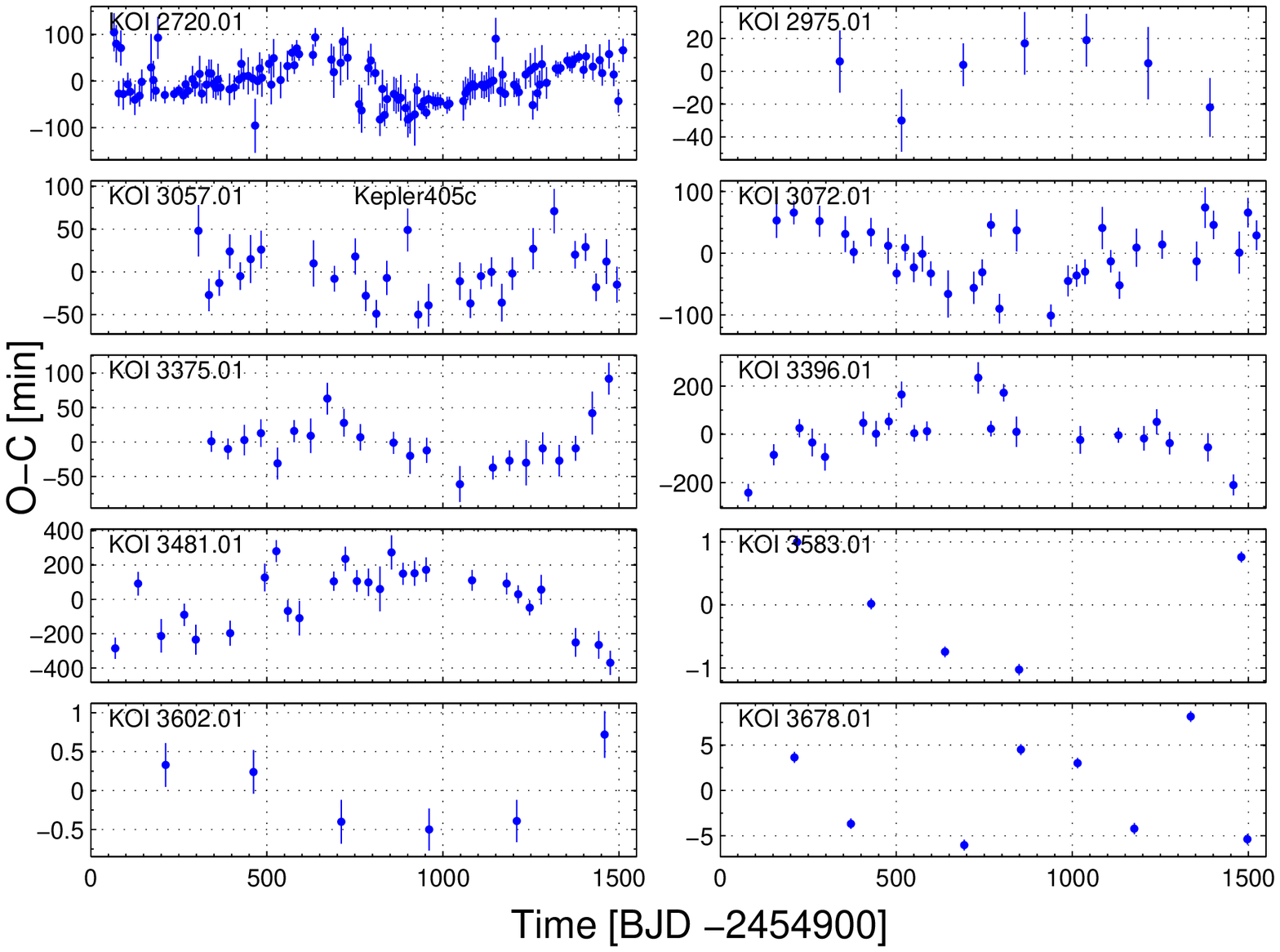}}
\caption{KOIs with significant long-term TTVs.
%See Figure~\ref{TTV1} for details.
}
\label{TTV24}
\end{figure*}
%-------------------------------------------------------------

%---------------------------
% Figure  1+25
%---------------------------
\begin{figure*}[p]
\centering
%\resizebox{16cm}{11cm}
\resizebox{16cm}{11cm} {\includegraphics{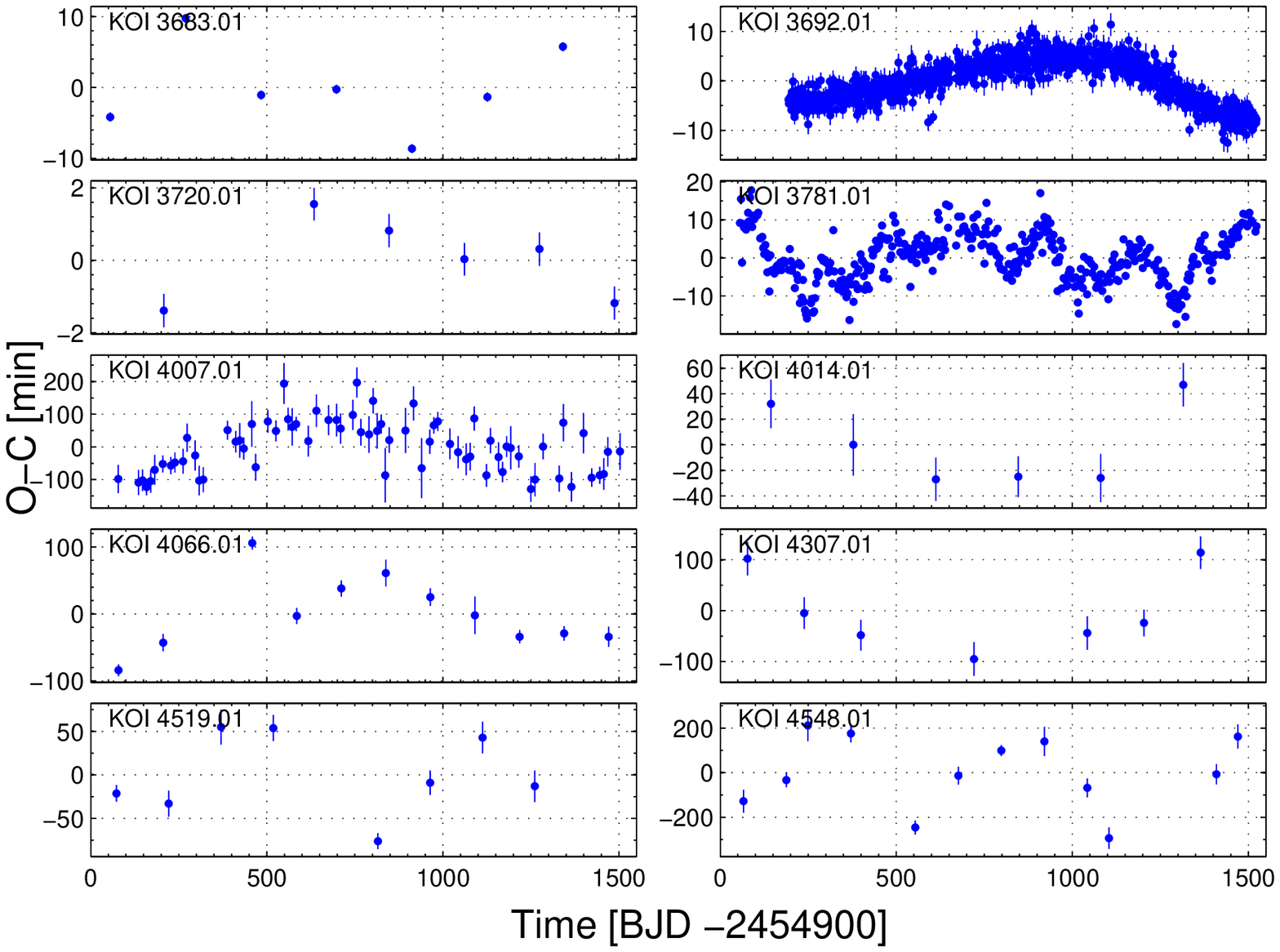}}
\caption{KOIs with significant long-term TTVs.
%See Figure~\ref{TTV1} for details.
}
\label{TTV25}
\end{figure*}
%-------------------------------------------------------------

%---------------------------
% Figure  1+26
%---------------------------
\begin{figure*}[p]
\centering
%\resizebox{16cm}{11cm}
\resizebox{16cm}{11cm} {\includegraphics{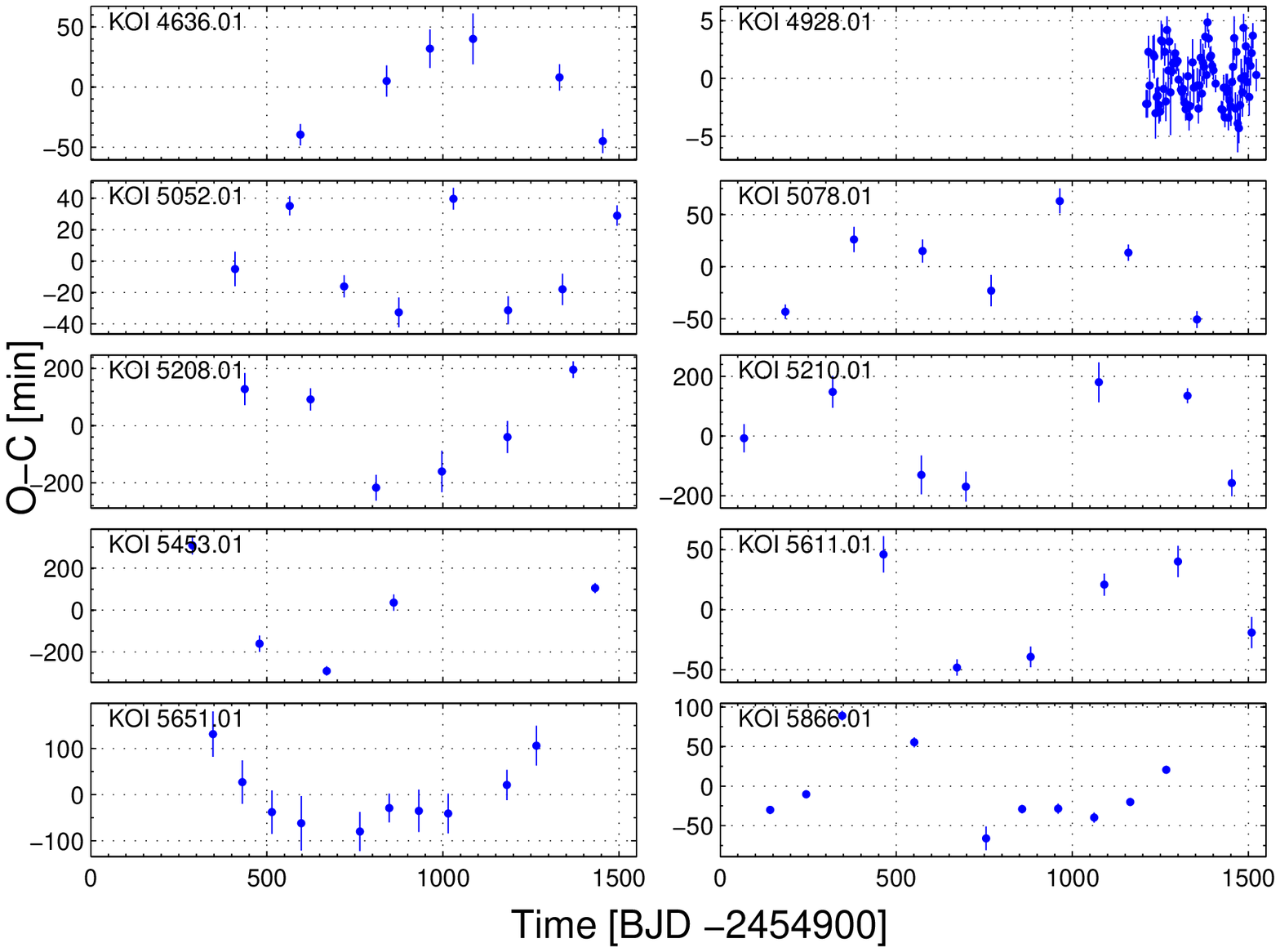}}
\caption{KOIs with significant long-term TTVs.
%See Figure~\ref{TTV1} for details.
}
\label{TTV26}
\end{figure*}
%-------------------------------------------------------------

%---------------------------
% Figure  1+12:1
%---------------------------
\begin{figure*}
\centering
%\resizebox{16cm}{11cm}
\resizebox{16cm}{11cm}
{\includegraphics{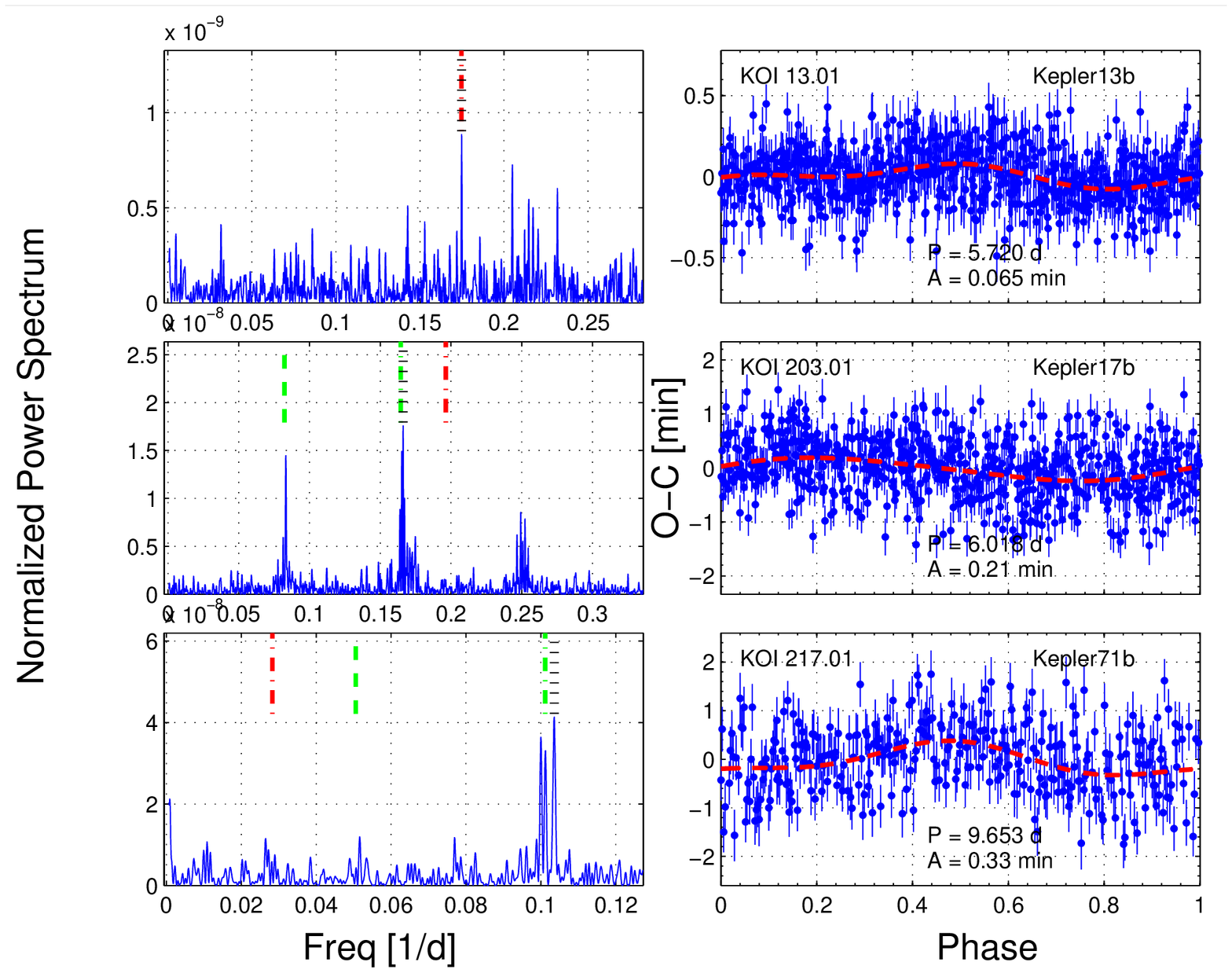}}
\caption{KOIs with short-period TTVs.  For each KOI, the plot shows the power spectrum
periodogram and the phase-folded O-Cs with the period corresponding to the highest peak (marked by a dotted black line). 
To emphasize the modulation, we slightly reduced the $y$-range of the plots of the folded light curves. 
The dashed green line represents
the stellar activity frequency or one of its aliases, if
present in the stellar light curve, and the dashed-dotted red
line represents the frequency induced by the sampling. 
The phase-folded light-curve panels include a two-harmonic fit.
Period and amplitude are listed in each of the right panels.}
\label{TTV_short1}
\end{figure*}
%-------------------------------------------------------------

%---------------------------
% Figure  1+12:2
%---------------------------
\begin{figure*}
\centering
%\resizebox{16cm}{11cm}
\resizebox{16cm}{11cm}
{\includegraphics{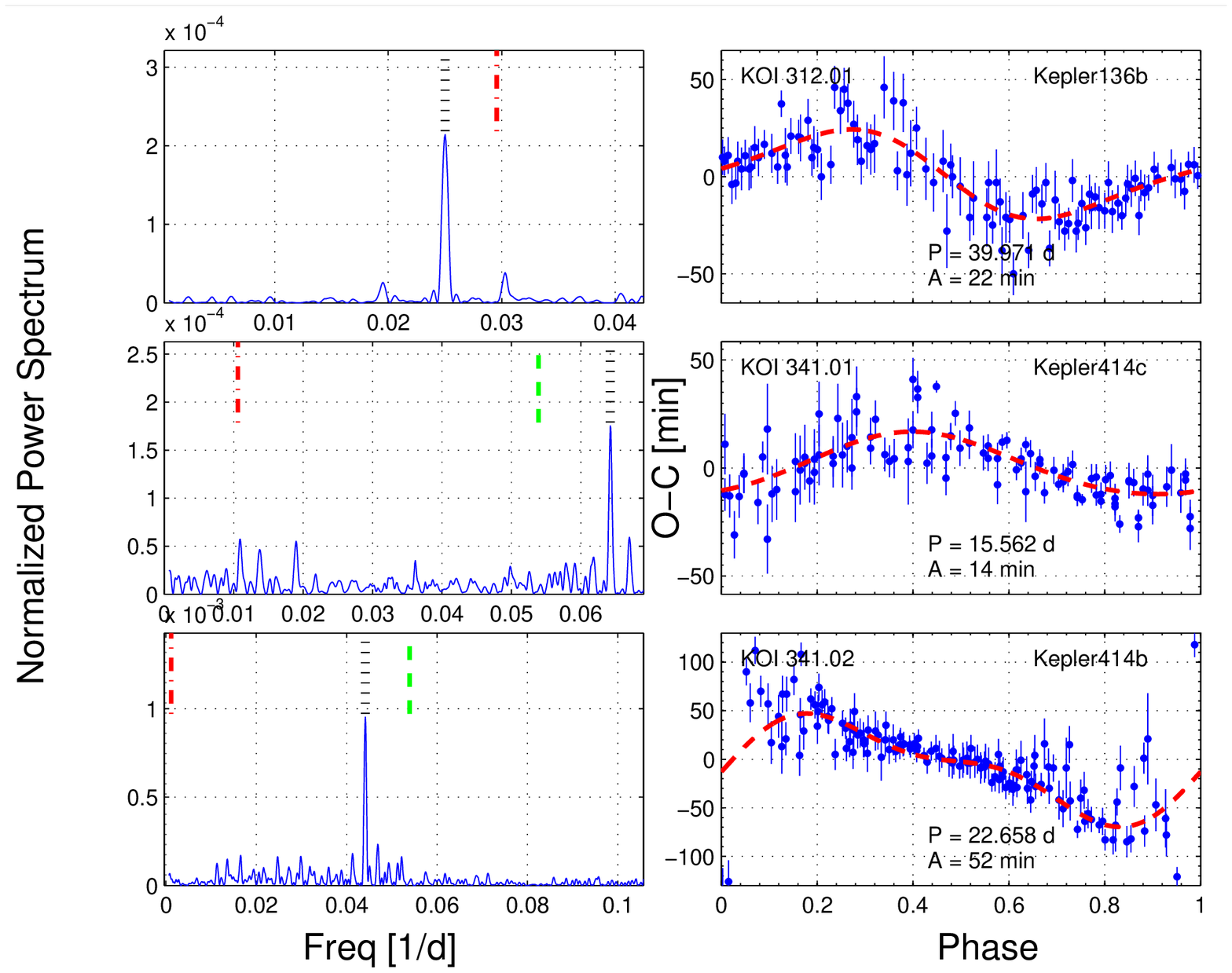}}
\caption{The KOIs with short-period TTVs.  For each KOI, the power spectrum
periodogram
and the phase-folded O-Cs are plotted (see Figure~\ref{TTV_short1} for
details).
}
\label{TTV_short2}
\end{figure*}
%-------------------------------------------------------------

%---------------------------
% Figure  1+12:3
%---------------------------
\begin{figure*}
\centering
%\resizebox{16cm}{11cm}
\resizebox{16cm}{11cm}
{\includegraphics{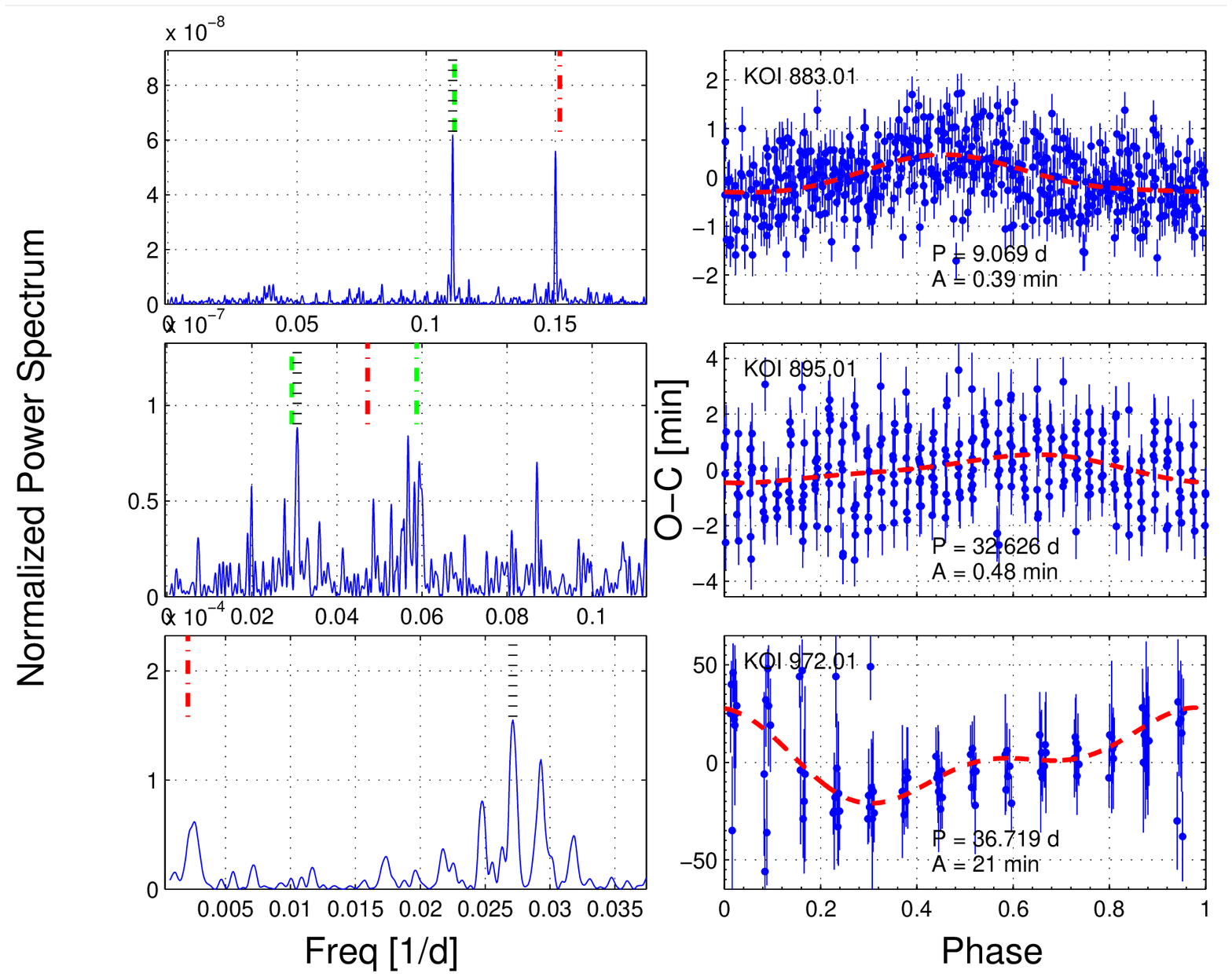}}
\caption{The KOIs with short-period TTVs.  For each KOI, the power spectrum periodogram
and the phase-folded O-Cs are plotted  (see Figure~\ref{TTV_short1} for
details).
}
\label{TTV_short3}
\end{figure*}
%-------------------------------------------------------------

%---------------------------
% Figure  1+12:4
%---------------------------
\begin{figure*}
\centering
%\resizebox{16cm}{11cm}
\resizebox{16cm}{11cm}
{\includegraphics{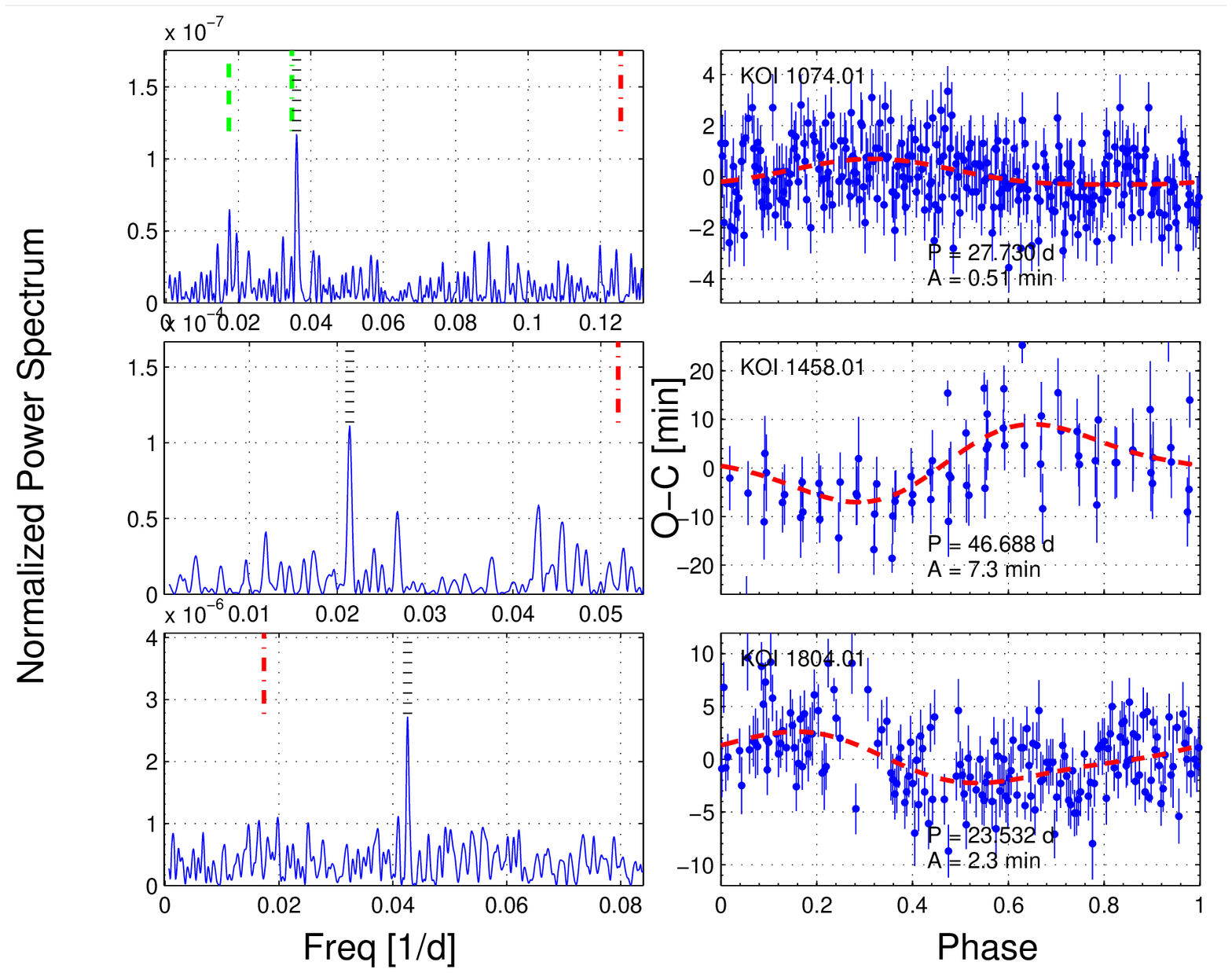}}
\caption{The KOIs with short-period TTVs.  For each KOI, the power spectrum periodogram
and the phase-folded O-Cs are plotted  (see Figure~\ref{TTV_short1} for
details).
}
\label{TTV_short4}
\end{figure*}
%-------------------------------------------------------------

%---------------------------
% Figure  1+12:5
%---------------------------
\begin{figure*}
\centering
%\resizebox{16cm}{11cm}
\resizebox{16cm}{11cm}
{\includegraphics{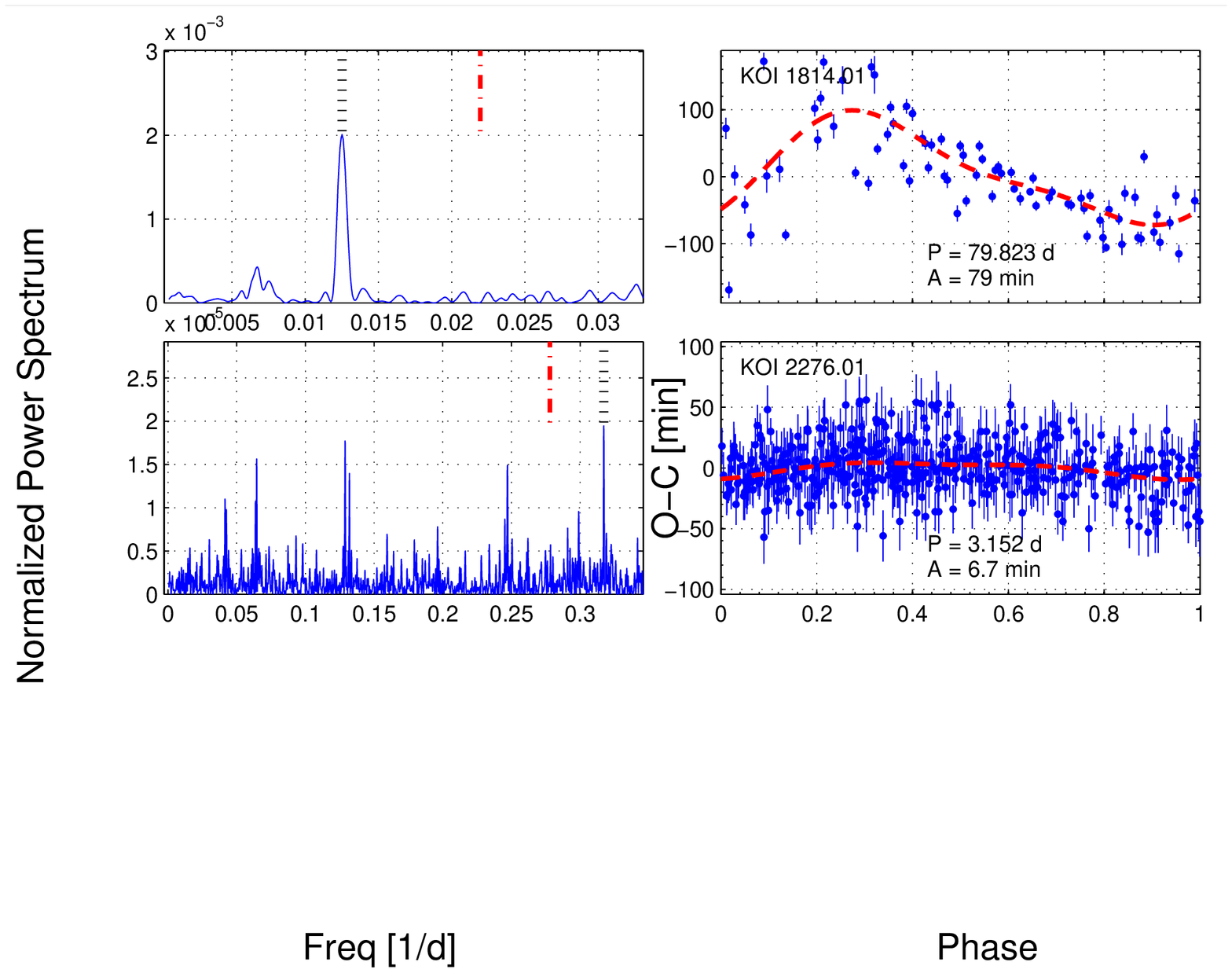}}
\caption{The KOIs with short-period TTVs.  For each KOI, the power spectrum periodogram
and the phase-folded O-Cs are plotted  (see Figure~\ref{TTV_short1} for
details).
}
\label{TTV_short5}
\end{figure*}
%-------------------------------------------------------------

%==============================
%Table 7

\begin{table}[]

\scriptsize  \caption{KOI multies}
\begin{tabular}{|rrrr|rr|rrrrrr|}

\hline \hline
KOI  ~& KOI     & $P_1$\tablenotemark{a} \ \ & $P_2$\tablenotemark{b}  \ \ & $\Delta$\tablenotemark{c} \ \  &    j:j-1\tablenotemark{d} & 
 ${\rm P}_{\rm TTV,1}$\tablenotemark{e}  & $\sigma_{\rm P_1}$\tablenotemark{f}  & 
  ${\rm P}_{\rm TTV,2}$\tablenotemark{g}  &  
$\sigma_{\rm P_2}$\tablenotemark{h}  &
 P$^j$\tablenotemark{i} &  References\\
(Inner) & (Outer) & (Inner)   & (Outer)   &                &           &    (Inner)        &        &   (Outer)  &             &   & \\
        &            & (day)  \ \    & (day)   \ \    &                &           & (day)  	        &  (day)             & (day)  	    &   (day)               &  (day)  ~ &  \\
\hline
$ 137.01 $ & $ 137.02 $ & $ 7.6416 $ & $ 14.8589 $ & $ 0.0278 $ & $ 2:1 $ & $ 269 $ & $ 1 $ & $ 268 $ & $ 1 $ & $ 270 $ &  \tablenotemark{1,19,20}~~~~~~Kepler18 \\ 
$ 152.02 $ & $ 152.01 $ & $ 27.4024 $ & $ 52.0908 $ & $ 0.0495 $ & $ 2:1 $ & $ 963 $ & $ 47 $ & $ 615 $ & $ 13 $ & $ 530 $ & \tablenotemark{2,3,13,20}~~~~~~~~Kepler79 \\ 
$ 152.01 $ & $ 152.04 $ & $ 52.0908 $ & $ 81.0643 $ & $ 0.0375 $ & $ 3:2 $ & $ 614 $ & $ 14 $ & $ 741 $ & $ 38 $ & $ 720 $ & \tablenotemark{3,13,20}~~~~~~Kepler79 \\ 
$ 152.03 $ & $ 152.02 $ & $ 13.4846 $ & $ 27.4024 $ & $ 0.0161 $ & $ 2:1 $ & $ 849 $ & $ 35 $ & $ 963 $ & $ 47 $ & $ 850 $ &  \tablenotemark{2,13,20}~~~~~~Kepler79 \\ 
$ 157.06 $ & $ 157.01 $ & $ 10.3040 $ & $ 13.0249 $ & $ 0.0112 $ & $ 5:4 $ & \nodata & \nodata & $ 229 $ & $ 3 $ & $ 230 $ &  \tablenotemark{4,14,20}~~~~~~Kepler11 \\ 
$ 157.03 $ & $ 157.04 $ & $ 31.9954 $ & $ 46.6857 $ & $ 0.0272 $ & $ 3:2 $ & $ 183 $ & $ 2 $ & $ 558 $ & $ 12 $ & $ 570 $ & \tablenotemark{4,14,20}~~~~~~Kepler11 \\ 
$ 157.02 $ & $ 157.03 $ & $ 22.6872 $ & $ 31.9954 $ & $ 0.0577 $ & $ 4:3 $ & \nodata & \nodata & $ \tablenotemark{j} 139$ & $ 2 $ & $ 140 $ & \tablenotemark{4,14,20}~~~~~~Kepler11 \\ 
$ 157.02 $ & $ 157.04 $ & $ 22.6872 $ & $ 46.6857 $ & $ 0.0289 $ & $ 2:1 $ & \nodata & \nodata & $ 558 $ & $ 11 $ & $ 810 $ & \tablenotemark{4,14,20}~~~~~~Kepler11 \\ 
$ 168.03 $ & $ 168.01 $ & $ 7.1070 $ & $ 10.7425 $ & $ 0.0077 $ & $ 3:2 $ & $ 452 $ & $ 12 $ & $ 443 $ & $ 6 $ & $ 470 $ & \tablenotemark{5,19,20}~~~~~~Kepler23 \\ 
$ 244.02 $ & $ 244.01 $ & $ 6.2385 $ & $ 12.7204 $ & $ 0.0195 $ & $ 2:1 $ & $ 326 $ & $ 3 $ & $ 329 $ & $ 4 $ & $ 330 $ &  \tablenotemark{6,19,20}~~~~~~Kepler25 \\ 
$ 248.01 $ & $ 248.02 $ & $ 7.2039 $ & $ 10.9127 $ & $ 0.0099 $ & $ 3:2 $ & $ 368 $ & $ 3 $ & $ 370 $ & $ 4 $ & $ 370 $ & \tablenotemark{7,20}~~~Kepler49 \\ 
$ 250.01 $ & $ 250.02 $ & $ 12.2830 $ & $ 17.2512 $ & $ 0.0534 $ & $ 4:3 $ & $ 751 $ & $ 13 $ & $ 734 $ & $ 18 $ & $ 81 $ & \tablenotemark{6}Kepler26 \\ 
$ 262.01 $ & $ 262.02 $ & $ 7.8128 $ & $ 9.3766 $ & $ 0.0001 $ & $ 6:5 $ & $ 1060 $ & $ 24 $ & $ 664 $ & $ 11 $ & $ 10000 $ & \tablenotemark{7}Kepler50 \\ 
$ 271.03 $ & $ 271.02 $ & $ 14.4360 $ & $ 29.3933 $ & $ 0.0181 $ & $ 2:1 $ & \nodata & \nodata & $ 1086 $ & $ 65 $ & $ 810 $ & \tablenotemark{3}Kepler127 \\ 
$ 274.01 $ & $ 274.02 $ & $ 15.0894 $ & $ 22.8039 $ & $ 0.0075 $ & $ 3:2 $ & $ 954 $ & $ 29 $ & $ 922 $ & $ 22 $ & $ 1000 $ & \tablenotemark{2,20}~~~Kepler128 \\ 
$ 277.02 $ & $ 277.01 $ & $ 13.8487 $ & $ 16.2321 $ & $ 0.0047 $ & $ 7:6 $ & $ 439 $ & $ 2 $ & $ 449 $ & $ 1 $ & $ 500 $ & \tablenotemark{8}Kepler36 \\ 
$ 314.03 $ & $ 314.01 $ & $ 10.3120 $ & $ 13.7811 $ & $ 0.0023 $ & $ 4:3 $ & \nodata & \nodata & $ 1204 $ & $ 89 $ & $ 1500 $ & \tablenotemark{3,9,20}~~~~~Kepler138 \\ 
$ 377.01 $ & $ 377.02 $ & $ 19.2452 $ & $ 38.9568 $ & $ 0.0121 $ & $ 2:1 $ & $ 1351 $ & $ 0 $ & $ 1353 $ & $ 0 $ & $ 1600 $ & \tablenotemark{10,20}~~~~Kepler9 \\ 
$ 500.01 $ & $ 500.02 $ & $ 7.0535 $ & $ 9.5216 $ & $ 0.0124 $ & $ 4:3 $ & $ 190 $ & $ 2 $ & $ 193 $ & $ 1 $ & $ 190 $ & \tablenotemark{2,17}~~~Kepler80 \\ 
$ 500.04 $ & $ 500.01 $ & $ 4.6453 $ & $ 7.0535 $ & $ 0.0123 $ & $ 3:2 $ & \nodata & \nodata & $ 190 $ & $ 2 $ & $ 190 $ & \tablenotemark{2,17}~~~Kepler80 \\ 
$ 500.04 $ & $ 500.02 $ & $ 4.6453 $ & $ 9.5216 $ & $ 0.0249 $ & $ 2:1 $ & \nodata & \nodata & $ 193 $ & $ 1 $ & $ 190 $ & \tablenotemark{2,17}~~~Kepler80 \\ 
$ 520.01 $ & $ 520.03 $ & $ 12.7594 $ & $ 25.7526 $ & $ 0.0092 $ & $ 2:1 $ & $ 1294 $ & $ 54 $ & $ 1306 $ & $ 70 $ & $ 1400 $ & \tablenotemark{3,20}~~~Kepler176 \\ 
$ 620.01 $ & $ 620.03 $ & $ 45.1554 $ & $ 85.3168 $ & $ 0.0553 $ & $ 2:1 $ & $ 789 $ & $ 12 $ & $ 1084 $ & $ 67 $ & $ 770 $ & \tablenotemark{7,20}~~~Kepler51 \\ 
$ 620.03 $ & $ 620.02 $ & $ 85.3168 $ & $ 130.1781 $ & $ 0.0172 $ & $ 3:2 $ & $ 1084 $ & $ 67 $ & \nodata & \nodata & $ 2500 $ & \tablenotemark{7,11,20}~~~~~~Kepler51 \\ 
$ 730.01 $ & $ 730.03 $ & $ 14.7869 $ & $ 19.7257 $ & $ 0.0005 $ & $ 4:3 $ & $ 1292 $ & $ 79 $ & $ 1690 $ & $ 100 $ & $ 10000 $ & \tablenotemark{3}Kepler223 \\ 
$ 730.02 $ & $ 730.01 $ & $ 9.8479 $ & $ 14.7869 $ & $ 0.0010 $ & $ 3:2 $ & \nodata & \nodata & $ 1292 $ & $ 79 $ & $ 5000 $ & \tablenotemark{3}Kepler223 \\ 
$ 730.04 $ & $ 730.01 $ & $ 7.3840 $ & $ 14.7869 $ & $ 0.0013 $ & $ 2:1 $ & \nodata & \nodata & $ 1292 $ & $ 79 $ & $ 6000 $ & \tablenotemark{3}Kepler223 \\ 
$ 730.02 $ & $ 730.03 $ & $ 9.8479 $ & $ 19.7257 $ & $ 0.0015 $ & $ 2:1 $ & \nodata & \nodata & $ 1690 $ & $ 100 $ & $ 6500 $ & \tablenotemark{3}Kepler223 \\ 
$ 775.02 $ & $ 775.01 $ & $ 7.8774 $ & $ 16.3848 $ & $ 0.0400 $ & $ 2:1 $ & $ 206 $ & $ 1 $ & \nodata & \nodata & $ 200 $ & \tablenotemark{7,20}~~~Kepler52 \\ 

\hline 

\end{tabular}
%**********

\label{tab:multi}
\end{table}

\addtocounter{table}{-1}
\begin{table}[]
\scriptsize \caption{KOI multies}
\begin{tabular}{|rrrr|rr|rrrrrr|}

\hline \hline
KOI  ~& KOI     & $P_1$\tablenotemark{a} \ \ & $P_2$\tablenotemark{b}  \ \ & $\Delta$\tablenotemark{c} \ \  &    j:j-1\tablenotemark{d} & 
 ${\rm P}_{\rm TTV,1}$\tablenotemark{e}  & $\sigma_{\rm P_1}$\tablenotemark{f}  & 
  ${\rm P}_{\rm TTV,2}$\tablenotemark{g}  &  
$\sigma_{\rm P_2}$\tablenotemark{h}  &
 P$^j$\tablenotemark{i} &  References\\
(Inner) & (Outer) & (Inner)   & (Outer)   &                &           &    (Inner)        &        &   (Outer)  &             &   & \\
        &            & (day)  \ \    & (day)   \ \    &                &           & (day)  	        &  (day)             & (day)  	    &   (day)               &  (day)  ~ &  \\
\hline
$ 806.03 $ & $ 806.02 $ & $ 29.3235 $ & $ 60.3251 $ & $ 0.0286 $ & $ 2:1 $ & $ 985 $ & $ 1 $ & $ 969 $ & $ 3 $ & $ 1100 $ & \tablenotemark{12,15,20}~~~~~~~Kepler30 \\ 
$ 829.01 $ & $ 829.03 $ & $ 18.6493 $ & $ 38.5579 $ & $ 0.0338 $ & $ 2:1 $ & $ 554 $ & $ 20 $ & $ 496 $ & $ 15 $ & $ 570 $ & \tablenotemark{7,20}~~~Kepler53 \\ 
$ 834.01 $ & $ 834.05 $ & $ 23.6537 $ & $ 50.4472 $ & $ 0.0664 $ & $ 2:1 $ & $ 382 $ & $ 7 $ & \nodata & \nodata & $ 380 $ & \tablenotemark{2}Kepler238 \\ 
$ 841.01 $ & $ 841.02 $ & $ 15.3354 $ & $ 31.3302 $ & $ 0.0215 $ & $ 2:1 $ & $ 723 $ & $ 11 $ & $ 717 $ & $ 8 $ & $ 730 $ & \tablenotemark{6,20}~~~Kepler27 \\ 
$ 869.03 $ & $ 869.02 $ & $ 17.4608 $ & $ 36.2755 $ & $ 0.0388 $ & $ 2:1 $ & \nodata & \nodata & $ 597 $ & $ 8 $ & $ 470 $ & \tablenotemark{3}Kepler245 \\ 
$ 870.01 $ & $ 870.02 $ & $ 5.9123 $ & $ 8.9858 $ & $ 0.0132 $ & $ 3:2 $ & $ 230 $ & $ 2 $ & $ 230 $ & $ 3 $ & $ 230 $ &  \tablenotemark{6,19,20}~~~~~~Kepler28 \\ 
$ 877.01 $ & $ 877.02 $ & $ 5.9549 $ & $ 12.0399 $ & $ 0.0109 $ & $ 2:1 $ & \nodata & \nodata & $ 535 $ & $ 12 $ & $ 550 $ & \tablenotemark{2,20}~~~Kepler81 \\ 
$ 880.01 $ & $ 880.02 $ & $ 26.4445 $ & $ 51.5383 $ & $ 0.0255 $ & $ 2:1 $ & $ 969 $ & $ 34 $ & $ 1213 $ & $ 13 $ & $ 1000 $ & \tablenotemark{2,20}~~~Kepler82 \\ 
$ 886.01 $ & $ 886.02 $ & $ 8.0108 $ & $ 12.0715 $ & $ 0.0046 $ & $ 3:2 $ & $ 850 $ & $ 5 $ & $ 852 $ & $ 6 $ & $ 870 $ & \tablenotemark{7,20}~~~Kepler54 \\ 
$ 935.01 $ & $ 935.02 $ & $ 20.8602 $ & $ 42.6341 $ & $ 0.0219 $ & $ 2:1 $ & $ 940 $ & $ 13 $ & $ 1162 $ & $ 42 $ & $ 970 $ & \tablenotemark{12,20}~~~~~Kepler31 \\ 
$ 935.02 $ & $ 935.03 $ & $ 42.6341 $ & $ 87.6476 $ & $ 0.0279 $ & $ 2:1 $ & $ 1162 $ & $ 42 $ & \nodata & \nodata & $ 1600 $ &  \tablenotemark{12,20}~~~~~Kepler31  \\ 
$ 952.01 $ & $ 952.02 $ & $ 5.9013 $ & $ 8.7521 $ & $ 0.0113 $ & $ 3:2 $ & $ 267 $ & $ 4 $ & \nodata & \nodata & $ 260 $ & \tablenotemark{12,19,20}~~~~~~~Kepler32 \\ 
$ 1102.02 $ & $ 1102.01 $ & $ 8.1451 $ & $ 12.3335 $ & $ 0.0095 $ & $ 3:2 $ & $ 440 $ & $ 8 $ & $ 413 $ & $ 7 $ & $ 430 $ & \tablenotemark{5,19,20}~~~~~~Kepler24 \\ 
$ 1203.01 $ & $ 1203.03 $ & $ 31.8838 $ & $ 48.6457 $ & $ 0.0171 $ & $ 3:2 $ & \nodata & \nodata & $ 821 $ & $ 38 $ & $ 950 $ & \tablenotemark{2,20}~~~Kepler276 \\ 
$ 1215.01 $ & $ 1215.02 $ & $ 17.3240 $ & $ 33.0067 $ & $ 0.0474 $ & $ 2:1 $ & $ 348 $ & $ 8 $ & \nodata & \nodata & $ 350 $ & \tablenotemark{2,20}~~~Kepler277 \\ 
$ 1236.01 $ & $ 1236.03 $ & $ 35.7353 $ & $ 54.4117 $ & $ 0.0151 $ & $ 3:2 $ & $ 1194 $ & $ 23 $ & $ 1254 $ & $ 21 $ & $ 1200 $ & \tablenotemark{2,18}~~~Kepler279 \\ 
$ 1241.02 $ & $ 1241.01 $ & $ 10.5012 $ & $ 21.4057 $ & $ 0.0192 $ & $ 2:1 $ & $ 542 $ & $ 6 $ & \nodata & \nodata & $ 560 $ & \tablenotemark{7,20}~~~Kepler56 \\ 
$ 1270.01 $ & $ 1270.02 $ & $ 5.7293 $ & $ 11.6092 $ & $ 0.0131 $ & $ 2:1 $ & \nodata & \nodata & $ 456 $ & $ 5 $ & $ 440 $ & \tablenotemark{7,20}~~~Kepler57 \\ 
$ 1426.01 $ & $ 1426.02 $ & $ 38.8684 $ & $ 74.9285 $ & $ 0.0361 $ & $ 2:1 $ & $ 1040 $ & $ 20 $ & \nodata & \nodata & $ 1000 $ & \tablenotemark{3,16,20}~~~~~~Kepler297 \\ 
$ 1426.02 $ & $ 1426.03 $ & $ 74.9287 $ & $ 150.0254 $ & $ 0.0011 $ & $ 2:1 $ & \nodata & \nodata & $ 1322 $ & $ 60 $ & $ 60000 $ & \tablenotemark{3,16,20}~~~~~~Kepler297 \\ 
$ 1529.02 $ & $ 1529.01 $ & $ 11.8682 $ & $ 17.9770 $ & $ 0.0098 $ & $ 3:2 $ & \nodata & \nodata & $ 618 $ & $ 9 $ & $ 610 $ & \tablenotemark{7}Kepler59 \\ 
$ 1576.01 $ & $ 1576.02 $ & $ 10.4158 $ & $ 13.0842 $ & $ 0.0050 $ & $ 5:4 $ & $ 517 $ & $ 24 $ & $ 481 $ & $ 20 $ & $ 530 $ &  \tablenotemark{2,20}~~~Kepler307 \\ 
$ 1599.02 $ & $ 1599.01 $ & $ 13.6141 $ & $ 20.4117 $ & $ 0.0005 $ & $ 3:2 $ & \nodata & \nodata & $ 1487 $ & $ 37 $ & $ 15000 $ &  \\ 
$ 1783.01 $ & $ 1783.02 $ & $ 134.4793 $ & $ 284.0423 $ & $ 0.0561 $ & $ 2:1 $ & $ 1390 $ & $ 140 $ & \nodata & \nodata & $ 2500 $ &  \\ 
$ 1955.04 $ & $ 1955.02 $ & $ 26.2349 $ & $ 39.4572 $ & $ 0.0027 $ & $ 3:2 $ & $ 563 $ & $ 28 $ & \nodata & \nodata & $ 5000 $ & \tablenotemark{3}Kepler342 \\ 
$ 2038.01 $ & $ 2038.02 $ & $ 8.3053 $ & $ 12.5136 $ & $ 0.0045 $ & $ 3:2 $ & $ 1091 $ & $ 42 $ & $ 1008 $ & $ 28 $ & $ 930 $ & \tablenotemark{2,18,20}~~~~~~Kepler85 \\ 
$ 2038.02 $ & $ 2038.04 $ & $ 12.5136 $ & $ 25.2177 $ & $ 0.0076 $ & $ 2:1 $ & $ 1008 $ & $ 28 $ & \nodata & \nodata & $ 1700 $ & \tablenotemark{3,18,20}~~~~~~Kepler85 \\ 
$ 2086.01 $ & $ 2086.02 $ & $ 7.1316 $ & $ 8.9190 $ & $ 0.0005 $ & $ 5:4 $ & \nodata & \nodata & $ 715 $ & $ 28 $ & $ 3600 $ &  \tablenotemark{7}Kepler60 \\

\hline 

\end{tabular}
%**********

\label{tab:multi}
\end{table}

\addtocounter{table}{-1}
\begin{table}[]
\scriptsize  \caption{ --continued}
\begin{tabular}{|rrrr|rr|rrrrrr|}

\hline \hline
KOI  ~& KOI     & $P_1$\tablenotemark{a} \ \ & $P_2$\tablenotemark{b}  \ \ & $\Delta$\tablenotemark{c} \ \  &    j:j-1\tablenotemark{d} & 
 ${\rm P}_{\rm TTV,1}$\tablenotemark{e}  & $\sigma_{\rm P_1}$\tablenotemark{f}  & 
  ${\rm P}_{\rm TTV,2}$\tablenotemark{g}  &  
$\sigma_{\rm P_2}$\tablenotemark{h}  &
 P$^j$\tablenotemark{i} &  References\\
(Inner) & (Outer) & (Inner)   & (Outer)   &                &           &    (Inner)        &        &   (Outer)  &             &   & \\
        &            & (day)  \ \    & (day)   \ \    &                &           & (day)  	        &  (day)             & (day)  	    &   (day)               &  (day)  ~ &  \\
\hline

$ 2092.01 $ & $ 2092.03 $ & $ 57.6899 $ & $ 77.0870 $ & $ 0.0022 $ & $ 4:3 $ & $ 1280 $ & $ 130 $ & \nodata & \nodata & $ 9000 $ & \tablenotemark{3}Kepler359 \\ 
$ 2672.02 $ & $ 2672.01 $ & $ 42.9935 $ & $ 88.5081 $ & $ 0.0293 $ & $ 2:1 $ & $ 1516 $ & $ 32 $ & $ 1303 $ & $ 11 $ & $ 1500 $ & \tablenotemark{2,18}~~~Kepler396 \\ 
\hline 
\end{tabular}
%**********
\tablecomments{
$^a$Orbital period of the inner planet. 
$^b$Orbital period of the outer planet. 
$^c$Normalized distance to resonance defined as $\Delta \equiv \frac{P_2}{P_1} \frac{j-1}{j} -1$ \citep{lithwick12}.
$^d$Resonance type.
$^e$The TTV period of the inner planet (found by modeling the data). 
$^f$The TTV period uncertainty of the inner planet. 
$^g$The TTV period of the outer planet (found by modeling the data). 
$^h$The TTV period uncertainty of the inner planet.
$^i$The TTV super-period infered from the orbital periods: $ P^j \equiv \frac{1}{\left | j/P_2 - (j-1)/P_1 \right |}$.
$^j$The second strongest TTV period.\\
Reference.
$^1$\citet{cochran11}.
$^2$\citet{xie14}.
$^3$\citet{rowe14}. 
$^4$\citet{lissauer11a}.
$^5$\citet{ford12a}.
$^6$\citet{steffen12a}.
$^7$\citet{steffen13}.
$^8$\citet{carter12}.
$^9$\citet{kipping14}.
$^{10}$\citet{holman10}.
$^{11}$\citet{masuda14}.
$^{12}$\citet{fabrycky12}.
$^{13}$\citet{hutter14}.
$^{14}$\citet{lissauer13}.
$^{15}$\citet{sanchis12}.
$^{16}$\citet{diamond15}.
$^{17}$\citet{ragozzine12}.
$^{18}$\citet{yang13}.
$^{19}$\citet{lithwick12}.
$^{20}$\citet{hadden14}.
}

\label{tab:multi}
\end{table}

\begin{thebibliography}{}

\bibitem[Agol et al.(2005)]{agol05}
Agol, E., Steffen, J., Sari, R., \& Clarkson, W.\ 2005, \mnras,
359, 567

\bibitem[Agol 
\& Deck(2015)]{agol15} Agol, E., \& Deck, K.\ 2015, American Astronomical Society Meeting Abstracts, 225, \#105.07 

\bibitem[Ballard et al.(2011)]{ballard11} Ballard, S., Fabrycky, D.,
Fressin, F., et al.\ 2011, \apj, 743, 200

%\bibitem[Batalha et al.(2012)]{batalha12}
%Batalha, N.~M., Rowe, J.~F., Bryson, S.~T., et al.\ 2012,
%arXiv:1202.5852

\bibitem[Batalha et al.(2013)]{batalha13} Batalha, N.~M., Rowe, 
J.~F., Bryson, S.~T., et al.\ 2013, \apjs, 204, 24 

\bibitem[Bonomo et al.(2012)]{bonomo12} Bonomo, A.~S.,
H{\'e}brard, G., Santerne, A., et al.\ 2012, \aap, 538, A96

\bibitem[Borucki et al.(2011)]{borucki11}
Borucki, W.~J., Koch, D.~G., Basri, G., et al.\ 2011, \apj, 736,
19

\bibitem[Brown et al.(2011)]{brown11} Brown, T.~M., Latham,
D.~W., Everett, M.~E., \& Esquerdo, G.~A.\ 2011, \aj, 142, 112

\bibitem[Carter et al.(2012)]{carter12} Carter, J.~A., Agol, E.,
Chaplin,
W.~J., et al.\ 2012, Science, 337, 556

\bibitem[Claret \& Bloemen(2011)]{claret11} Claret, A., \&
Bloemen, S.\ 2011, \aap, 529, A75

\bibitem[Cochran et al.(2011)]{cochran11}
Cochran, W.~D., Fabrycky, D.~C., Torres, G., et al.\ 2011, \apjs,
197, 7

\bibitem[Dawson et al.(2014)]{dawson14} Dawson, R.~I., Johnson, J.~A., Fabrycky, D.~C., et al.\ 2014, \apj, 791, 89 

\bibitem[Deleuil et al.(2014)]{deleuil14} Deleuil, M., Almenara, J.-M., Santerne, A., et al.\ 2014, \aap, 564, A56 

\bibitem[Desert et al.(2011)]{desert11} Desert, J.-M.,
Charbonneau, D., Demory, B.-O., et al.\ 2011, \apjs, 197, 14

\bibitem[Diamond-Lowe et al.(2015)]{diamond15} Diamond-Lowe, H., 
Stevenson, K.~B., Fabrycky, D., et al.\ 2015, American Astronomical Society 
Meeting Abstracts, 225, \#438.01 

\bibitem[Fabrycky et al.(2012)]{fabrycky12}
Fabrycky, D.~C., Ford, E.~B., Steffen, J.~H., et al.\ 2012, \apj,
750, 114

\bibitem[Fabrycky et al.(2014)]{fabrycky14} 
Fabrycky, D.~C., Lissauer, J.~J., Ragozzine, D., et al.\ 2014, \apj, 790, 146 

\bibitem[Ford et al.(2011)]{ford11}     %TTV-I
Ford, E.~B., Rowe, J.~F., Fabrycky, D.~C., et al.\ 2011, \apjs,
197, 2

\bibitem[Ford et al.(2012a)]{ford12a} Ford, E.~B., Fabrycky,
D.~C., Steffen,
J.~H., et al.\ 2012a, \apj, 750, 113 % TTV-II

\bibitem[Ford et al.(2012b)]{ford12b}     %TTV-V
Ford, E.~B., Ragozzine, D., Rowe, J.~F., et al.\ 2012b, \apj, 756,
185

\bibitem[Foreman-Mackey et al.(2013)]{forman13} Foreman-Mackey, 
D., Conley, A., Meierjurgen Farr, W., et al.\ 2013, Astrophysics Source 
Code Library, 1303.002 

\bibitem[Goodman \& Weare (2010)]{goodman10} Goodman, J. \& Weare, J., 2010, Comm. App. Math. Comp. Sci., 5, 65

\bibitem[Hadden \& Lithwick(2014)]{hadden14} Hadden, S., \& Lithwick, Y.\ 2014, \apj, 787, 80 

\bibitem[Holczer et al.(2015)]{holczer15} Holczer, T., Shporer, 
A., Mazeh, T., et al.\ 2015, arXiv:1504.04028 


\bibitem[Holman \& Murray(2005)]{holman05}
Holman, M.~J., \& Murray, N.~W.\ 2005, Science, 307, 1288
%
\bibitem[Holman et al.(2010)]{holman10}
Holman, M.~J., Fabrycky, D.~C., Ragozzine, D., et al.\ 2010, Science, 330, 51

\bibitem[Howell et al.(2010)]{howell10} Howell, S.~B., Rowe, J.~F., Sherry, W., et al.\ 2010, \apj, 725, 1633

\bibitem[Howell et al.(2014)]{howell14} Howell, S.~B., Sobeck, 
C., Haas, M., et al.\ 2014, \pasp, 126, 398 


\bibitem[Jontof-Hutter et al.(2014)]{hutter14} 
Jontof-Hutter, D., Lissauer, J.~J., Rowe, J.~F., \& Fabrycky, D.~C.\ 2014, \apj, 785, 15 

\bibitem[Kipping et al.(2014)]{kipping14} Kipping, D.~M., Nesvorny, D., Buchhave, L.~A., et al.\ 2014, \apj, 784, 28 

\bibitem[Lagarais, Reeds \& Wright(1998)]{lagarais98} Lagarais,
J.C., J. A. Reeds, M. H. Wright, \& P. E. Wright, "Convergence
Properties of the Nelder-Mead Simplex Method in Low Dimensions,"
SIAM Journal of Optimization, Vol. 9 Number 1, pp. 112-147, 1998.


\bibitem[Lissauer et al.(2011a)]{lissauer11a}
Lissauer, J.~J., Fabrycky, D.~C., Ford, E.~B., et al.\ 2011a,
\nat, 470, 53

\bibitem[Lissauer et al.(2011b)]{lissauer11b}
Lissauer, J.~J., Ragozzine, D., Fabrycky, D.~C., et al.\ 2011b,
\apjs, 197, 8

\bibitem[Lissauer et al.(2013)]{lissauer13} Lissauer, J.~J., 
Jontof-Hutter, D., Rowe, J.~F., et al.\ 2013, \apj, 770, 131 

\bibitem[Lithwick et al.(2012)]{lithwick12} Lithwick, Y., Xie,
J., \& Wu, Y.\ 2012, \apj, 761, 122

\bibitem[Mandel \& Agol(2002)]{mandel02}
Mandel, K., \& Agol, E.\ 2002, \apjl, 580, L171
​\bibitem[Masuda(2014)]{masuda14} Masuda, K.\ 2014, \apj, 783, 53 

\bibitem[Marcy et al.(2014)]{marcy14} Marcy, G.~W., Isaacson, H., Howard, A.~W., et al.\ 2014, \apjs, 210, 20 


\bibitem[Mazeh et al.(2012)]{mazeh12}
Mazeh, T., Nachmani, G., Sokol, G., Faigler, S., \& Zucker, S.\ 2012, \aap, 541, A56

\bibitem[Mazeh et al.(2013)]{mazeh13} Mazeh, T., Nachmani, G., 
Holczer, T., et al.\ 2013, \apjs, 208, 16 

\bibitem[Mazeh et al.(2015a)]{mazeh15a} Mazeh, T., Holczer, T., 
\& Shporer, A.\ 2015, \apj, 800, 142 

\bibitem[Mazeh et al.(2015b)]{mazeh15b} Mazeh, T., Perets, H.~B., 
McQuillan, A., \& Goldstein, E.~S.\ 2015, \apj, 801, 3 

\bibitem[McQuillan, Aigrain \& Mazeh (2013)]{mcquillan13}
McQuillan, A., Aigrain, S., \& Mazeh, T.\ 2013, \mnras, 432, 1203

\bibitem[Nesvorn{\'y} et al.(2012)]{nesvorny12} 
Nesvorn{\'y}, D., Kipping, D.~M., Buchhave, L.~A., et al.\ 2012, Science, 336, 1133

\bibitem[Nesvorn{\'y} et al.(2013)]{nesvorny13} 
Nesvorn{\'y}, D., Kipping, D., Terrell, D., et al.\ 2013, \apj, 777, 3 
​\bibitem[Nesvorn{\'y} et al.(2014)]{nesvorny14} 
Nesvorn{\'y}, D., Kipping, D., Terrell, D., \& Feroz, F.\ 2014, \apj, 790, 31 

\bibitem[Ofir et al.(2014)]{ofir14} Ofir, A., Dreizler, S., Zechmeister, M., \& Husser, T.-O.\ 2014, \aap, 561, A103 

\bibitem[Raetz et al.(2014)]{raetz14} Raetz, S., Maciejewski, G., Ginski, C., et al.\ 2014, \mnras, 444, 1351 

%\bibitem[Raetz et al.(2013)]{raetz13} Raetz, S., Ginski, C., 
%Mugrauer, M., et al.\ 2013, Protostars and Planets VI Posters, 46 

\bibitem[Ragozzine \& Kepler Team(2012)]{ragozzine12} Ragozzine, D., \& Kepler Team
2012, AAS/Division for Planetary Sciences Meeting Abstracts, 44,
\#200.04

\bibitem[Rowe et al.(2014)]{rowe14} Rowe, J.~F., Bryson, S.~T., Marcy, G.~W., et al.\ 2014, \apj, 784, 45 

\bibitem[Rowe et al.(2015)]{rowe15} 
Rowe, J.~F., Coughlin, J.~L., Antoci, V., et al.\ 2015, \apjs, 217, 16 

\bibitem[Rowe \& Thompson(2015)]{rowe_thompson15} 
Rowe, J.~F., \& Thompson, S.~E.\ 2015, arXiv:1504.00707

\bibitem[Sanchis-Ojeda et al.(2012)]{sanchis12}
Sanchis-Ojeda, R., Fabrycky, D.~C., Winn, J.~N., et al.\ 2012,
\nat, 487, 449

\bibitem[Santerne et al.(2012)]{santerne12} Santerne, A.,
D{\'{\i}}az, R.~F., Moutou, C., et al.\ 2012, \aap, 545, A76

\bibitem[Shporer et al.(2011)]{shporer11} Shporer, A., Jenkins,
J.~M., Rowe, J.~F., et al.\ 2011, \aj, 142, 195

\bibitem[Steffen et al.(2010)]{steffen10} Steffen, J.~H.,
Batalha, N.~M., Borucki, W.~J., et al.\ 2010, \apj, 725, 1226

\bibitem[Steffen et al.(2011)]{steffen11} Steffen, J.~H., Quinn,
S.~N., Borucki, W.~J., et al.\ 2011, \mnras, 417, L31

\bibitem[Steffen et al.(2012a)]{steffen12a}
Steffen, J.~H., Fabrycky, D.~C., Ford, E.~B., et al.\ 2012a,
\mnras, 421, 2342  %TTV-III

\bibitem[Steffen et al.(2012b)]{steffen12b}
Steffen, J.~H., Ford, E.~B., Rowe, J.~F., et al.\ 2012b, \apj,
756, 186 %TTV-VI

\bibitem[Steffen et al.(2013)]{steffen13} Steffen, J.~H.,
Fabrycky, D.~C., Agol, E., et al.\ 2013, \mnras, 428, 1077  %TTV-VII

\bibitem[Szab{\'o} et al.(2011)]{szabo11}
Szab{\'o}, G.~M., Szab{\'o}, R., Benk{\H o}, J.~M., et al.\ 2011,
\apjl,
736, L4

\bibitem[Szab{\'o} et al.(2012)]{szabo12mnras}
Szab{\'o}, G.~M., P{\'a}l, A., Derekas, A., et al.\ 2012, \mnras,
421, L122

\bibitem[Szab{\'o} et al.(2013)]{szabo13} 
Szab{\'o}, R., Szab{\'o}, G.~M., D{\'a}lya, G., et al.\ 2013, \aap, 553, A17 

\bibitem[Tamuz, Mazeh \& North (2006)]{tamuz06}
Tamuz, O., Mazeh, T., \& North, P.\ 2006, \mnras, 367, 1521

\bibitem[Tingley et al.(2011)]{tingley11} Tingley, B., Palle, E.,
Parviainen, H., et
al.\ 2011, \aap, 536, L9

\bibitem[Van Eylen et al.(2014)]{van-eylen14} Van Eylen, V., Lund, M.~N., Silva Aguirre, V., et al.\ 2014, \apj, 782, 14

\bibitem[Va{\v n}ko et al.(2013)]{vavnko13} Va{\v n}ko, M., Maciejewski, G., Jakub{\'{\i}}k, M., et al.\ 2013, \mnras, 432, 944 ​

\bibitem[Wang et al.(2012)]{wang12} Wang, S., Ji, J.,
\& Zhou, J.-L.\ 2012, \apj, 753, 170

\bibitem[Wang et al.(2014)]{wang14} Wang, J., Xie, J.-W., Barclay, T., \& Fischer, D.~A.\ 2014, \apj, 783, 4

\bibitem[Weiss et al.(2013)]{weiss13} Weiss, L.~M., Marcy, G.~W., Rowe, J.~F., et al.\ 2013, \apj, 768, 14 

\bibitem[Winn(2011)]{winn11}
Winn, J.~N.\ 2011, European Physical Journal Web of Conferences,
11, 5002

%\bibitem[Xie(2012)]{xie12} Xie, J.-W.\ 2012, arXiv:1208.3312
\bibitem[Xie(2013)]{xie13} 
Xie, J.-W.\ 2013, \apjs, 208, 22

\bibitem[Xie(2014)]{xie14} Xie, J.-W.\ 2014, \apjs, 210, 25 

%\bibitem[Xie et al.(2014)]{xie14b} Xie, J.-W., Wu, Y., \& Lithwick, Y.\ 2014, \apj, 789, 165 

\bibitem[Yang et al.(2013)]{yang13} 
Yang, M., Liu, H.-G., Zhang, H., Yang, J.-Y., \& Zhou, J.-L.\ 2013, \apj, 778, 110 

\end{thebibliography}
\end{document}